\documentclass[10pt,letterpaper]{article}
\usepackage[top=0.85in,left=2.75in,footskip=0.75in,marginparwidth=2in]{geometry}
\pdfoutput=1

\usepackage[utf8]{inputenc}

\usepackage{cite}

\usepackage{nameref,hyperref}

\usepackage{microtype}
\DisableLigatures[f]{encoding = *, family = * }

\raggedright
\usepackage{changepage}

\usepackage[aboveskip=1pt,labelfont=bf,labelsep=period,singlelinecheck=off]{caption}

\usepackage{lastpage,fancyhdr,graphicx}
\usepackage{epstopdf}
\fancyheadoffset[L]{2.25in}
\fancyfootoffset[L]{2.25in}

\usepackage{color}

\definecolor{Gray}{gray}{.25}

\usepackage{graphicx}

\usepackage{sidecap}

\usepackage{wrapfig}
\usepackage[pscoord]{eso-pic}
\usepackage[fulladjust]{marginnote}
\reversemarginpar

\usepackage{amsmath}
\usepackage{amssymb}
\usepackage{siunitx}
\usepackage{enumitem}

\DeclareSIUnit\micron{\micro\metre}
\DeclareSIUnit\adu{adu}
\DeclareSIUnit\photon{photon}

\newcommand*{\figfont}{\fontfamily{ptm}\selectfont}

\setlist[enumerate,1]{label=\arabic*.,ref=\arabic*}
\setlist[enumerate,2]{label=(\alph*),ref=\arabic{enumi} (\alph*)}

\begin{document}
\vspace*{0.35in}

% title goes here:
\begin{flushleft}
{\Large
\textbf\newline{Continuous Diffraction of Molecules and Disordered Molecular Crystals.}
}
\newline
% authors go here:
\\
Henry N. Chapman\textsuperscript{1,2,3,*},
Oleksandr M. Yefanov\textsuperscript{1},
Kartik Ayyer\textsuperscript{1},
Thomas A. White\textsuperscript{1},
Anton Barty\textsuperscript{1},
Andrew Morgan\textsuperscript{1},
Valerio Mariani\textsuperscript{1}
Dominik Oberthuer\textsuperscript{1}
Kanupriya Pande\textsuperscript{1}
\\
\bigskip
\bf{1} Center for Free-Electron Laser Science, DESY, 22607 Hamburg, Germany
\\
\bf{2} Department of Physics University of Hamburg, 22761 Hamburg, Germany
\\
\bf{3} Centre for Ultrafast Imaging, 22607 Hamburg, Germany
\\
\bigskip
* henry.chapman@desy.de

\end{flushleft}

\section*{Abstract}
  The diffraction pattern of a single non-periodic compact object, such as a molecule, is
  continuous and is proportional to the square modulus of the Fourier transform of that
  object.  When arrayed in a crystal, the coherent sum of the continuous diffracted
  wave-fields from all objects gives rise to strong Bragg peaks that modulate the
  single-object transform.  Wilson statistics describe the distribution of continuous
  diffraction intensities to the same extent that they apply to Bragg diffraction.  The
  continuous diffraction obtained from translationally-disordered molecular crystals
  consists of the incoherent sum of the wave-fields from the individual rigid units (such
  as molecules) in the crystal, which is proportional to the incoherent sum of the
  diffraction from the rigid units in each of their crystallographic orientations.  This
  sum over orientations modifies the statistics in a similar way that crystal twinning
  modifies the distribution of Bragg intensities.  These statistics are applied to
  determine parameters of continuous diffraction such as its scaling, the beam coherence,
  and the number of independent wave-fields or object orientations contributing.
  Continuous diffraction is generally much weaker than Bragg diffraction and may be
  accompanied by a background that far exceeds the strength of the signal.  Instead of
  just relying upon the smallest measured intensities to guide the subtraction of the
  background it is shown how all measured values can be utilised to estimate the
  background, noise, and signal, by employing a modified ``noisy Wilson'' distribution
  that explicitly includes the background.  Parameters relating to the background and
  signal quantities can be estimated from the moments of the measured intensities.  The
  analysis method is demonstrated on previously-published \cite{Ayyer:2016} continuous
  diffraction data measured from imperfect crystals of photosystem II.

\section{Introduction}
\label{sec:intro}
\begin{figure}
\vspace{.5cm} % set vertical space between text and figure
\begin{adjustwidth}{-2in}{0in}
  \setlength{\unitlength}{1mm}
  \begin{picture}(120,110)(0,0)  %1.5 column figure
    \put(0,60){
      \includegraphics[width=5cm,keepaspectratio]{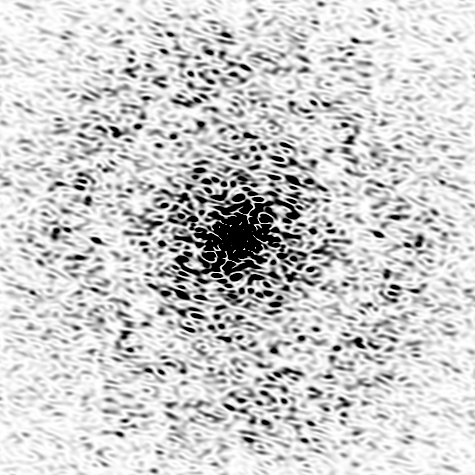}}
    \put(0,5){
      \includegraphics[width=5cm,keepaspectratio]{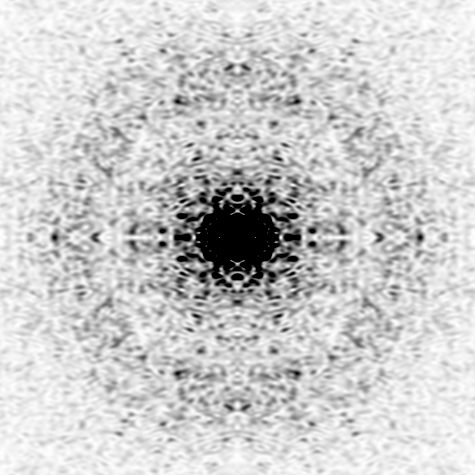}}
    \put(53,55){
      \includegraphics[height=5.5cm,keepaspectratio]{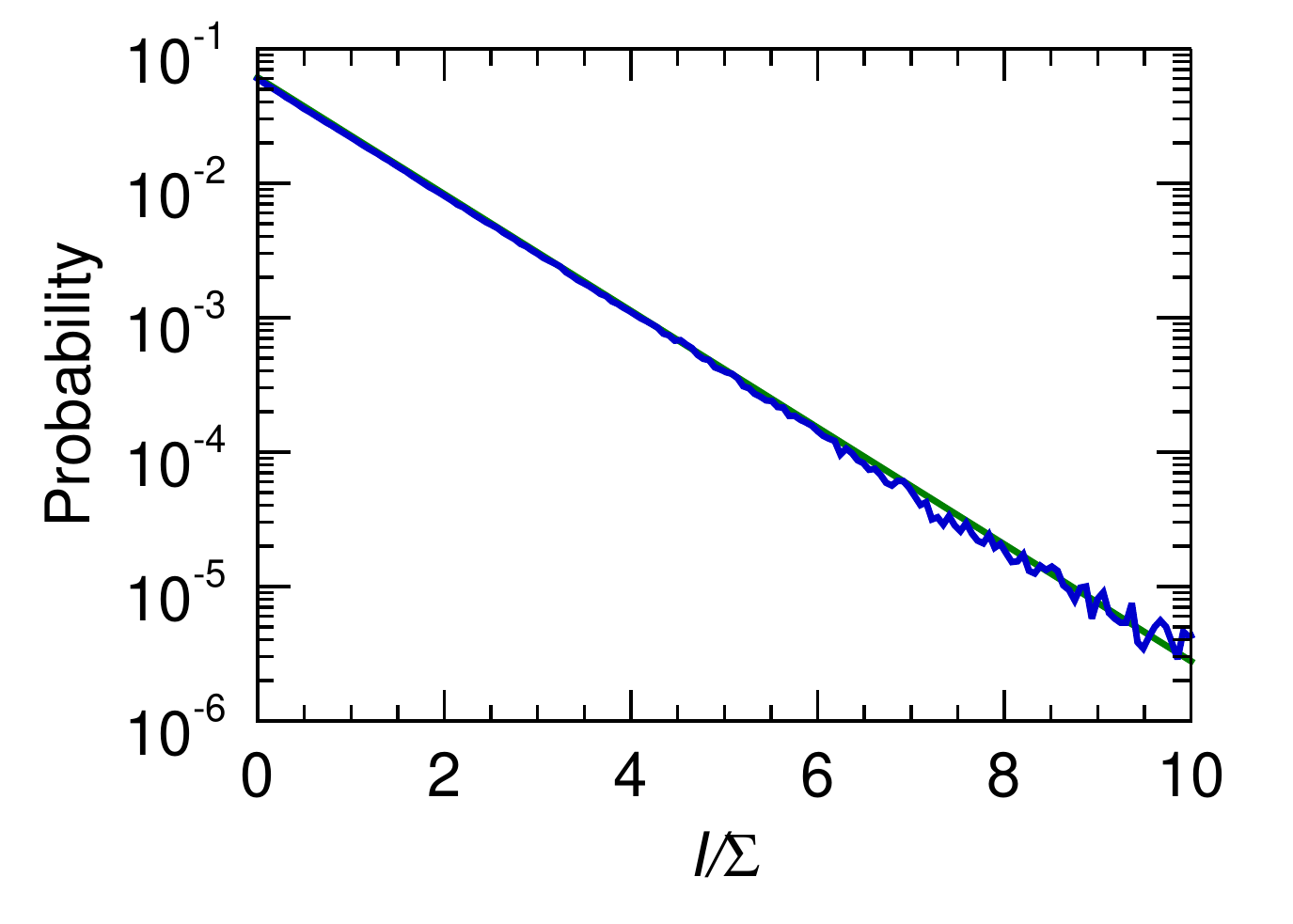}}
    \put(53,0){
      \includegraphics[height=5.5cm,keepaspectratio]{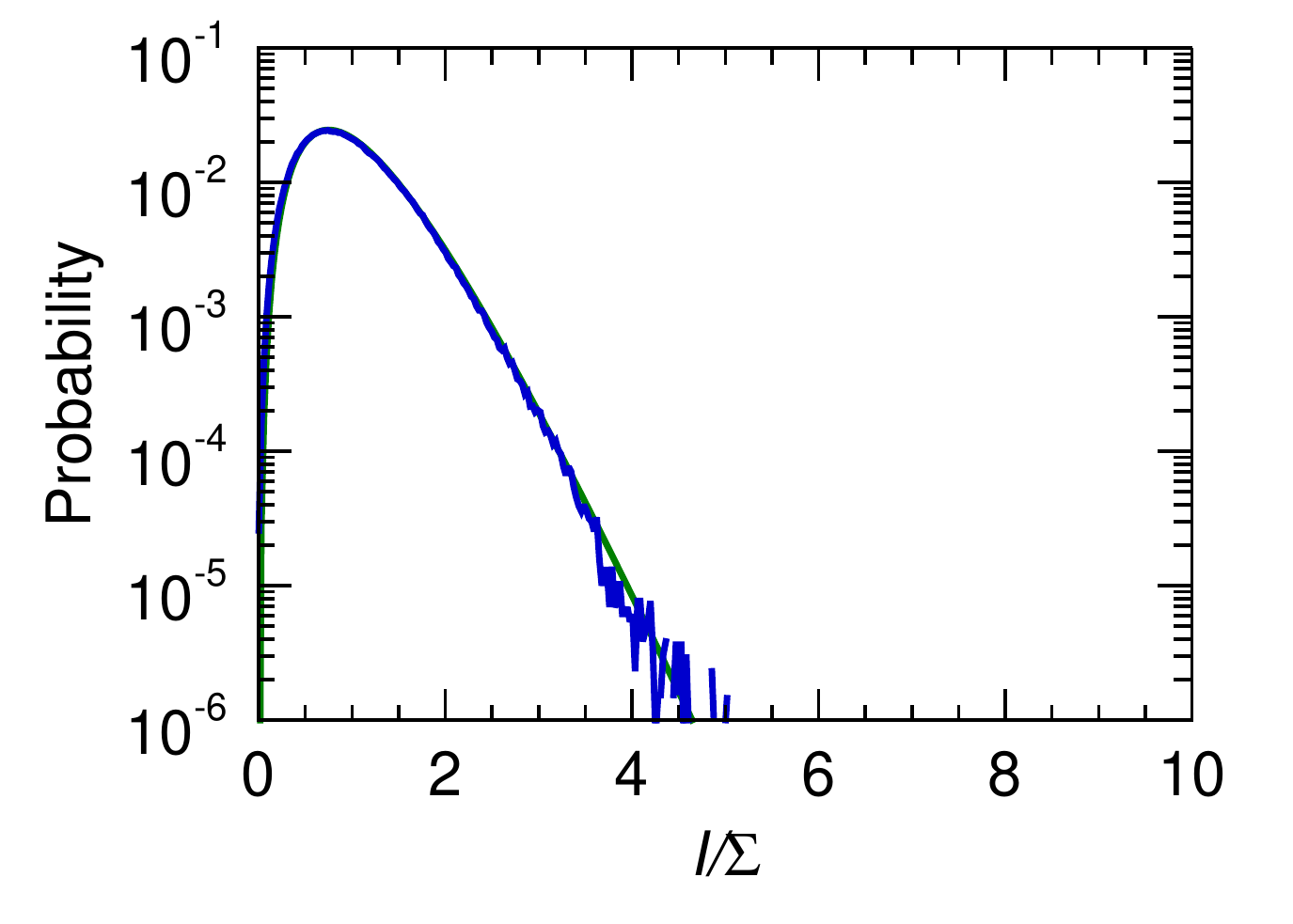}}
   \put(4,103){\figfont(\textit{a})}
    \put(116,102){\figfont(\textit{b})}
    \put(4,48){\figfont(\textit{c})}
    \put(116,47){\figfont(\textit{d})}
 \end{picture}
  \caption{(a) A central section of diffraction intensities of PS II complex from a
    calculation at $475^3$ points in a three-dimensional array of reciprocal space,
    centered on the origin.  The resolution at the center edge of the simulated array is
    \SI{0.33}{\per\angstrom}, and intensity samples are spaced by
    \SI{0.0014}{\per\angstrom}, which is twice the Nyquist sampling of the intensities of
    an object \SI{178}{\angstrom} wide.  (b) Histogram of intensity values at samples in a
    shell of reciprocal space of width \SI{0.0325}{\per\angstrom} and centered at
    $q = \SI{0.23}{\per\angstrom}$ (blue), and the negative exponential of
    Eqn.~(\ref{eq:1}) in green.  (c) A central section from the same 3D array after
    applying symmetry operations of the 222 point group, displayed on the same colour map
    as (a), which ranges from zero counts in white to maximum counts in black.  The section is only perpendicular to one two-fold axis, which is
    horizontal in this view.  Visually, the pattern has lower contrast than (a) which is
    confirmed in the histogram of intensity values (d) which has a smaller width
    (i.e. smaller variance) for the same reciprocal space shell as for (b). The Gamma
    distribution $p(I; 4)$ of Eqn. (\ref{eq:3}) is given in green. \textcopyright The Authors licensed under CC BY
4.0}
  \label{fig:1}
\end{adjustwidth}
\end{figure}
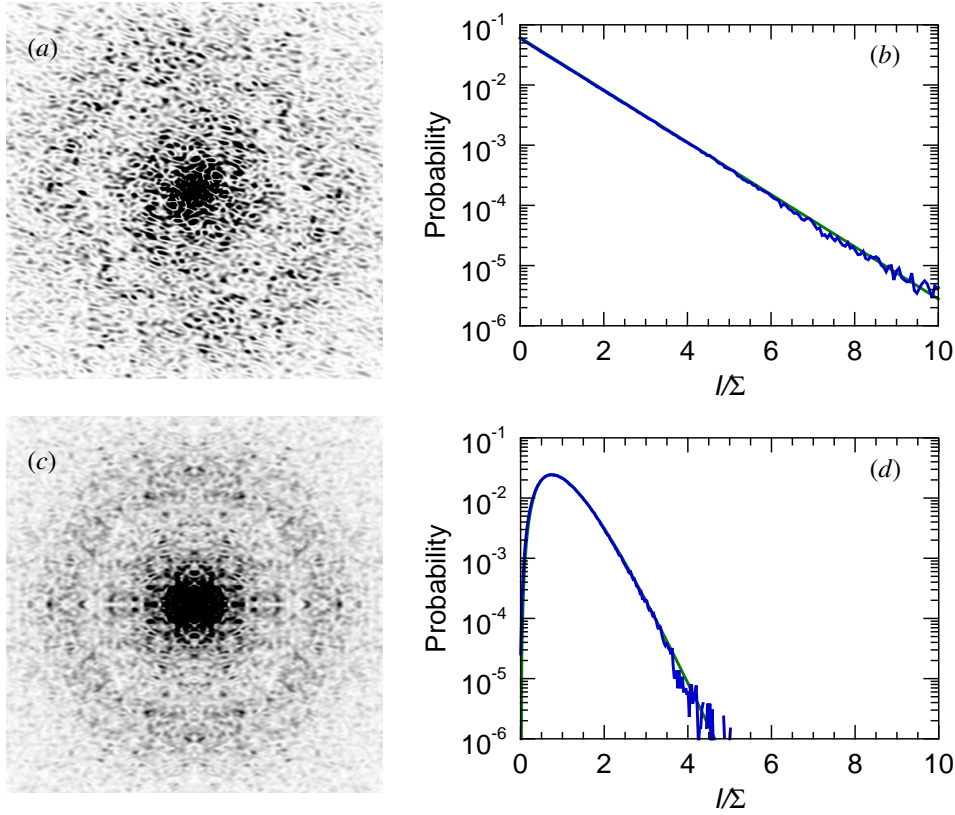

The statistics of diffraction intensities in protein crystallography have guided data
analysis and data verification, such as providing a basis for a treatment of negative
diffraction intensities \cite{French:1978}, and the identification of crystal symmetries
\cite{Wilson:1949,Rogers:1950} and crystal twinning \cite{Rees:1980}.  The probability
distribution of Bragg intensities in the X-ray diffraction pattern of a molecular crystal
was first considered by Wilson \cite{Wilson:1949}, now referred to in the field as Wilson
statistics.  The assumptions on which the derivation of these statistics depend, namely
that the atoms in the molecule are random and independent, equally apply to the case of
the continuous coherent diffraction of a single molecule \cite{Huldt:2003}.  Such
diffraction does not contain Bragg peaks since the object is not periodic, but instead is
proportional to the square modulus of the continuous Fourier transform of the molecule,
such as the computed patterns shown in Fig.~\ref{fig:1}.  These are also known as speckle
patterns, and are similar to patterns that can be observed by shining an optical laser
beam on a uniformly rough surface such as a painted wall.  In both cases the contrast of
the speckles is high and their size, which is roughly homogeneous, is inversely
proportional to the size of the illuminated object (the diameter of the molecule or laser
beam). This similarity holds for the statistical description of the intensities.  Speckle
patterns of laser beams reflected from rough surfaces also obey Wilson statistics and a
body of literature, parallel to that of macromolecular crystallography, has presented
derivations of intensity distributions and their experimental verifications, explored
methods to reduce the contrast of speckle in cases where it is considered a nuisance, and
utilised the statistics to determine coherence or roughness properties 
\cite{Dainty:1976,Goodman:1985,Goodman:2007}.  Although optical speckle patterns 
were observed in the
nineteenth century and explained by Laue \cite{vonLaue:1914} and
Lord Rayleigh \cite{Rayleigh:1918}, it was not until the invention of the laser that detailed
examination was taken up.  It is interesting that there was no such hindrance in X-ray
crystallography, where X-ray sources provided beams that could be made coherent
enough---using collimating devices---to give rise to coherent diffraction patterns
(consisting of Bragg peaks obeying Wilson statistics) from molecular crystals.

With the more recent measurements and studies of continuous diffraction patterns of
molecular samples, such as X-ray or electron diffraction of aligned molecules
\cite{Kupper:2014,Hensley:2012}, single un-oriented molecules and viruses 
\cite{Seibert:2011,Aquila:2015}, or translationally-disordered crystals \cite{Ayyer:2016}, an understanding of
the distribution of continuous diffraction intensities is required for the same reasons as
mentioned above for crystallography.  The motivation for these studies is clear: the
continuous diffraction, when sampled at or beyond its Nyquist frequency, provides a
complete description of the diffracted wave-field, and directly gives access to the full
unaliased autocorrelation function of the object.  Under most conditions, the information
content of the measurable diffraction intensities exceeds that required to completely
describe the electron density of the sample, allowing iterative algorithms to directly
retrieve the diffraction phases without the need for prior knowledge about the object, a
method known as diffractive imaging \cite{Thibault:2010}.

Perhaps the most crucial aspect of primary data analysis of continuous diffraction
measurements is to determine the contribution of any incoherent background to the pattern.
Unlike the narrow Bragg peaks in the diffraction patterns of crystals, which can be
distinguished from a slowly-varying background reasonably well, the morphology of a
continuous speckle pattern makes this discrimination less straight forward.  In addition,
without the ``coherency gain'' that concentrates intensity into Bragg peaks \cite{Sayre:1995},
the continuous diffraction is much weaker per pixel than Bragg diffraction and any incoherent
background may far surpass the strength of the signal of interest.  As we shall see, the
common assumption that local minima of the diffraction pattern should be zero and thus can
be attributed to background is not always correct, especially when particles are oriented
in several discrete orientations as can be the case for continuous diffraction of a
translationally-disordered crystal.

Indeed, the stimulus for this work was to better treat the continuous X-ray diffraction
patterns that were measured from imperfect crystals of photosystem II (PS II)
\cite{Ayyer:2016}.  The presence of Bragg peaks in a diffraction pattern indicates a high
degree of correlation of objects in a regular arrangement, over a number of objects that
is inversely proportional to the width of the Bragg peak and to a precision determined by
the highest scattering angles (the highest resolution) to which those Bragg peaks exist.
Crystals of large protein complexes, such as membrane proteins, often only give Bragg
diffraction to a limited resolution, such as \SI{5}{\angstrom}, for example.  The lack of correlation
beyond this limit may be due to the objects being different from one another at this
length scale, or that the objects are structurally alike but randomly displaced from the
regular lattice.  Combinations of these effects may also occur, including rotational
disorder of rigid objects or translational disorder of conformationally-varying molecules,
but it is the lack of translational correlation that causes the Bragg peaks to vanish
beyond a certain resolution. Scattering from the crystal still occurs beyond the
highest-angle Bragg peaks, since the atomic scattering factors and numbers of atoms do not
change just because the crystal is not ordered.  In the case of translational disorder of
repeating rigid units, the diminishing Bragg efficiency with increasing resolution is
balanced by an increase of the incoherent sum of the continuous diffraction patterns of
those units.  This continuous diffraction is identical to that from a gas of molecules
that are oriented in the discrete crystallographic orientations, and is amenable to direct
phasing using iterative methods as demonstrated by Ayyer \emph{et al.} \cite{Ayyer:2016}.

In this paper we review, in Sec.~\ref{sec:stats-mol}, the statistics of continuous
speckle patterns of ensembles of molecules aligned in one or several discrete orientations
for both centric and acentric structures.  These statistics follow familiar distributions
of Bragg intensities of twinned crystals, since the intensities arise in both cases from
an incoherent sum of independent coherent diffraction patterns.  As is also well known in
the field of speckle metrology, this incoherent sum reduces the contrast of the pattern
and increases the most common intensity value from zero.  The consideration of coherence
is more critical for continuous diffraction than for Bragg diffraction, since partial
coherence reduces the contrast of speckle patterns and modifies their statistics, with
consequences on the ability to phase them.  We verify the predicted distribution of
intensities of partially coherent diffraction patterns by simulation.  Even when using a
coherent source such as an X-ray free-electron laser, the finite size of pixels in the
diffraction detector gives the same effect, by reciprocity, as partial coherence.

In Sec.~\ref{sec:stats-crystal} we consider the statistics of the continuous diffraction
of translationally- disordered crystals or other collections of molecules in discrete
orientations.  While this continuous diffraction follows the point-group symmetry of the
crystal, as does the Bragg diffraction, the distribution of intensities may be different
to that of the Bragg intensities due to the incoherent addition of diffraction from rigid
units, compared with the coherent addition of scattering from the entire contents of the
unit cell that gives rise to Bragg diffraction.  We consider some special central sections
of reciprocal space that are perpendicular to symmetry axes of the point group and which
possess distributions that do not occur in diffraction of twinned crystals, and give some
examples to highlight how the statistics of the continuous diffraction could indicate or
verify the symmetry of the rigid unit in a translationally-disordered crystal.

We derive the distributions of diffraction intensities consisting of the continuous
diffraction of discretely-oriented structures accompanied by an incoherent background, in
Sec.~\ref{sec:noisy-Wilson}.  While the implications of subtracting background from Bragg
data and the aforementioned treatment of negative intensities have been long considered
\cite{French:1978}, there has not been a detailed investigation of the case where the
standard deviation of the background is a significant fraction of or is larger than the
diffraction signal.  We consider first the case where the background is normally
distributed and give an explicit expression for the distribution of the intensities.
Although we cannot obtain a similar expression for the case of photon-counting
measurements, where the background follows a Poisson distribution, we determine the
moments of intensities for both cases, and show that the parameters of the background
(mean and variance in the case of the Normal distribution) and the scaling of the signal
can be solved from the moments of the measured intensities, given that the number of
independent orientations of the diffracting structures is known.  In Sec.~\ref{sec:analysis-patterns} we apply this
estimation of parameters in shells of reciprocal space for diffraction patterns measured
from translationally-disordered PS II crystals, as previously published \cite{Ayyer:2016}.
In that work, background was estimated simply from the average intensities in circular
rings of constant $q$, but we show here improved results obtained by estimating the background level
and the diffraction signal scaling from the moments of the intensities.  In
Sec.~\ref{sec:analysis-3D}, we examine the statistics of the aggregated three-dimensional
continuous diffraction from the scaled and background-corrected PS II patterns and find
that the acentric intensities fit a distribution corresponding to the incoherent summation
of four independent structures, consistent with the four crystallographic orientations of
the PS II dimer.  Finally, in Sec.~\ref{sec:comparison} an improved cross correlation is
observed between the background-corrected continuous diffraction and simulated diffraction
from an atomic model of a PS II dimer.

\section{Statistics of diffraction intensities of aligned molecules}
\label{sec:stats-mol}
The distribution of intensities measured in the diffraction pattern of a molecular
structure can be derived by considering the coordinates $\boldsymbol{x}_i$ of atoms in the
object to be rationally independent or random \cite{Schmueli:1995}.  Under those
conditions, for a structure that is not centrosymmetric, and for a large number of atoms
with approximately equal atomic scattering factors, the contributions of atoms to the
phases of the diffraction amplitudes, $\theta_i=2 \pi \boldsymbol{q} \cdot \boldsymbol{x}_i$, are
uniformly distributed (between 0 and $2\pi$) for any given photon momentum transfer
$\boldsymbol{q}$.  The distribution of the magnitudes of the diffraction amplitudes, each a
sum over the contributions from each atom, can then be estimated by analogy to a random
walk in the complex plane or by application of the central limit theorem
\cite{Schmueli:1995,Dainty:1976}.  Using the latter approach and for the case of
unpolarised radiation, it is seen that at a particular $q$ (=$|\boldsymbol{q}|$) shell (where atomic scattering
factors do not vary), the real and imaginary parts of the complex-valued diffraction
amplitudes are both normally distributed with a mean of zero and mean square proportional
to $\langle \sin^2 \theta_i \rangle = 1/2$ or $\langle \cos^2 \theta_i \rangle = 1/2$.  The
diffraction intensities, $I$, are equal to the sum of the squares of the real and imaginary
parts.  The distribution of a sum of squares of $k$ independent standard normal random
variables is given by the chi-squared distribution of order $k$, which can be scaled to any
particular variance \cite{Papoulis:1991}.  Thus the intensities $I$ in a given shell of $\boldsymbol{q}$ are
distributed according to the scaled chi-squared distribution of order 2, with a
probability distribution function given by
\begin{equation}
  \label{eq:1}
  p(I) = \frac{1}{\Sigma} \, \exp(-I/\Sigma), \;\; I > 0
\end{equation}
The mean of the intensity is $\Sigma$, which was set by the choice of the variance of the
individual normal distributions.  The variance of this distribution is $\Sigma^2$ and the most
common value (the mode) of $I$ is zero. (This distribution is also referred to as a negative
exponential distribution of scale $\Sigma$, an Erlang distribution with shape parameter 1 and
rate $1/\Sigma$, or a Gamma distribution with shape parameter 1 and scale $\Sigma$.  In
the notation of statistics, $I \sim \mathrm{Gamma}(1, \Sigma)$, meaning that the random
variable $I$ has the probability distribution of $\mathrm{Gamma}(1, \Sigma)$.)

When the structure is real and centrosymmetric, then the phases of the diffraction amplitudes take
on values of 0 or $\pi$, which is to say that the imaginary parts of the diffraction
amplitudes are zero.  This is true also for diffraction amplitudes on a central section
(or zone) of reciprocal space perpendicular to any projection of the structure that is
centrosymmetric, such as a projection along the 2-fold symmetry axis of a crystal.  By the
Fourier slice theorem, the
Fourier transform of a real-space projection, an integration along a real-space direction
of an object, is equal to the central section perpendicular to that direction of the
three-dimensional transform of the object.  The real parts of the diffraction amplitudes
of the centrosymmetric object or projection will still follow a normal distribution, and
thus the intensities, equal to their squares, will have a scaled chi-squared distribution
of order 1 (which can also be derived from the normal distribution by a change of
variable), given by
\begin{equation}
  \label{eq:2}
  p(I_C) = \frac{1}{\sqrt{2 \pi I_C \Sigma}} \, \exp(-I_C/2\Sigma), \;\; I_C > 0
\end{equation}
with a mean of $\Sigma$, variance $2\Sigma^2$, and mode of zero.  The intensities $I_C$
are referred to as centric.

Equations (\ref{eq:1}) and (\ref{eq:2}) are the well-known Wilson statistics applicable
to crystals of P1 symmetry and $\mathrm{P}\bar{1}$ symmetry, respectively
\cite{Schmueli:1995}, also referred to as Rayleigh statistics in the field of speckle
metrology \cite{Dainty:1976, Goodman:2007}.  The derivation of these statistics make no
assumption of crystallinity of the sample and hence are equally applicable to the
continuous diffraction of a single object \cite{Huldt:2003} (such as the calculated
single-molecule diffraction of a PS II complex shown in Fig.~\ref{fig:1} (a)) as they are to the Bragg
diffraction from a protein crystal, coherent diffraction from atomic glasses
\cite{Hruszkewycz:2012}, or that resulting from the reflection of a monochromatic and
polarised laser beam from a random rough surface \cite{Dainty:1976,Goodman:2007}.  The
applicability of Wilson statistics to single-molecule diffraction is demonstrated in
Fig.~\ref{fig:1} (b), where the distribution of simulated point-sampled intensities in a
shell of reciprocal space is plotted for the single PS II complex.

Equations (\ref{eq:1}) and (\ref{eq:2}) predict that the most common intensity value for
molecular diffraction is zero.  This is consistent with the view of a single-molecule
diffraction pattern as made up of speckles that are surrounded by low values, such as seen
in Fig.~\ref{fig:1} (a).  The speckle nature of the diffraction is less easily observed in Bragg
diffraction, but is certainly true given that the diffraction pattern of a crystal can be
described as a modulation of the continuous diffraction of the unit cell with the
reciprocal lattice.  A difference to Bragg diffraction, however, is that single-molecule
diffraction can be more readily affected by the spatial coherence of the illumination or,
equivalently, the detector pixel shape function, as discussed below in
Sec.~\ref{sec:coherence}. 

\begin{figure}
  \setlength{\unitlength}{1mm}
  \begin{picture}(120,80)(0,0)  %1.5 Column figure
    \put(0,40){
      \includegraphics[width=5.8cm, keepaspectratio]{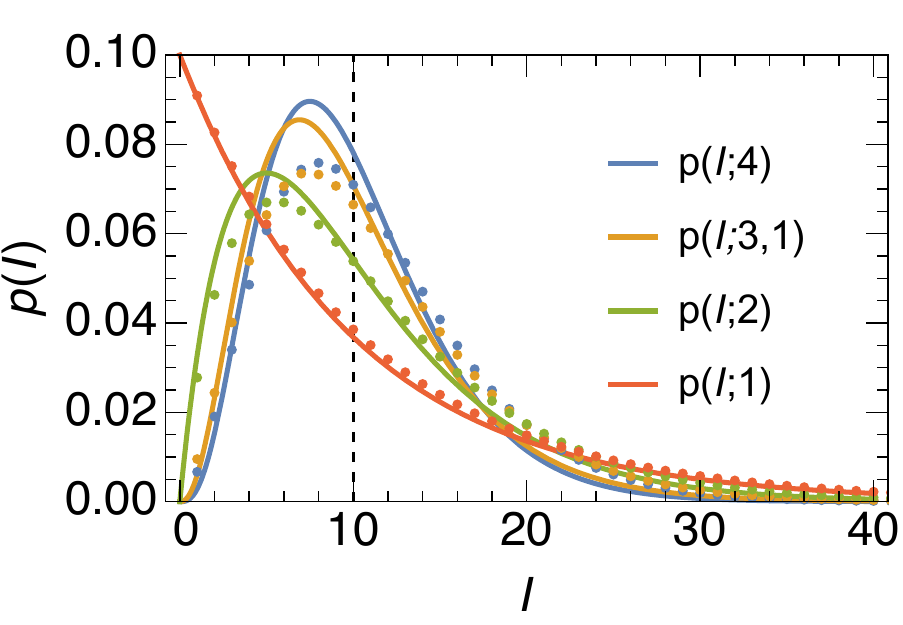}
    }
    \put(60,40){
       \includegraphics[width=5.8cm, keepaspectratio]{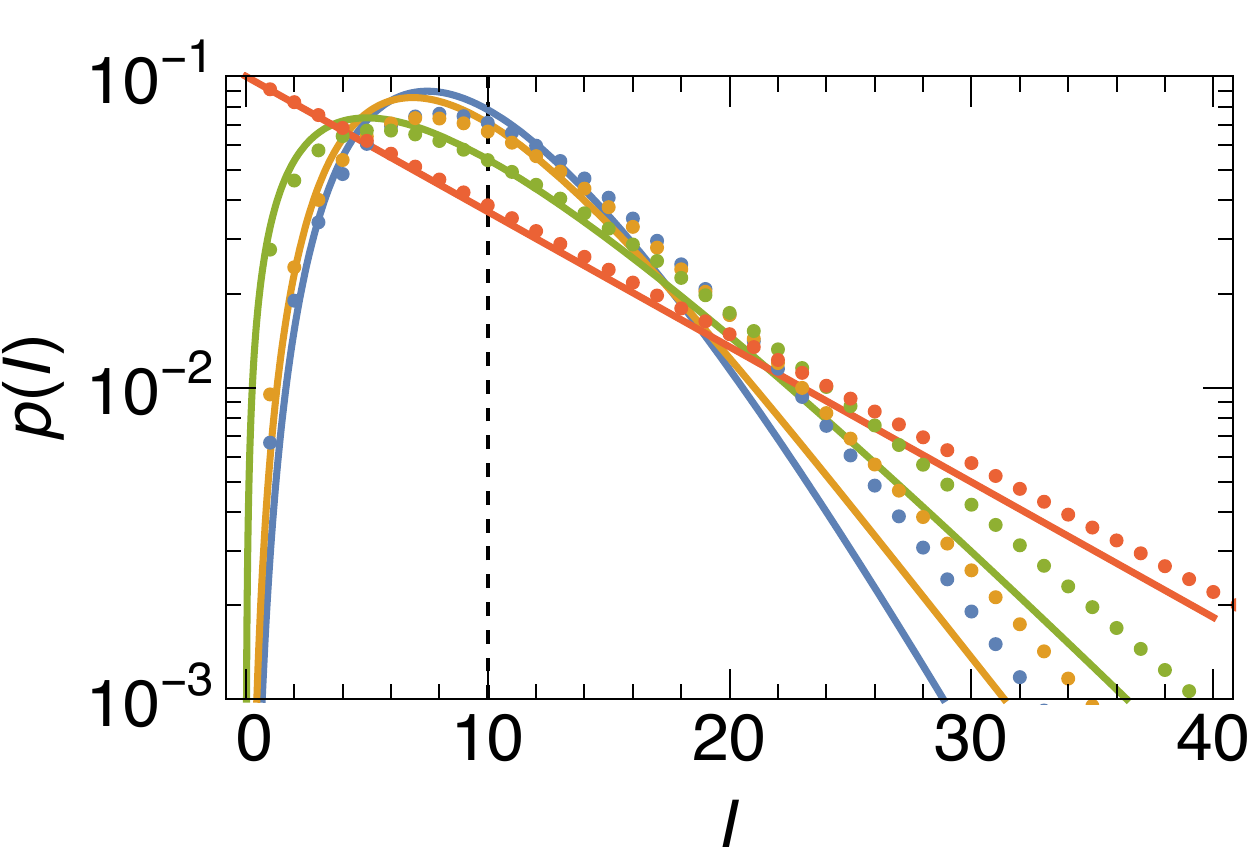}
    }
    \put(0,0){
      \includegraphics[width=5.8cm, keepaspectratio]{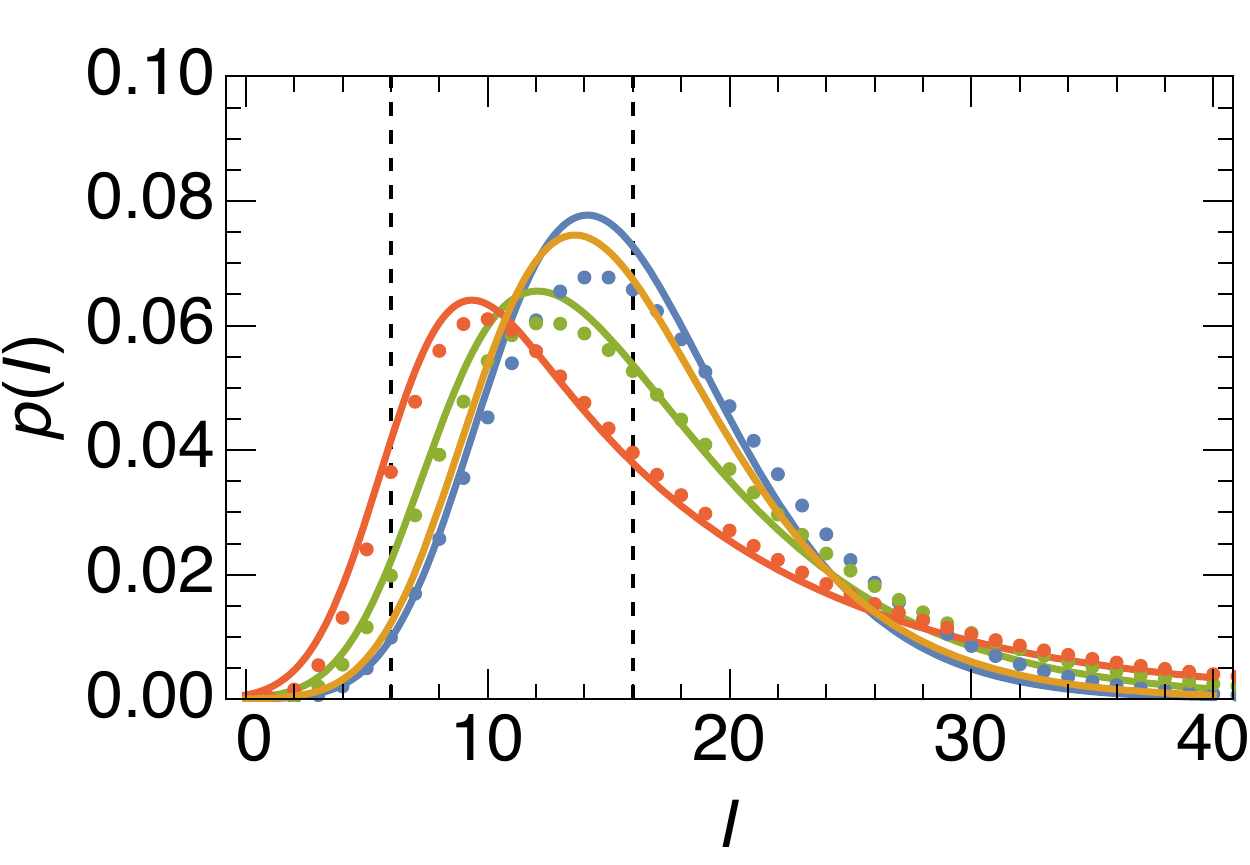}
    }
    \put(60,0){
        \includegraphics[width=5.8cm, keepaspectratio]{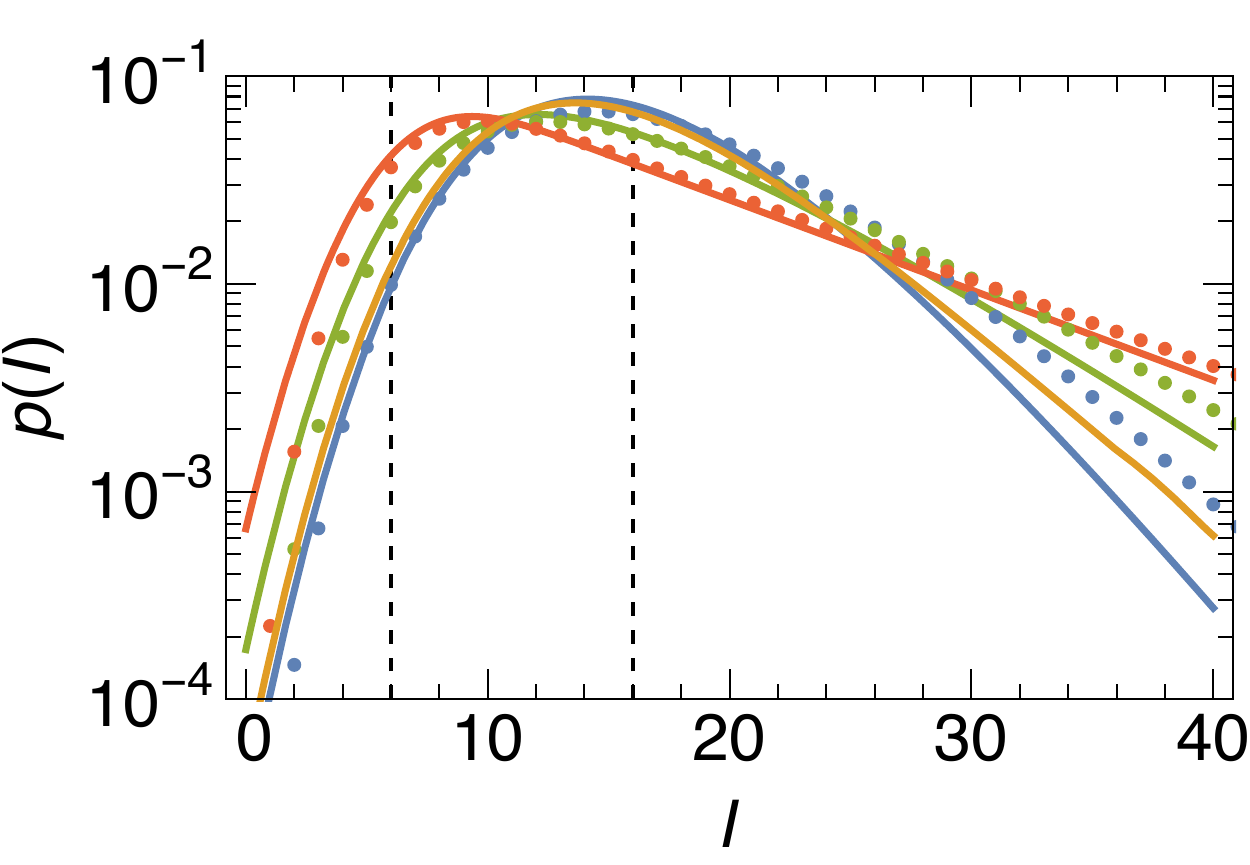}
    }
    \put(52,73){\figfont(\textit{a})}
    \put(112,73){\figfont(\textit{b})}
    \put(52,33){\figfont(\textit{c})}
    \put(112,33){\figfont(\textit{d})}
  \end{picture}
  \caption{Plots of the distributions of diffraction intensities of disordered crystals
    with $\mathrm{P}2_12_12_1$ symmetry with a mean of $\Sigma = 10$ for
    continuously-distributed values (lines) and for photon counting (dots).  Linear (a)
    and logarithmic (b) plots of $p(I;4)$, $p(I;3,1)$, $p(I;2)$, and $p(I,1)$, all without
    background, corresponding to the distributions for acentric continuous diffraction
    intensities, partly-centric continuous diffraction intensities, centric continuous
    diffraction intensities (on central sections normal to crystallographic two-fold
    axes), and acentric Bragg intensities, respectively. Linear (c) and logarithmic (d)
    plots of the “noisy Wilson” distributions for the same cases with a background of
    $\mu = 6$ and $\sigma = 2.45$.  The dashed vertical lines correspond to the value of
    the mean signal, $\Sigma$, in (a) and (b) and the mean background, $\mu$, and
    background plus signal, $\mu+\Sigma$, in (c) and (d). \textcopyright The Authors licensed under CC BY
4.0}
  \label{fig:distributions}
\end{figure}
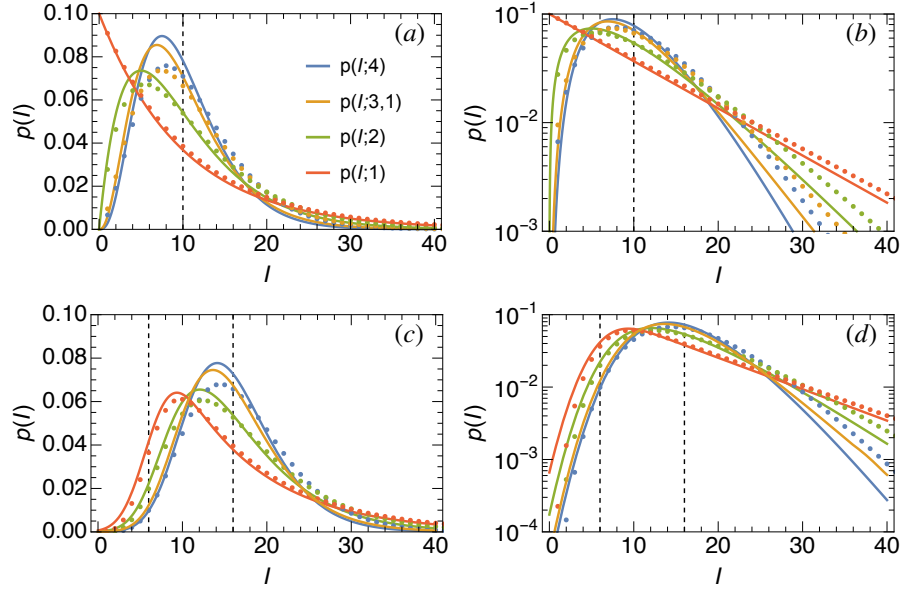

One insightful application of Wilson statistics is to identify the presence of crystal
twinning purely from observations of diffraction intensities \cite{Rees:1980,Rees:1982}.
The same tests can be carried out on diffraction of oriented single molecules.  For
example, alignment of molecules with an AC laser field gives rise to equal populations of
molecules aligned parallel and anti-parallel to a lab-frame vector \cite{Spence:2004b}.
As with the case of merohedral twinning of a crystal, the diffraction intensities of the
two populations sum incoherently.  As long as the intensities at
$\boldsymbol{R} \cdot \boldsymbol{q}$ are independent to those at $\boldsymbol{q}$ for the rotation
operator $\boldsymbol{R}$ describing the twinning, then the distribution of the summed
intensities follows the sum of two scaled chi-square distributions of order 2 (for a
non-centrosymmetric object), which from the definition of a chi-square distribution is a
chi-square distribution of order 4.  In general, diffraction intensities from $N$ equal
twin fractions (each with mean $\Sigma/N$) is given by a scaled chi-square distribution of
order $2N$, also equivalent to $I \sim \mathrm{Gamma}(N, \Sigma/N)$, with a probability
distribution function
\begin{equation}
  \label{eq:3}
  p(I;N) = \frac{N^N \, I^{N-1}}{\Sigma^N \, \Gamma(N)} \, \exp(-NI/\Sigma), \;\; I> 0
\end{equation}
where $\Gamma$ is the Euler Gamma function, equal to $(N-1)!$ for whole numbers of $N$.
Some plots of $p(I;N)$ are given in Fig.~\ref{fig:distributions} (a).  The most common
value for the continuous diffraction intensity for $N > 1$ orientations is not zero, but $(N -1)
\Sigma/N$.  The mean of this distribution is $\Sigma$ and the variance is reduced compared to the
single object to a value of $\Sigma^2/N$ (see Table~\ref{tab:1}).  The reduction in variance is quite
noticeable in the simulated diffraction intensities shown in Fig.~\ref{fig:1} (c), which is the
calculation of the incoherent sum of diffraction of PS II complexes oriented in the four
different orientations of the 222 space group (any orientation is related to
another through a rotation of \ang{180} about one of the three orthogonal axes).  For the same
mean, the standard deviation is halved in this case and the distribution of the simulated
intensities agrees with Eqn. (\ref{eq:3}) for $N = 4$ as seen in Fig.~\ref{fig:1} (d).

In the case of centrosymmetric objects in $N$ unique orientations (whether due to crystal
symmetry or twinning), the probability
distribution will be given by the sum of random variables with scaled chi-square
distributions of order 1, which is a scaled chi-square distribution of order $N$,
$I \sim \mathrm{Gamma}(N/2, 2\Sigma/N)$, with a probability distribution function
\begin{equation}
  \label{eq:4}
  p(I_C; N) = \frac{(N/2)^{N/2} \, I_C^{N/2-1}}{\Sigma^{N/2} \, \Gamma(N/2)} \, \exp(-NI_C/2\Sigma),
    \;\; I_C > 0.
\end{equation}
The distribution of Eqn.~\ref{eq:4} has mean $\Sigma$, variance $2\Sigma^2/N$, and is
equal to $p(I; N/2)$ when $N$ is even. This will be the case for central sections of $\boldsymbol{q}$
that are perpendicular to a two-fold rotation axis of a dimer, for example. In the limit
of an infinite number of orientations, such as the case of solution scattering of
unoriented molecules, it can be found through the central limit theorem that $p(I;N)$ and
$p(I_C;N)$ both approach a normal distribution with a mean and variance both equal to $\Sigma$
\cite{Siegrist:2015}, which is also the limit of Poisson statistics.

We note that Eqns. (\ref{eq:1}) to (\ref{eq:4}) hold for any scaling of intensities
$I$ whether they be recorded as photon counts or ``detector units'' referred here as adu.  For example, for a
detector gain $a$, the intensities in detector units $I = a\bar{I}$ for the continuous diffraction of $N$
orientations are distributed as $I \sim \mathrm{Gamma}(N, a\Sigma/N)$, giving the
probability distribution $p(I) = a p(\bar{I})$ with a mean $a\Sigma$ and
variance $a^2 \Sigma^2/N$.

A case that breaks the independence between diffraction intensities at
$\boldsymbol{R} \cdot \boldsymbol{q}$ and $\boldsymbol{q}$ is when the electron density of a single
molecule is real-valued, so that the diffraction intensities are centrosymmetric, and the
operator $\boldsymbol{R}$ is a rotation by \ang{180}.  Under this condition, the diffraction
intensities $I(\boldsymbol{q})$ of any object will be equal to $I(\boldsymbol{R} \cdot \boldsymbol{q})$
for values of $\boldsymbol{q}$ in a central section that is perpendicular to the
two-fold rotation axis.  In this reciprocal-space plane it would appear that the number of
orientations of rigid objects is reduced by half or, equivalently, that the number of
orientations did not change but the object’s projection is centrosymmetric.  These
diffraction intensities can thus be considered centric, even though the object itself is
not centrosymmetric.  Although similar to the case mentioned above of a crystal with a
two-fold rotation axis, there is a difference in that the projection of the structure of
the crystal along the two-fold axis is centrosymmetric, whereas it is the incoherent sum
of the projections of aligned and anti-aligned molecules that is centrosymmetric.  In the
case of $N$ equally-populated alignment fractions, the centric reflections are those in
central sections perpendicular to any two-fold rotation axes in the point group of the
alignments.  In those planes, for a real-valued structure, the distribution of diffraction
intensities will be given by Eqn. (\ref{eq:4}) since the number of independent normal distributions
being summed is reduced by half.

\subsection{Discrete Distribution}
\label{sec:discrete} %
Many of today’s X-ray detectors are sensitive to single photons, and diffraction
measurements made with them are therefore governed by counting statistics.  It is well
appreciated that this discretisation leads to a signal described by Poisson statistics.
For example, the counts in a particular pixel on the detector in a diffraction
experiment of a static object illuminated with a beam of constant flux will follow the
probability distribution
\begin{equation}
  \label{eq:Poisson}
  p(\bar{I}) = \frac{\bar{\mu}^{\bar{I}}\, e^{\bar{\mu}}}{\bar{I}!}
\end{equation}
for a mean number of photons $\bar{\mu}$, and where $\bar{I}$ are the discrete numbers of
photons per pixel (here the bar indicates values in photon counts). One feature of this
distribution is that the variance is equal to the mean, $\bar{\sigma}^2 = \bar{\mu}$.  For
large values of $\bar{I}$ this distribution approaches the Normal distribution with
$\sigma^2 = \mu$.  The statistics of the discrete diffraction of a molecule, measured at a
particular $q$ shell, is found by selecting a random variable from the appropriate Gamma
distribution (e.g. Eqn. (\ref{eq:3})) and then realising a particular value of that
variable by feeding it as the mean value of a Poisson distribution.  This is known as a
mixture distribution and is conceptually quite different from the distribution of the sum
of random variables discussed above.  The mixture distribution of photon counts, where the
Poisson mean is distributed according to $\mathrm{Gamma}(N,\bar{\Sigma}/N)$ for $N$ equal
twin fractions, is given by the negative binomial distribution
$\mathrm{NegativeBionomial}(N, N/(N+\bar{\Sigma}))$
\begin{equation}
  \label{eq:discrete-Wilson}
  p(\bar{I}; N) = \left(\frac{N}{N+\bar{\Sigma}} \right)^N \, \left(
    \frac{\bar{\Sigma}}{N+\bar{\Sigma}} \right)^{\bar{I}} \,
  \frac{(N-1+\bar{I})!}{(N-1)!\,\bar{I}!}, \;\; \bar{I} \geq 0
\end{equation}
with a mean $\bar{\Sigma}$ and variance $\bar{\Sigma}(N+\bar{\Sigma})/N$
\cite{Siegrist:2015} (Chapter 10.4), \cite{Goodman:2007}.  Thus, this distribution approaches the
Poisson distribution for large $N$ and the variance is larger than for the non-discrete
distribution of Eqn. (\ref{eq:3}).  Some plots of the distributions are given in
Figs.~\ref{fig:distributions} (a) and (b) for the case of $\bar{\Sigma} = 10$ counts.

\subsection{Linear Polarisation}
\label{sec:polarisation}
In the above we have assumed that the incident radiation is unpolarised.  In that case,
atomic scattering factors are dependent only on the magnitude of the photon momentum
transfer $q$, giving rise to diffraction intensities that follow a given Gamma or negative
binomial distribution for detector pixels located on a shell of constant $q$.  Radiation at
synchrotron and FEL facilities is usually linearly polarised, which modifies the
diffraction intensities by a factor equal to the square of the dot product of the electric
field vectors of the incident and scattered rays (which themselves are perpendicular to
the direction of propagation of the rays).  For example, for horizontally ($x$) polarised
radiation, the intensity pattern is modulated by
\begin{equation}
  \label{eq:polarisation}
  P(k_x, k_y)=1-\left(\frac{\lambda}{2 \pi}\right)^2 k_x^2
\end{equation}
where $k_x$ and $k_y$ are the scattered wave-vector components in the detector plane.  The
measured intensities of each diffraction pattern $I(k_x,k_y)$ can be corrected by dividing
by $P(k_x,k_y)$.  For measurements of a non-discrete diffraction signal, this correction
will have the intended consequence of generating signals that follow the statistics of the
Gamma distribution on a particular $q$ shell.  For the photon counting measurements this is
not the case, since multiplying counts by a variable correction will alter the variance by
a different factor than the mean.  After correcting for the polarisation, counts in a $q$
shell of the diffraction of an unstructured object will no longer obey Poisson statistics.

In our analysis below we utilise the variance to determine parameters such as scaling and
background of diffraction patterns.  For discrete measurements we must first account for
the polarisation in this analysis, as follows: First the polarisation-corrected
diffraction pattern is averaged in thin shells of $q$ (or $k$) from which a 2D
polarisation-uncorrected average is regenerated
\begin{equation}
  \label{eq:polarisation-average}
  I_{av}(k_x,k_y) = \langle I(k_x,k_y)/P(k_x,k_y)\rangle_{|k|}\,P(k_x,k_y)
\end{equation}
This average no longer contains any speckles but it can be contoured to find sets of
detector pixels (or coordinates $k_x$, $k_y$) with equal mean counts $\Sigma$ in the
polarisation-uncorrected measurement.  These contoured regions are then used instead of
shells of equal $q$ to determine the distribution of intensities.  This approach will
account for any signal or background originating from elastic scattering from a region
near the sample, but will not account for so-called detector dark noise, X-ray
fluorescence, or scattering from sources far upstream or downstream of the sample.

\subsection{Spatial Coherence and Pixel Size}
\label{sec:coherence}
\begin{wrapfigure}{l}{75mm}
   \setlength{\unitlength}{0.88mm}
  \begin{picture}(85,122)(0,0)  %One-column figure
    \put(0,62){
       \includegraphics[width=7.5cm, keepaspectratio]{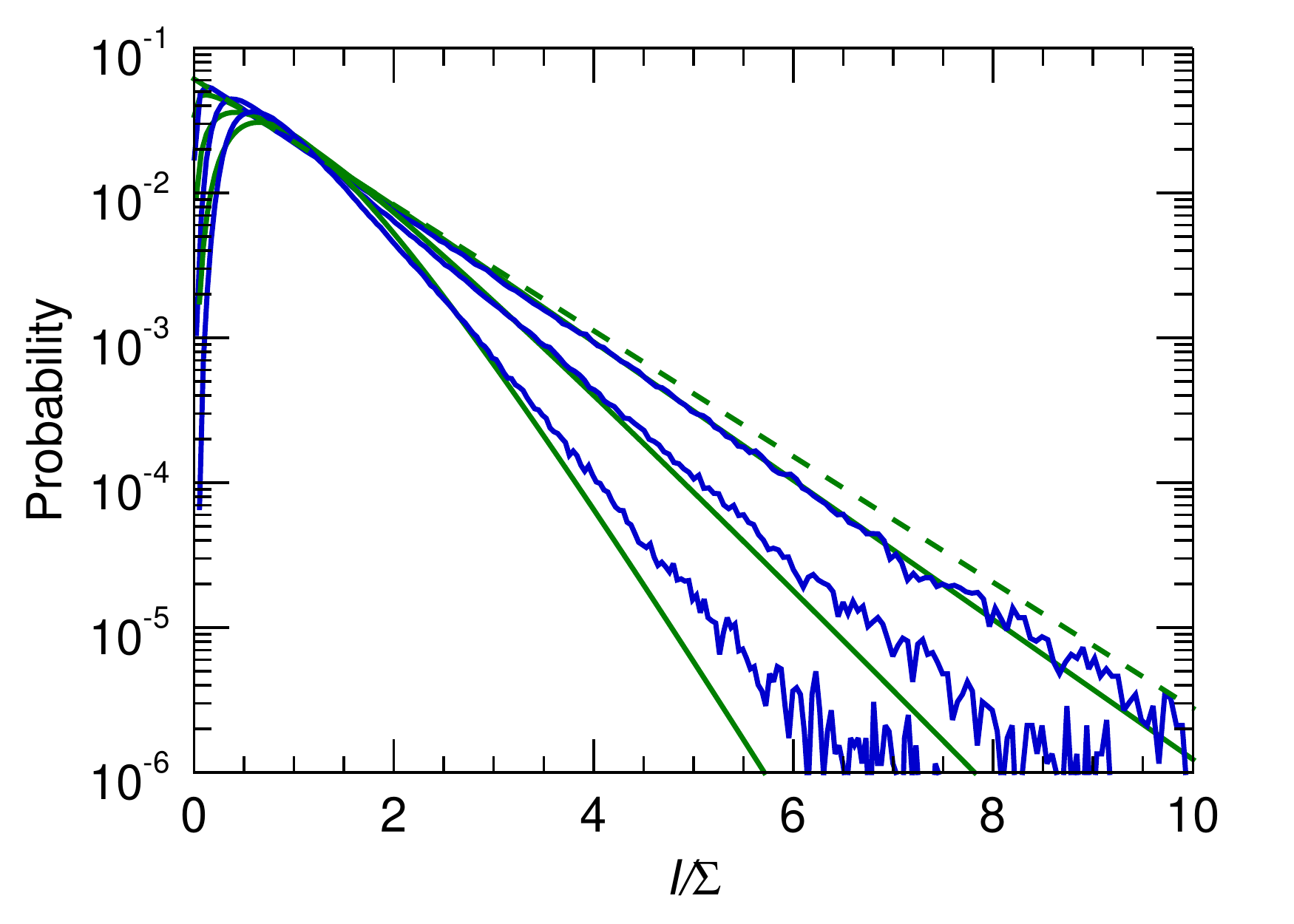}
    }
    \put(0,0){
       \includegraphics[width=7.5cm, keepaspectratio]{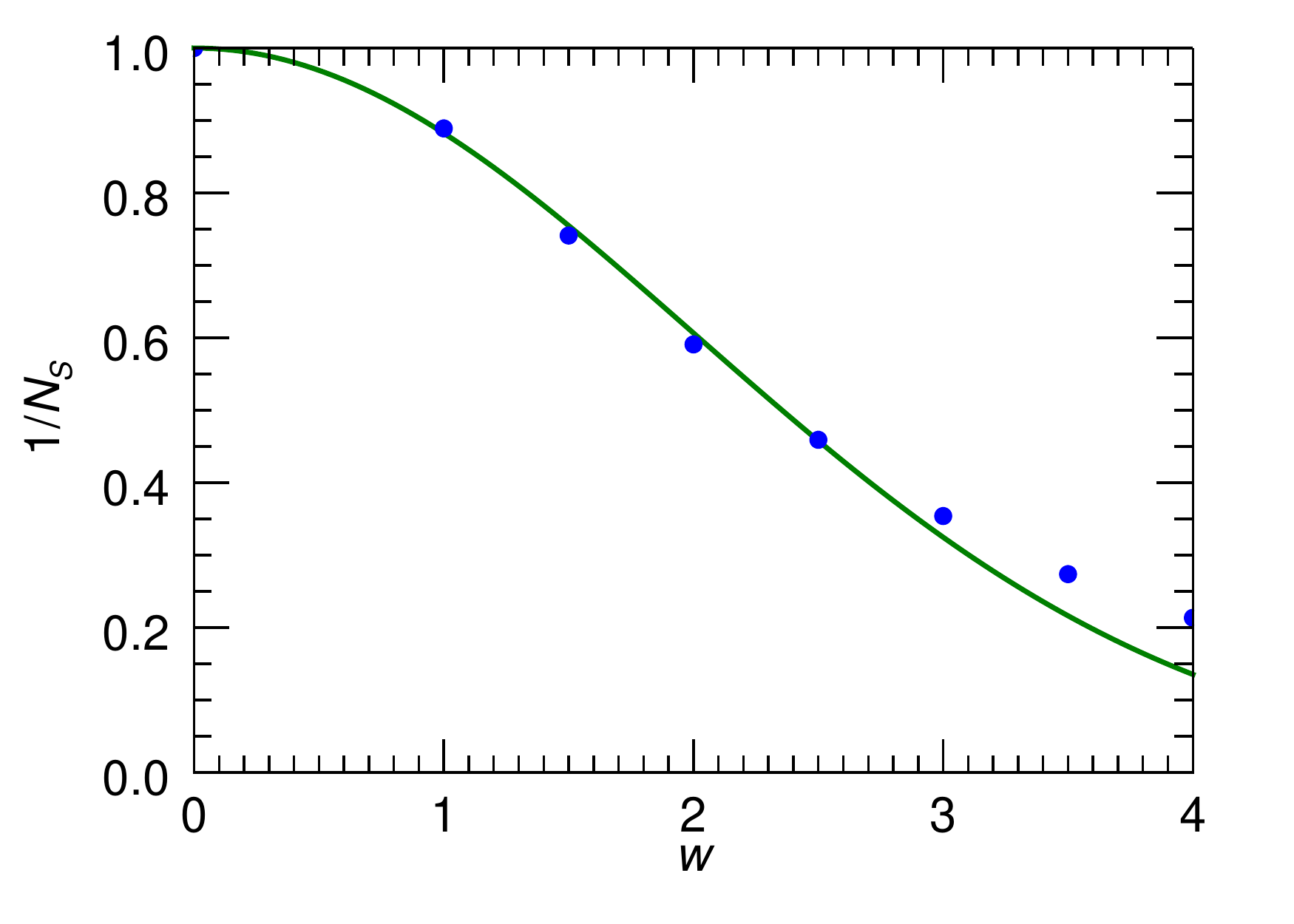}
    }
    \put(72,114){\figfont(\textit{a})}
    \put(72,52){\figfont(\textit{b})}
  \end{picture}
 \captionsetup{labelformat=empty} % makes sure dummy caption is blank
  \caption{} % add dummy caption - otherwise \label won't work and figure numbering will not count up
 \label{fig:coherence}
\end{wrapfigure}

The continuous diffraction from single objects can be sampled arbitrarily finely, unlike
the discrete locations of Bragg peaks.  The simulated diffraction intensities shown in
Fig.~\ref{fig:1} (a) were calculated at twice the Nyquist sampling rate required to fully
describe the continuous intensity wave-field, which is to say four times the sampling
density in each dimension as would be obtained from Bragg peaks of a P1 crystal in which
the molecules were packed in the smallest possible P1 unit cell (see
e.g. \cite{Thibault:2010}).  At high sampling rates there are obviously correlations
between neighbouring intensities, since they are likely to be sampling the same speckle.
Even so, this ``oversampling'' does not affect the statistics in the limit of randomly
positioned atoms.  The simulation does however differ from actual measurements of a
diffraction pattern, in that intensities will not in reality be sampled at points, but
will be averaged over the active areas of the detector pixels, described by the
convolution
\begin{equation}
  \label{eq:source-convolution}
  I_m(k_x,k_y)=I(k_x,k_y)\otimes s(k_x,k_y)
\end{equation}
\marginpar{
\vspace{0.7cm}
\textbf{Figure \ref{fig:coherence}.} (a) Distributions of the simulated intensities of a single PS II complex after
    convolving the 3D reciprocal-space array of diffraction intensities with cubic voxels
    of widths 1, 2, and 3 times the Nyquist sampling rate of the continuous diffraction
    intensity. The negative exponential distribution of the point-sampled intensities is
    shown with the dashed line.  (b) Plot of $1/N_S = \mathrm{Var}[I]/\mathrm{Mean}[I]^2$
    versus the voxel width, $w$.  The voxel width is normalised to the Nyquist sampling
    distance.  Shown in green is a Gaussian of width 2. \textcopyright The Authors licensed under CC BY
4.0
}
where $s(k_x,k_y)$ is the pixel response function.  The statistics of the measurements $I_m$
will clearly differ from those of $I$ if the pixel is larger than a speckle size. The
blurring by the pixel response reduces the contrast of the speckles, eliminating zeroes
and raising low intensity values as well as reducing the peak intensities.  This can be
seen in the plots of Fig.~\ref{fig:coherence} (a), where histograms of the simulated intensities of a PS II
complex are given after first convolving the patterns with cubic voxels of varying sizes.
The distributions become more truncated and narrower, with an appearance similar to the
distributions of the incoherent sum of $N$ independent patterns as shown in Fig.~\ref{fig:distributions}.
% \begin{figure}
%   \setlength{\unitlength}{1mm}
%   \begin{picture}(85,122)(0,0)  %One-column figure
%     \put(0,62){
%        \includegraphics[width=8.5cm, keepaspectratio]{chapman-fig3a.pdf}
%     }
%     \put(0,0){
%        \includegraphics[width=8.5cm, keepaspectratio]{chapman-fig3b.pdf}
%     }
%     \put(72,114){\figfont(\textit{a})}
%     \put(72,52){\figfont(\textit{b})}
%   \end{picture}
%   \caption{(a) Distributions of the simulated intensities of a single PS II complex after
%     convolving the 3D reciprocal-space array of diffraction intensities with cubic voxels
%     of widths 1, 2, and 3 times the Nyquist sampling rate of the continuous diffraction
%     intensity. The negative exponential distribution of the point-sampled intensities is
%     shown with the dashed line.  (b) Plot of $1/N_S = \mathrm{Var}[I]/\mathrm{Mean}[I]^2$
%     versus the voxel width, $w$.  The voxel width is normalised to the Nyquist sampling
%     distance.  Shown in green is a Gaussian of width 2. }
%   \label{fig:coherence}
% \end{figure}

The effect on the intensity statistics of a continuous speckle pattern convolved with a
pixel shape was examined by Dainty \cite{Dainty:1976}, who showed that the variance of the
intensities is reduced from the ideal value of $\Sigma^2$ by a factor given by the ratio
of the speckle size (equal to the inverse of the width of the autocorrelation function of
the object) divided by the pixel area in $q$ space.  In particular, Dainty posited that
the intensities $I_m$ of Eqn. (\ref{eq:source-convolution}) can be expressed as a weighted
sum of independent random variables, and that the distribution can in fact be approximated
(for a wide range of pixel response functions and molecule autocorrelation functions) by
the Gamma distribution of Eqn. (\ref{eq:3}).  In this case $N = N_S$, the number of
speckles per pixel, need not be a whole number. In Fig.~\ref{fig:coherence} (a), these
distributions are additionally plotted and can be compared with the histograms of the
convolved simulated diffraction patterns.  The distributions show a good agreement with
the simulations by setting $1/N_S = \mathrm{Var}[I] / \mathrm{Mean}[I]^2$.  This
parameter, referred to as the ``speckle contrast'' \cite{Dainty:1976}, can be considered
as the degree of purity of the measurement in a detector pixel, or in other words an
indicator of the degree of coherence, as discussed below.  A plot of $1/N_S$ versus the
voxel width of the recording process is given in Fig.~\ref{fig:coherence} (b) and is found
to decrease with width roughly as a Gaussian.  Here, the width is normalised to the
Nyquist sampling width of the diffraction intensities (the inverse of twice the
width of the molecule), which is about half the width of a speckle.  The width of the
Gaussian plotted in Fig.~\ref{fig:coherence} (b) is 2, equal to a speckle width.

The speckle contrast in a diffraction pattern can be used as a measure of coherence
\cite{Dainty:1976}.  Reducing the spatial coherence of the illumination will reduce the
variance of the diffracted intensities.  This can be quite clearly understood in the
Gauss-Schell model of partial coherence \cite{Mandel:1995} where a partially-coherent beam
is equivalent to one produced by an incoherent source of finite extent.  In this model,
any point in the source gives rise to a fully coherent beam that produces a fully coherent
diffraction pattern positioned relative to the axis defined by the line joining that
source point and some arbitrary but common point in the object.  For a small enough
angular extent of the source, the pattern from each source point will be an identical but
shifted version of that produced by any other source point.  The patterns produced by each
source point will be mutually incoherent, so in the limit of a small angular extent of the
source the resulting diffraction pattern will be a convolution of the coherent pattern
with a function describing the angular distribution of the source intensity.  Thus,
Eqn. (\ref{eq:source-convolution}) also represents the case of partially coherent
diffraction, where $s$ describes the angular extent of the source, also equal to the
Fourier transform of the mutual coherence function \cite{Goodman:1985}.  Again, this
coherence length $w$ can be expressed in terms of the parameter $N_S$, equal to the
fractional number of speckles that lie in the angular extent of the source.

In the case of $N$ independent orientations of molecule measured with $N_S$ speckles per
detector pixel or coherence area, the distribution will be modified in a similar way as
for a single orientation. For a given mean $\Sigma$, the variance will be modified from
$\Sigma^2/N$ by an additional division by $N_S$ to $\Sigma^2/N'$, where $N' = N\,N_S$ and
the distribution of intensities will be approximated by
$I \sim \mathrm{Gamma}(N',\Sigma/N')$ as per Eqn. (\ref{eq:3}).

The phasing of continuous diffraction patterns using iterative algorithms depends
critically on accurate sampling of the intensities.  Any reduction in contrast or addition
of a constant will eliminate intensity zeroes and cause discontinuities of phased
amplitudes, a situation that is inconsistent with diffraction arising from a compact
object.  Much progress in diffractive imaging was made recently by accounting for the
decrease in contrast in continuous diffraction caused by partial coherence
\cite{Whitehead:2009}. The
coherence width, or equivalently the detector pixel width, is usually required as a fixed 
parameter in schemes of partially-coherent diffractive imaging, and measurements of the
coherence properties of the beam must often be made to carry out these schemes
\cite{Flewett:2009,Chen:2012}.  For macromolecular diffractive imaging, where the object
is typically less than several hundred \aa ngstr\"{o}ms in width, achieving the necessary
coherence width of the beam, equal to double the object width \cite{Spence:2004}, is
routinely achieved, and the necessary sampling density and pixel width can be determined
by examining the autocorrelation of the object.  Nevertheless, the beam coherence, pixel
width, sample heterogeneity, and errors in aggregating data from many diffraction patterns may all give rise
to an effective degree of coherence that can be determined directly from the intensity
statistics if the number of object orientations are known.  A variation of the
determined coherence as a function of $q$ may indicate rotational disorder of the
molecules, or an alignment error in aggregating data from many single-molecule
diffraction snapshots.

\section{Statistics of diffraction intensities of translationally disordered crystals}
\label{sec:stats-crystal}
The diffraction pattern of a crystal exhibiting a degree of translational disorder
consists of Bragg peaks, modulated by a $q$-dependent Debye-Waller factor, and continuous
diffraction that arises contrariwise to the decrease in Bragg intensities.
Ayyer \emph{et al.} \cite{Ayyer:2016} consider a disordered finite crystal consisting of a particular
(and unique) rigid object that is repeated $M$ times in different orientations and
positions according to the crystal symmetry, in each of $K$ unit cells of the
crystal. (Crystals consisting of more than one kind of rigid object can also be
considered.) The 3D diffraction pattern of such a crystal with identical rigid units that
are randomly displaced from their ideal crystallography positions in each direction
following a normal distribution of variance $\sigma_\Delta^2$, is then given by
\begin{multline}
  \label{eq:diffraction}
  I(\boldsymbol{q}) =  K \left[ \sum_{m=1}^M \left| F(\boldsymbol{R}_m \cdot \boldsymbol{q}) \right
    |^2\right ] (1 - e^{-4 \pi^2 \sigma_\Delta^2 q^2}) \\
  + \left | \sum_{m=1}^M
    F(\boldsymbol{R}_m \cdot \boldsymbol{q})\, e^{2 \pi i \boldsymbol{q} \cdot \boldsymbol{t}_m}\right |^2
    \,  e^{-4 \pi^2 \sigma_\Delta^2 q^2} \sum_{j=1}^K \sum_{k=1}^K e^{2\pi i (\boldsymbol{a}_j
    - \boldsymbol{a}_k)\cdot \boldsymbol{q}}
\end{multline}
where $F(\boldsymbol{q})$ is the complex-valued Fourier transform of the density of the single
rigid unit, $\boldsymbol{R}_m$ and $\boldsymbol{t}_m$ are the rotation and translation operators
for the $m$th rigid unit, and $\boldsymbol{a}_k$ are the real-space lattice positions of the
crystal. (Pre-factors in Eqn. (\ref{eq:diffraction}) are ignored.)  Note that the
isotropic mean
square displacement of rigid units in 3D space is equal to $3\sigma_\Delta^2$. The second term of
Eqn. (\ref{eq:diffraction}) is the square modulus of the Fourier transform of the entire
unit cell, modulated by a Debye-Waller factor and by the double product that gives the
reciprocal lattice.  The first term is markedly different, and is given by the incoherent
sum of the Fourier transforms of the rigid object in each of its crystallographic
orientations, all modulated by a function, the complementary Debye-Waller factor, that
monotonically increases with $q$.  This term is not multiplied by a reciprocal lattice,
and is thus continuous and proportional to the single molecule diffraction when there is
only one orientation of the rigid randomly-translated object per unit cell.  In the more
general case, it is equal to the incoherent sum of the diffraction from $N$ unique
orientations of the rigid unit.  The number of unique orientations of the rigid unit may
be a subset of those given by the point group of the crystal if the rigid unit itself is
crystallographically symmetric (i.e.\ not non-crystallographic), or there may be more
orientations than dictated by the crystal symmetry if the rigid units are oriented
according to non-crystallographic symmetry.  An example of the former situation is given
below for PS I crystals in the space group P$6_3$.

The Bragg intensities are proportional to the coherent diffraction of the entire unit cell
of the crystal (indicated by the square modulus outside the sum in the second term of
Eqn. (\ref{eq:diffraction})) and hence depend on the space-group symmetry of the crystal.  The continuous
diffraction is proportional to the incoherent sum of the diffraction of the independent
rigid objects in the crystal, subject to the (possibly reduced) point-group symmetry of
the crystal given by the number of unique orientations of the (possibly symmetric) rigid
objects.  The statistics of the Bragg and continuous diffraction intensities are therefore
different, depending on these symmetries.  The Bragg reflections obey Wilson statistics
with a symmetry dependence examined by Rogers \cite{Rogers:1950}.  The continuous diffraction will
obey Wilson statistics such as given by Eqns. (\ref{eq:3}) or (\ref{eq:4}), subject to the symmetry of the
rigid unit and on the number of unique orientations of that rigid unit.  Eqns. (\ref{eq:3}) and (\ref{eq:4})
assume equal populations of objects in each of the orientations.  In some crystals this
will not be true, in which case the distributions can be derived from sums of squares of
normally-distributed random variables with different variances \cite{Rees:1982}.  In general,
for a rigid unit with $N_R$ non-crystallographic rotation operations and $N_C$ crystal point
group operations the centric intensities corresponding to any one of the
non-crystallographic symmetries will be given by the incoherent sum of the centric
diffraction from those objects in the particular orientation and the acentric diffraction
from the rest of the objects.  Assuming that the populations of rigid units in each
crystallographic orientation are equal, these intensities will have a distribution $I \sim
\mathrm{Gamma}(N_C-1, \Sigma/N_C) + \mathrm{Gamma}(1/2, 2\Sigma/N_C)$.  The
probability distribution function of the sum of two random variables is equal to the
convolution of their distributions, which can be calculated through the product of their
inverse Fourier transforms.  In statistics these are referred to as the characteristic
functions \cite{Papoulis:1991,Schmueli:1995}.  

As an example, consider a PS II crystal in space group P$2_12_12_1$. This consists of four
dimers in unique orientations found by rotating any one of them by \ang{180} about each of the
three orthogonal axes of the orthorhombic cell.  The two-fold rotation symmetry of the
dimer is non-crystallographic in this case, and the axis is not aligned along any of the
crystallographic axes.  The Bragg intensities are therefore in general acentric, given by
Eqn. (\ref{eq:1}).  In central sections of reciprocal space perpendicular to the orthogonal crystal
axes, however, the Bragg intensities are centric since projections of the crystal
structure down those axes will be centrosymmetric, with a distribution given by Eqn. (\ref{eq:2}).
The projection of the crystal structure down the dimer two-fold axis of any of the four
dimers will not be centrosymmetric, however, since the crystal as a whole does not share
this symmetry.  The continuous diffraction of a PS II crystal with translational disorder
will be governed by the incoherent sum of diffraction of equal populations of dimers in
each of four orientations, given by Eqn. (\ref{eq:3}) with $N = 4$, and will thus exhibit
$mmm$ symmetry.  If the rigid unit is the dimer
then central sections perpendicular to the dimer two-fold axis should include diffraction
from the one quarter of all dimers whose projections are centrosymmetric in that view.
The statistics in that case will be determined by a sum given by three parts acentric random
variables and one part centric random variables, resulting in $I \sim
\mathrm{Gamma}(3,\Sigma/4) + \mathrm{Gamma}(1/2,\Sigma/2)$ which has the distribution
\begin{multline}
  \label{eq:p31}
  p(I;3,1)=\frac{4}{\sqrt{\pi}\Sigma^3} \Biggl[  
  F_D \left( \sqrt{\frac{2I}{\Sigma}} \right) 
      (16I^2+8\Sigma I + 3\Sigma^2) \\
   -  \sqrt{2 \Sigma I} \, (4I + 3\Sigma) \Biggl]
     \,
     \exp(-2I/\Sigma), \;\; I > 0
\end{multline}
where $F_D$ is the integral
\begin{equation}
  \label{eq:fd}
  F_D(x) = \exp(-x^2)\,\int_0^x \exp(y^2)\,dy = \frac{\sqrt{\pi}}{2}\exp(-x^2) \, \mathrm{Erfi(x)}
\end{equation}
and Erfi is the imaginary error function, $\mathrm{Erfi}(x)=\mathrm{Erf}(ix)/i$.  Plots of
$p(I;4)$ and $p(I;3,1)$ are given in Fig.~\ref{fig:distributions} (a) and (b), showing that the dimer symmetry
causes a higher probability of high intensities compared with the completely acentric
reflections.  It therefore should be possible to detect non-crystallographic symmetry of
the rigid object from deviations of the statistics in particular central sections of
reciprocal space.

The central sections perpendicular to the three orthogonal crystal axes of PS II are all
perpendicular to a two-fold rotation axis and so, to the extent that the structure is
real-valued, these intensities will be centric, with a distribution given by Eqn. (\ref{eq:4}) with
$N = 4$.  This is equal to the acentric distribution with $N = 2$, shown in Fig.~\ref{fig:distributions}.

As another example we consider a crystal of photosystem I, which has a hexagonal space
group P$6_3$.  The structure consists of trimers with 3-fold rotational symmetry located in
alternating layers where the trimers are rotated by \ang{60} about this 3-fold axis and
translated perpendicular to it.  The trimer symmetry is crystallographic.  If the rigid
object was hypothetically the entire trimer then the continuous diffraction arising from
translational disorder would consist of the incoherent sum of the trimer in only these two
orientations.  Thus, in general, the probability distribution of the continuous
diffraction intensities in any given $q$ shell will be equal to $p(I;2)$ (Eqn. (\ref{eq:3}) with $N =
2$).  The continuous diffraction of the trimer will have 3-fold rotational symmetry and, if
the electron density of the trimer is real-valued, will be centrosymmetric.  In the
central section perpendicular to the 3-fold axis the diffraction from the \ang{60}-rotated
real-valued trimer will be identical, and hence in this plane of reciprocal space it will
appear as if there is only one object contributing to the diffraction (or two
centrosymmetric objects).  These intensities can therefore be considered as centric, with
a distribution in a given $q$ shell equal to $p(I;1)$ (Eqn. (\ref{eq:3}) with $N = 1$ or Eqn. (\ref{eq:4}) with $N
= 2$). The Bragg reflections in general positions will be centric ($N = 1$), or acentric in the $hk0$
zone.  Thus comparisons of the statistics of Bragg and continuous diffraction in centric and
acentric zones can be used to constrain the number of rigid-body units contributing to the
continuous diffraction. 

\section{Modified Statistics with Background Noise}
\label{sec:noisy-Wilson}
The continuous diffraction from a disordered crystal can be phased using iterative phasing
algorithms, as has been well established for coherent diffractive imaging of single
non-periodic objects.  One of the experimental issues that can arise in coherent
diffractive imaging is the incoherent addition of background intensity.  For diffraction
of disordered crystals, the diffuse scattering from the solvent adds incoherently to the
pattern.  This incoherent background must be estimated and subtracted, since otherwise
phasing cannot be reliably achieved---the intensity sum does not match to the square modulus
of the Fourier transform of an object of compact support.  The complication can be
appreciated by considering diffraction amplitudes that vary from positive to negative; for
example, phases that vary from $\pi$ to $-\pi$.  The diffraction amplitude must therefore pass
through zero, which cannot be satisfied if the measured intensity is everywhere greater
than zero due to a background.

When utilising Bragg peaks alone, the usual practise in crystallography, the background
can be reliably estimated from the measured intensity values surrounding the peak.  This
obviously cannot be done for the continuous diffraction.  In that case, background is
often estimated from a measurement without the sample in place.  In macromolecular
crystallography, the sample is usually surrounded by solvent, which creates a diffuse
background with a characteristic profile (including the so-called ``water ring'').
However, the crystal itself contains solvent which may differ in composition from pure
buffer solution, so the amount of the background is not necessarily equal to the no-sample
pattern.  One way to estimate a smoothly-varying background is to fit a function to local
minima of the diffraction pattern.  This is a valid approach for an object of a single
orientation, whose distribution of diffraction intensities follows Eqn.~(\ref{eq:1}), but
not when $N > 1$.  A much better approach is to utilise all intensity values in a
reciprocal shell, not just the minima, and to fit the appropriate distribution to estimate
the background.  We explore this approach here by examining the properties of the
distribution expected of intensities in a reciprocal-space shell from a disordered crystal
with an incoherent background.  This is carried out first for the case of non-discrete
intensity measurements where the background in shells of $q$ are considered to be normally
distributed.  This is the limiting case for large photon counts per pixel, and allows
analytical expressions of the resulting distribution of the sum of the aligned-molecule
diffraction with the background. The case of discrete signals is presented in
Sec.~\ref{sec:Poisson} where the background is assumed to follow Poisson statistics.

\begin{table}
\begin{adjustwidth}{-1in}{0in} % comment out/remove adjustwidth environment if table fits in text column.
%\centering
\caption{Moments of the distribution of intensities obeying ``noisy Wilson'' statistics. }
\begin{tabular}{lccc}      % Alignment for each cell: l=left, c=center, r=right
\\[-6 pt]  %prevent caption from crowding equation headings
                                            & $p_{NW}(I;N)$ & $p_{NW}(I;N-1,1)$ &
                                                                                  $p_{DNW}(I;N)$      \\[6pt]
\hline
\\[-6pt]
Mean [$\mu_{NW}$]               & $\displaystyle\mu+\Sigma$   & $\displaystyle\mu+\Sigma$     & $\displaystyle\bar{\mu}+\bar{\Sigma}$   \\[6pt]
Variance [$\sigma_{NW}^2$]  & $\displaystyle\sigma^2+\frac{\Sigma^2}{N}$ &
                                                                $\displaystyle\sigma^2+\frac{\Sigma^2}{N}
                                                                + \frac{\Sigma^2}{N^2}$
                                                                                           & $\displaystyle\bar{\mu}
                                                                                             +\bar{\Sigma}
                                                                                             +\frac{\bar{\Sigma}^2}{N}$ \\[10pt]
Skewness [$s_{NW}$]                 & $\displaystyle\frac{2 \Sigma^3}{N^2(\Sigma^2/N+\sigma^2)^{3/2}}$    
                                                                  &   $\displaystyle\frac{2(3+N)
                                                                    \Sigma^3}{N^2(\Sigma^2/N+\Sigma^2/N^2+\sigma^2)^{3/2}}$ 
                                                                                           &
 $\displaystyle\frac{\bar{\mu}+\bar{\Sigma}+3\bar{\Sigma}^2/N+2\bar{\Sigma}^3/N^2}{(\bar{\mu}+\bar{\Sigma}+\bar{\Sigma}^2/N)^{3/2}}$     \\
\end{tabular}
\label{tab:1}
\end{adjustwidth}
\end{table}

\subsection{Non-discrete Intensities with a Normal-distributed Background}
\label{sec:normal}
We refer to the disibution of the incoherent sum of the non-discrete acentric diffraction
and a normally distributed background as the noisy Wilson distribution.  For a background
mean $\mu$ and variance $\sigma^2$, added to molecular diffraction of mean $\Sigma$ from
$N$ orientations the distribution is given by $p_{NW}(I)$ by Eqn.~(\ref{eq:pNW}) in Appendix 
\ref{appendix:noisy-Wilson}.  Some examples of the distribution are plotted in
Figs. \ref{fig:distributions} (c) and (d), where it is seen that $p_{NW}(I)$ is
skewed. This skewness is a property of the signal, following the Gamma distribution,
rather than the skew-less normal-distributed noise.  In situations of low signal to
background, this skewness can therefore indicate the presence of continuous diffraction
signal.  However, as we shall see in Sec.~\ref{sec:Poisson}, unlike the normal
distribution the Poisson distribution is skewed, with a skewness decreasing with the
inverse of the square root of the mean counts.  This is significant for mean counts
approaching almost 100 photons, so the application of the results here requires suitably
large signals or averages over many patterns.

As mentioned above, Equation (\ref{eq:pNW}) can also be evaluated by the Fourier transform of the
product of the characteristic functions of the Gamma and normal distributions.  Likewise,
it is possible to derive the moments of $p_{NW}(I)$ from the Fourier transform of the
derivatives of its characteristic function.  Such an analysis can also be carried out for
the partially centric intensities that arise due to non-crystallographic symmetry of the
rigid unit, such as for the probability distribution $p_{NW}(I;N-1,1)$---even though an
expression for the probability distribution cannot be readily derived.  Expressions for
these moments are given in Table~\ref{tab:1} for the acentric and partially centric intensities.
From the expressions in Table~\ref{tab:1} it is possible to solve for the parameters $\Sigma$, $\mu$, and $\sigma^2$
from the moments of the measured intensities in a given $q$ shell.  These parameters are
respectively the mean of the continuous diffraction, the mean of the background, and the
variance of the background in that shell.  This is far less computationally expensive than
fitting a probability distribution function to those intensities to obtain the parameters.
The solution to the simultaneous set of equations given by the first column of Table~\ref{tab:1}
yields the following expressions:
\begin{subequations}
\label{eq:solutions-1}
  \begin{align}
    & \Sigma = (N^2 s_{NW}/2)^{1/3} \sigma_{NW} \\
    & \sigma^2=\sigma_{NW}^2 \left(1-(N/4)^{1/3} s_{NW}^{2/3} \right) \\
    & \mu=\mu_{NW}-\Sigma
\end{align}
\end{subequations}
These estimates can be influenced by intensity values that do not conform to the expected
distribution $p_{NW}(I)$, such as from Bragg peaks or centric intensities. The procedure of
fitting the probability distribution function $p_{NW}(I)$ to the histogram of $I$ tends to avoid
the influence of outliers, but ideally any Bragg peaks should be identified and excluded
from the analysis.  The parameters $\Sigma$, $\mu$, and $\sigma^2$ can be estimated in a number of reciprocal
space shells (e.g. 50 equally-spaced shells) so that a smooth curve can be fit to each of
the parameters as a function of function of $q$.  In this way a radially-symmetric
background $\mu(q)$ can be subtracted from the diffraction pattern. The curve $\Sigma(q)$ can be used
to generate a Wilson plot of the continuous diffraction \cite{French:1978}, the
radially-weighted average of which can be used to scale each pattern before merging with
others to form a 3D array of intensities.  The error in the intensity measurements due to
background can be estimated from $\sigma(q)$.

\subsection{Discrete Intensities with a Poisson-distributed Background}
\label{sec:Poisson}
The Poisson distribution is given by Eqn. (\ref{eq:Poisson}). The variance of this
distribution is equal to the mean, $\bar{\sigma}^2 = \bar{\mu}$ and the skew is equal to
$\bar{\mu}^{-1/2}$, giving appreciable values of skew even for signals of tens of photons
and showing that the analysis of Sec.~\ref{sec:normal} is suitable only in the limit of large photon
counts.  The distribution of the sum of acentric diffraction and unstructured background
is given by $\bar{I} \sim \mathrm{NegativeBinomial}(N,N/(N+\bar{\Sigma}))+\mathrm{Poisson}(\bar{\mu})$.  We refer to this as the ``discrete noisy Wilson'' (DNW) distribution. An
analytic expression for this distribution cannot be readily determined, but the
probability distribution functions can be evaluated numerically using a program such as
Mathematica, as shown in Figs. \ref{fig:distributions} (c) and (d).  Additionally the moments can be found from
the characteristic functions of the distributions and are given in the third column of
Table~\ref{tab:1} for intensities measured in photon counts.  In this case, the expressions for the
mean and variance are analogous to those for the normal-distributed background (replacing
$\sigma^2$ for $\bar{\mu}$) but the skewness differs in that it has a contribution from the background in the
numerator.  Unlike the case of the normal-distributed background, the skew is not a unique
identifier of the presence of aligned-molecule diffraction. The mean acentric diffraction $\bar{\Sigma}$
and mean background $\bar{\mu}$ are determined by easily solving the two equations for
$\bar{\mu}_{DNW}$ and $\bar{\sigma}^2_{DNW}$ in Table~\ref{tab:1}:
\begin{subequations}
\label{eq:solutions-2}
  \begin{align}
    & \bar{\Sigma} = \sqrt{N(\bar{\sigma}^2_{DNW}-\bar{\mu}_{DNW})} \\
    & \bar{\mu}=\bar{\mu}_{NW}-\bar{\Sigma}
\end{align}
\end{subequations}
As compared with the continuous case, the presence of the signal is revealed by an excess
of the variance of the intensities over the mean. Equality of these quantities occurs if
the intensities followed a Poisson distribution, which would occur if no aligned-molecule
diffraction signal was present. If $\bar{\sigma}^2_{DNW} < \bar{\mu}_{DNW}$ then the best
estimate of $\bar{\Sigma}$ is zero.

The gain and offset of the detector can be estimated from a pattern recorded without any
aligned-molecule diffraction but only a Poisson-distributed background, such as scattering
from a liquid, or fluorescence.  For a detector gain $a$ and offset $b$, the mean
intensity in detector units is $a\bar{\mu}+b$ and variance $a^2\bar{\mu}$ .  A linear fit
to a plot of the sample variance as a function of the sample mean for different exposures
or shells of $q$, for example, will give a slope equal the gain $a$, and offset $b$,
assuming that the detector properties are the same over all pixels.  For patterns recorded
with linear incident polarisation, the procedure outlined in Sec.~\ref{sec:polarisation}
must be used to find groups of pixels in the polarisation-uncorrected pattern from which
to compute the mean and variance.  
%The detector gain and offset parameters can also be
%estimated in a low-exposure regime where the distribution of intensities shows peaks
%corresponding to integer numbers of photons \cite{Hart:2012}.

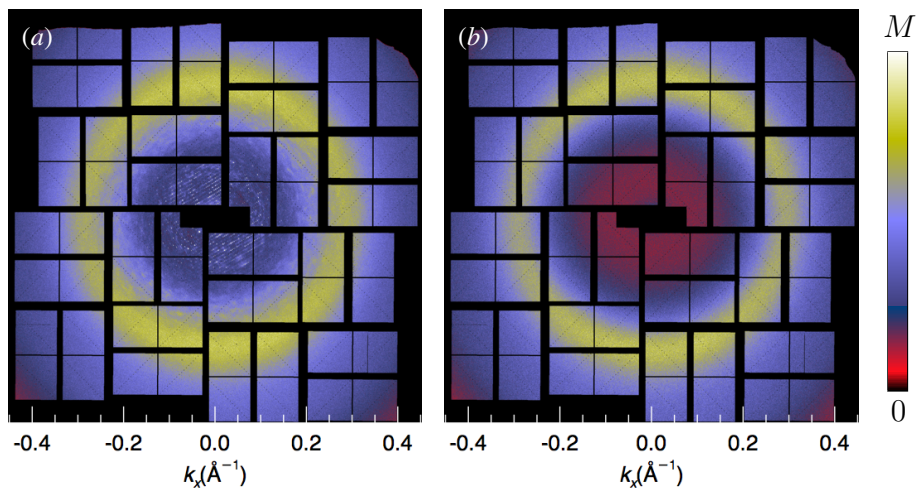
\begin{figure}
  \setlength{\unitlength}{1mm}
  \begin{picture}(120,63)(0,0)  %1.5 column width
    \put(0,0){
      %pdf files didn't compress well, so going with png
      \includegraphics[width=5.5cm, keepaspectratio]{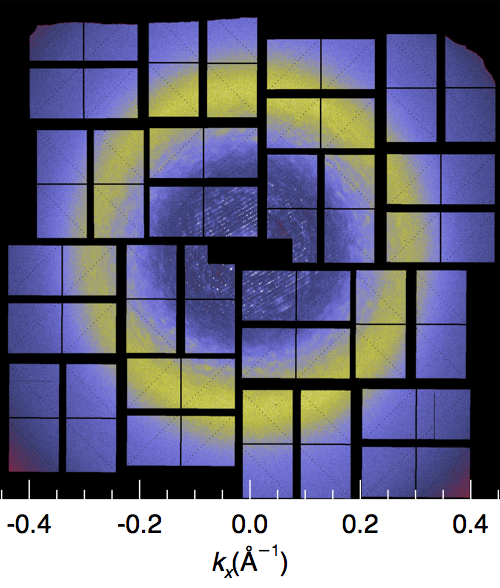}
    }
    \put(58,0){
       \includegraphics[width=5.5cm, keepaspectratio]{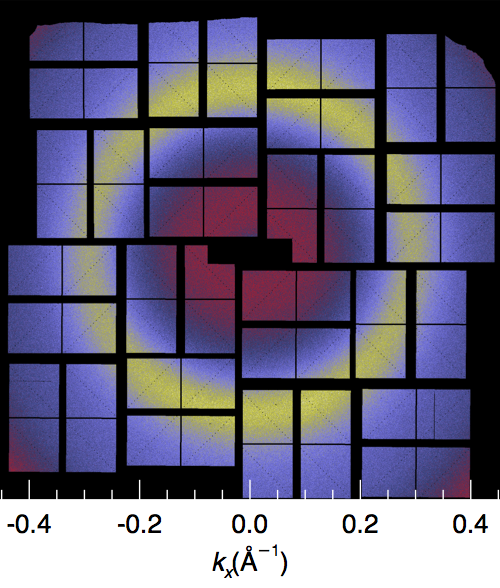}
    }
    \put(117,14){
      \frame{\includegraphics[width=2.5mm,height=4.5cm]{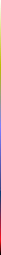}}
    }
    \put(118.5,10){\large$0$}
    \put(117.5,61){\large$M$}
%    \put(3,55){\textcolor{white}{\large\textbf{\textsf{a}}}}
%    \put(68,55){\textcolor{white}{\large\textbf{\textsf{b}}}}
    \put(3,60){\figfont\textcolor{white}{(\textit{a})}}
    \put(61,60){\figfont\textcolor{white}{(\textit{b})}}
  \end{picture}
\caption{(a) Single-pulse FEL snapshot diffraction pattern of PS II, showing Bragg peaks
  and continuous diffraction from the disordered crystal and diffuse scattering from the
  solvent medium.  (b) Single-pulse FEL pattern from the jet that was free of crystals,
  showing only the scattering from the liquid.  The colour scale spans 0 to $M = \SI{3000}{\adu}$
  for (a) and 0 to $M = \SI{2000}{\adu}$ for (b).  The incident beam was linearly polarised, in the
  horizontal direction in this view.  The patterns have not been corrected by the
  polarisation factor. \textcopyright The Authors licensed under CC BY
4.0}
\label{fig:snapshot}
\end{figure}

We note that for an integrating detector, the noise model could be improved by adding a
normal distribution corresponding to the detector noise.  For the CSPAD 
\cite{Philipp:2011,Hart:2012} used in the experiments described below, the standard deviation of the
detector noise is below a photon count, and thus only has a significant effect on the
computation of statistics for patterns with very low detector counts.  We ignore this
consideration here.

\section{Analysis of Continuous Diffraction Patterns}
\label{sec:analysis-patterns}
We demonstrate our analysis approaches on continuous diffraction patterns of PS II,
previously measured at an X-ray free-electron laser \cite{Ayyer:2016} by the method of
serial femtosecond crystallography.  Crystals in liquid suspension were jetted across the
focus of the X-ray beam while snapshot patterns were recorded on every X-ray pulse
\cite{Chapman:2011,Boutet:2012} on a CSPAD detector.  Measurements were carried out in
vacuum. The concentration of crystals in the jet was such that only a fraction of the
pulses hit a crystal, and a set of diffraction patterns was selected by searching for the
presence of Bragg peaks.

\subsection{Statistics of a single pattern}
\label{sec:single}
A typical snapshot pattern  (from a still, not rotating, crystal) is given in
Fig. \ref{fig:snapshot} (a), without any background subtraction or correction for the
polarisation of the incident beam.  The Bragg peaks obviously influence the statistics of
the intensities and must be excluded from our analysis of the continuous diffraction.  For
this, they must be first identified, which was done by comparing the pattern with a
version of itself that was modified by applying a median filter of width 9 pixels.  A mask
was defined by choosing pixels where the original values exceeded the median filtered
values by an amount equal to the mean intensity value in the shell.  This mask was then
dilated using a kernel that was 7 pixels wide. While this was quite aggressive in removing
regions around Bragg peaks, there were still a large number of pixels left to obtain
histograms of the continuous diffraction intensities.

A pattern free of crystal diffraction and showing scattering from the liquid jet that
carries the crystals is displayed in Fig. \ref{fig:snapshot} (b).  Following the analysis
procedure of Sec. \ref{sec:polarisation}, groups of pixels (excluding those that were masked) were determined by contouring
$I_{av}(k_x,k_y)$ of Eqn. (\ref{eq:polarisation}) at levels spaced by \SI{20}{\adu}. A linear regression of the
variances of the polarisation-uncorrected intensities within these groups to the means
showed a high degree of correlation (with a correlation coefficient of 0.998), giving a
detector gain of \SI{28.7}{\adu/\photon} and an offset of \SI{29}{\adu}.

\begin{figure}[h]
 \begin{adjustwidth}{-2in}{0in}
  \setlength{\unitlength}{1mm}
  \begin{picture}(160,70)(0,0)
    \put(0,0){
      \includegraphics[width=8cm, keepaspectratio]{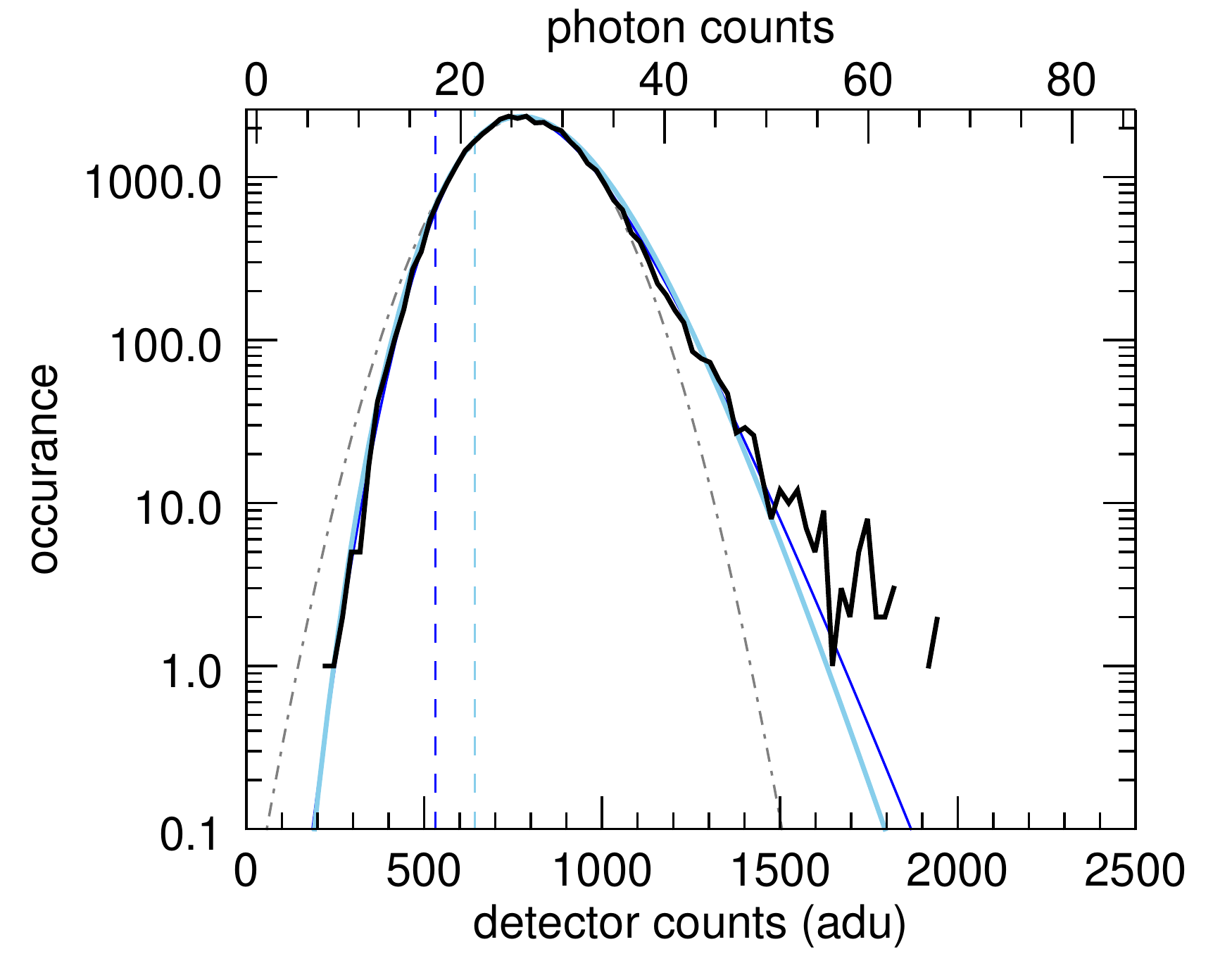}
    }
    \put(80,0){
      \includegraphics[width=8cm, keepaspectratio]{chapman-fig5a.pdf}
    }
    \put(20,51){\figfont(\textit{a})}
    \put(100,51){\figfont(\textit{b})}
  \end{picture}
  % \begin{picture}(85,138)(0,0)  %Single column
  %   \put(0,70){
  %     \includegraphics[width=7.5cm, keepaspectratio]{chapman-fig5a.pdf}
  %   }
  %   \put(0,0){
  %     \includegraphics[width=7.5cm, keepaspectratio]{chapman-fig5a.pdf f}
  %   }
  %   \put(21,124){\figfont(\textit{a})}
  %   \put(21,54){\figfont(\textit{b})}
  % \end{picture}
  \vspace{3mm}
 \caption{(a) Histogram (in black) of counts in a region of the pattern shown in
    Fig.~\ref{fig:snapshot} (a), but prior to polarisation correction, at a radius of
    approximately 260 pixels.  The region was obtained by contouring the
    polarisation-uncorrected radial average (see Eqn. (\ref{eq:polarisation-average})). A
    fit of $p_{NW}(I; 4)$ is shown in blue yielding a value of $\mu = \SI{532}{\adu} = 18.6$
    photons as indicated by the blue dashed line, and $\Sigma = \SI{269}{\adu} = 9.4$ photons.  The
    discrete noisy Wilson distribution obtained by applying Eqns. (\ref{eq:solutions-2}) is shown in sky
    blue and gives a higher estimate of the background with $\mu = 23$ photons and $\Sigma = 8$
    photons. (b) Histogram of counts in a region of the crystal-free pattern of Fig.~\ref{fig:snapshot}
    (b), with a fit of a Poisson distribution (sky blue) with a mean of 18 photons. The
    dash-dotted grey lines indicate Gaussian fits, shown to emphasise the skewness of the
    distributions. \textcopyright The Authors licensed under CC BY 4.0}
  \label{fig:histogram}
  \end{adjustwidth}
\end{figure} 

The distribution of intensities in a region of the pattern in Fig.~\ref{fig:snapshot} (a)
in a ring centred at about $q = \SI{0.15}{\per\angstrom}$ (260 pixel radius), is plotted in
Fig.~\ref{fig:histogram} (a), in addition to the fits of $p_{NW}$ and $p_{DNW}$ with
$N = 4$.  The parameters obtained from the fit of $p_{NW}$ were $\Sigma = 269$,
$\mu = 532$, and $\sigma = \SI{104}{\adu}$.  Thus, the intensities are dominated by the
background, as is obvious from Fig.~\ref{fig:snapshot} (a).  From the detector gain and
offset determined above, these correspond to $\Sigma = 9.4$, $\mu = 18.6$, and
$\sigma = 3.6$ photons.  The variance of the background $\sigma^2$ does not match the mean of
the background, suggesting that the model of normal-distributed background does not well
describe the data.  By applying Eqns. (\ref{eq:solutions-2}), the model of discrete
statistics, to the same region after first converting the detector signal into photon
counts we obtain the estimates of $\bar{\Sigma}= 5.9$ and $\bar{\mu}= 21$ photons.  That
is, the discrete model yields a larger estimate for the background and a smaller estimate
for the molecular diffraction.  The non-discrete analysis determines the magnitude of the
molecular diffraction signal based on the skew of the distribution, but photon counting
creates an inherent skew in any case.

\begin{figure}
 \begin{adjustwidth}{-2in}{0in}
  \setlength{\unitlength}{1mm}
  \begin{picture}(160,70)(0,0)
    \put(0,0){
      \includegraphics[width=8cm, keepaspectratio]{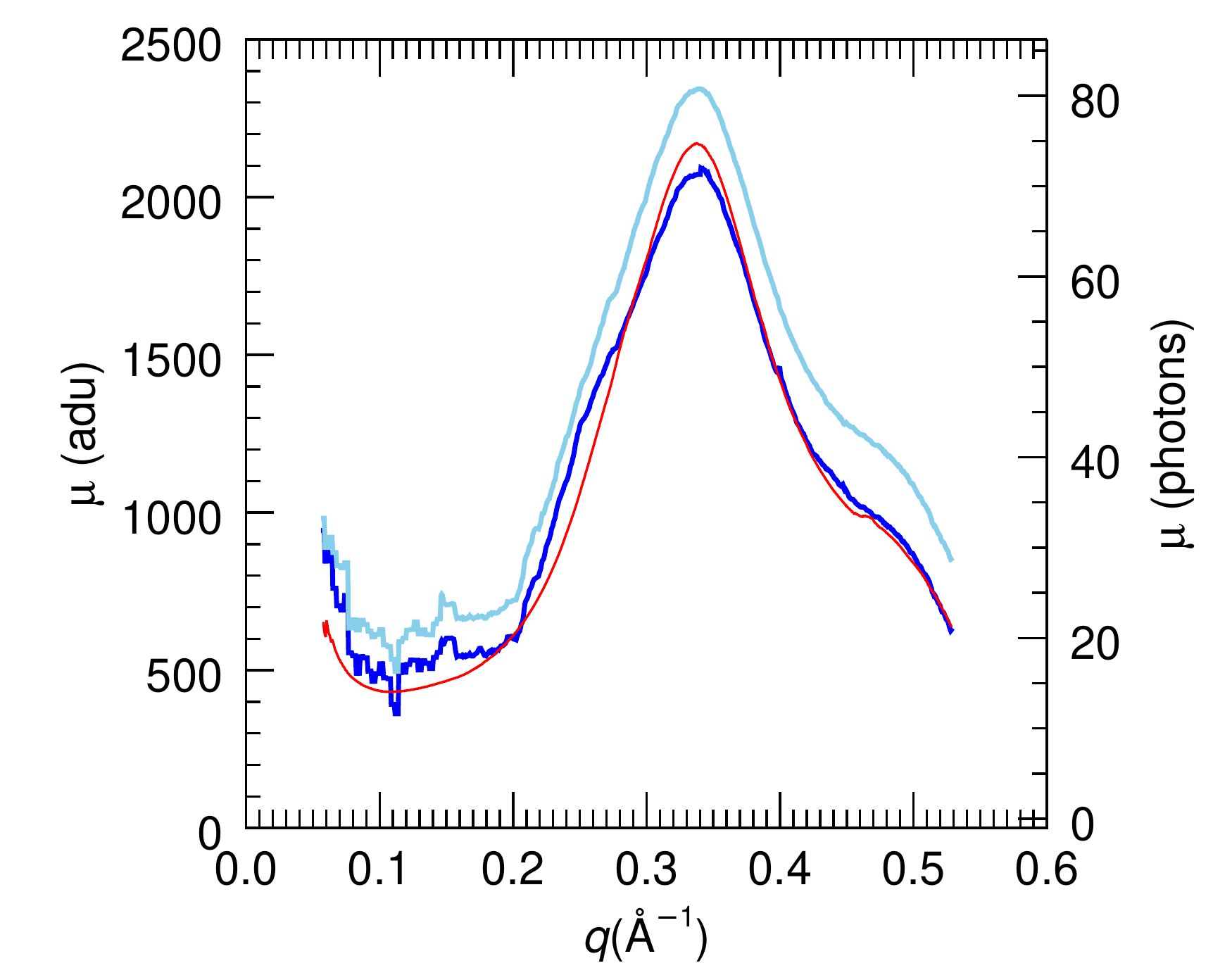}
    }
    \put(80,0){
      \includegraphics[width=8cm, keepaspectratio]{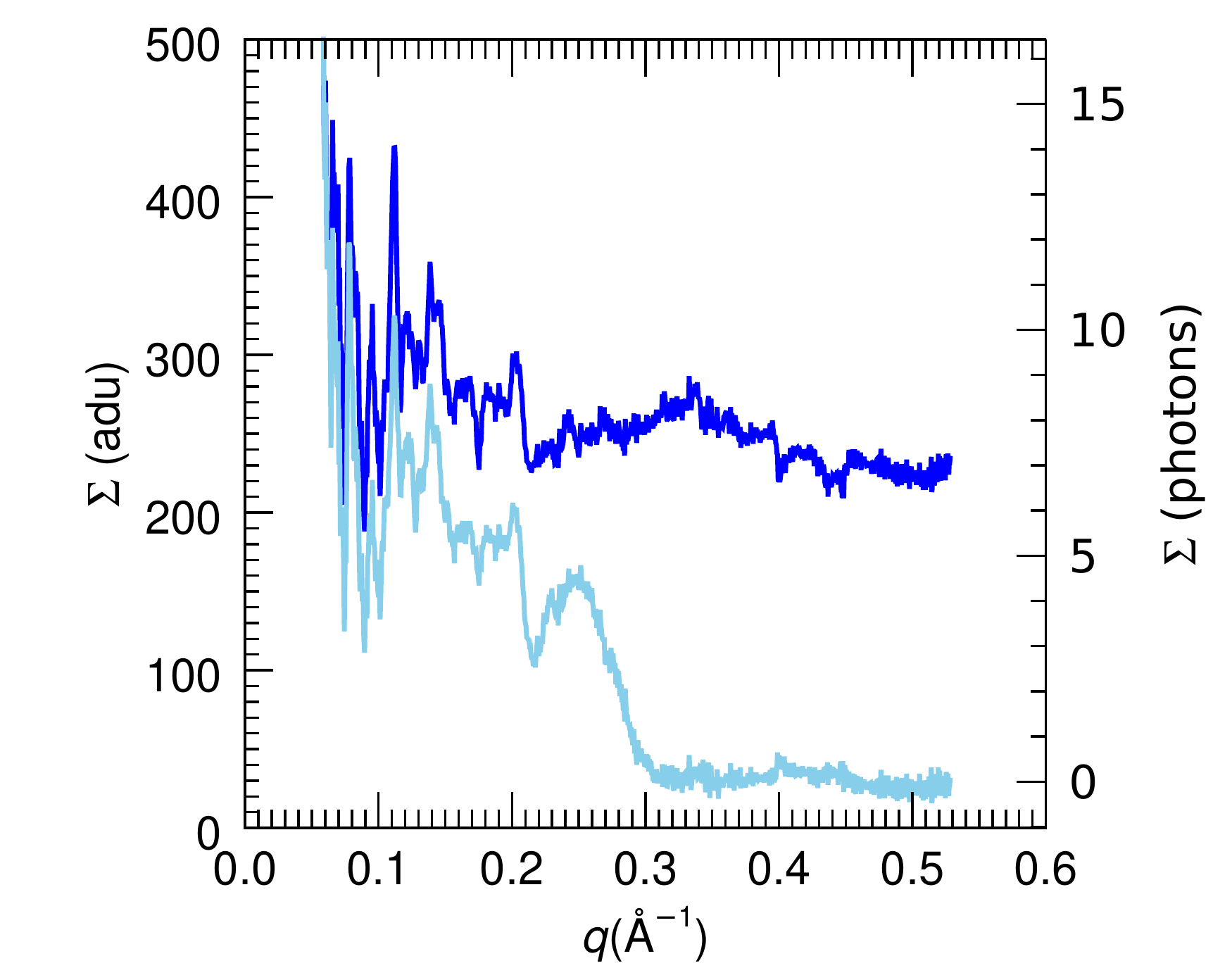}
    }
    \put(20,55){\figfont(\textit{a})}
    \put(140,55){\figfont(\textit{b})}
   \end{picture}
   \vspace{3mm}
  %  \begin{picture}(85,138)(0,0)   %Single column
  %   \put(0,70){
  %     \includegraphics[width=7.5cm, keepaspectratio]{chapman-fig6a.pdf}
  %   }
  %   \put(0,0){
  %     \includegraphics[width=7.5cm, keepaspectratio]{chapman-fig6b.pdf}
  %   }
  %   \put(66,128){\figfont(\textit{a})}
  %   \put(66,58){\figfont(\textit{b})} 
  % \end{picture}
 \caption{Plots of estimates of the background (a) and signal mean (b) for the pattern of
    Fig.~\ref{fig:snapshot} (a), obtained independently from contoured regions of the
    pattern obtained from the polarisation-uncorrected radial average.  Dark blue:
    $\mu(q)$ and $\Sigma(q)$ obtained by applying Eqns. (\ref{eq:solutions-1})
    (non-discrete statistics) to the moments of the intensity values. Sky blue: $\mu(q)$
    and $\Sigma(q)$ obtained by applying Eqns. (\ref{eq:solutions-2}) (discrete
    statistics) to the moments of the intensity values. Red: fitted radial average of a
    summed no-sample signal. \textcopyright The Authors licensed under CC BY 4.0}
  \label{fig:estimates}
  \end{adjustwidth}
\end{figure}

 A plot of the estimated background $\mu$, as a function of $q$, is given in
Fig.~\ref{fig:estimates} (a) for the pattern of Fig. \ref{fig:snapshot} (a).  The
values obtained from the moments of the intensity values using
Eqns. (\ref{eq:solutions-1}), assuming non-continuous statistics, are plotted in
blue, and those using Eqns. (\ref{eq:solutions-2}) are plotted in sky blue.  As with
the values shown in Fig.~\ref{fig:histogram} (a) at $q = \SI{0.15}{\per\angstrom}$, the use of
the discrete distributions consistently estimates a higher background.  In red the radial
average of the no-sample background is also plotted, scaled to fit the background
estimates.  The form of the background $\mu(q)$ matches the no-sample signal, but there
are some differences which could possibly be due to a different composition of the solvent
in the crystal to the buffer.  Plots of the estimated signal $\Sigma(q)$ are shown in
Fig.~\ref{fig:estimates} (b).  Here again, the estimate based on discrete statistics
appears more reasonable, decreasing to near zero at the highest values of $q$.  In that
region the variance of the photon counts was approximately equal the mean and hence
attributed to the Poisson-distributed background.  We expect, from
Eqn.~(\ref{eq:diffraction}), that the continuous diffraction should be zero at
$q = 0$ and modulated by $1-\exp(-4\pi^2 \sigma_\Delta^2 q^2)$.  Nevertheless, even with
this modulation that approaches unity at large $q$, the diffraction signal diminishes with $q$
due to the dependence of the atomic form factors and possibly due to conformational
variations in the molecules.

\begin{figure}
  \vspace{.5cm} % set vertical space between text and figure
  \begin{adjustwidth}{-2in}{0in}
  \setlength{\unitlength}{1mm}
  % \begin{picture}(130,130)(0,0)  %This is better laid out as at full page width, see below
  %                               %commented out
  %   \put(0,65){
  %     \includegraphics[width=6cm, keepaspectratio]{chapman-fig7a.png}
  %   }
  %   \put(65,65){
  %     \includegraphics[width=6cm, keepaspectratio]{chapman-fig7b.png}
  %   }
  %   \put(0,0){
  %     \includegraphics[width=6cm, keepaspectratio]{chapman-fig7c.png}
  %   }
  %  \put(130,70){
  %     \frame{\includegraphics[width=3mm,height=4.8cm]{color_bar.png}}
  %   } 
  %   \put(131.5,66){\fontfamily{phv}\fontsize{8pt}{8pt}\selectfont 0}
  %   \put(130.5,120){\fontfamily{phv}\fontsize{8pt}{8pt}\selectfont 600}
  %   \put(3,120){\textcolor{white}{\figfont(\textit{a})}}    
  %   \put(68,120){\textcolor{white}{\figfont(\textit{b})}}
  %   \put(3,55){\textcolor{white}{\figfont(\textit{c})}}
  % \end{picture}
  \begin{picture}(180,60)(0,0)  %Full page width
    \put(0,0){
      \includegraphics[width=5.7cm, keepaspectratio]{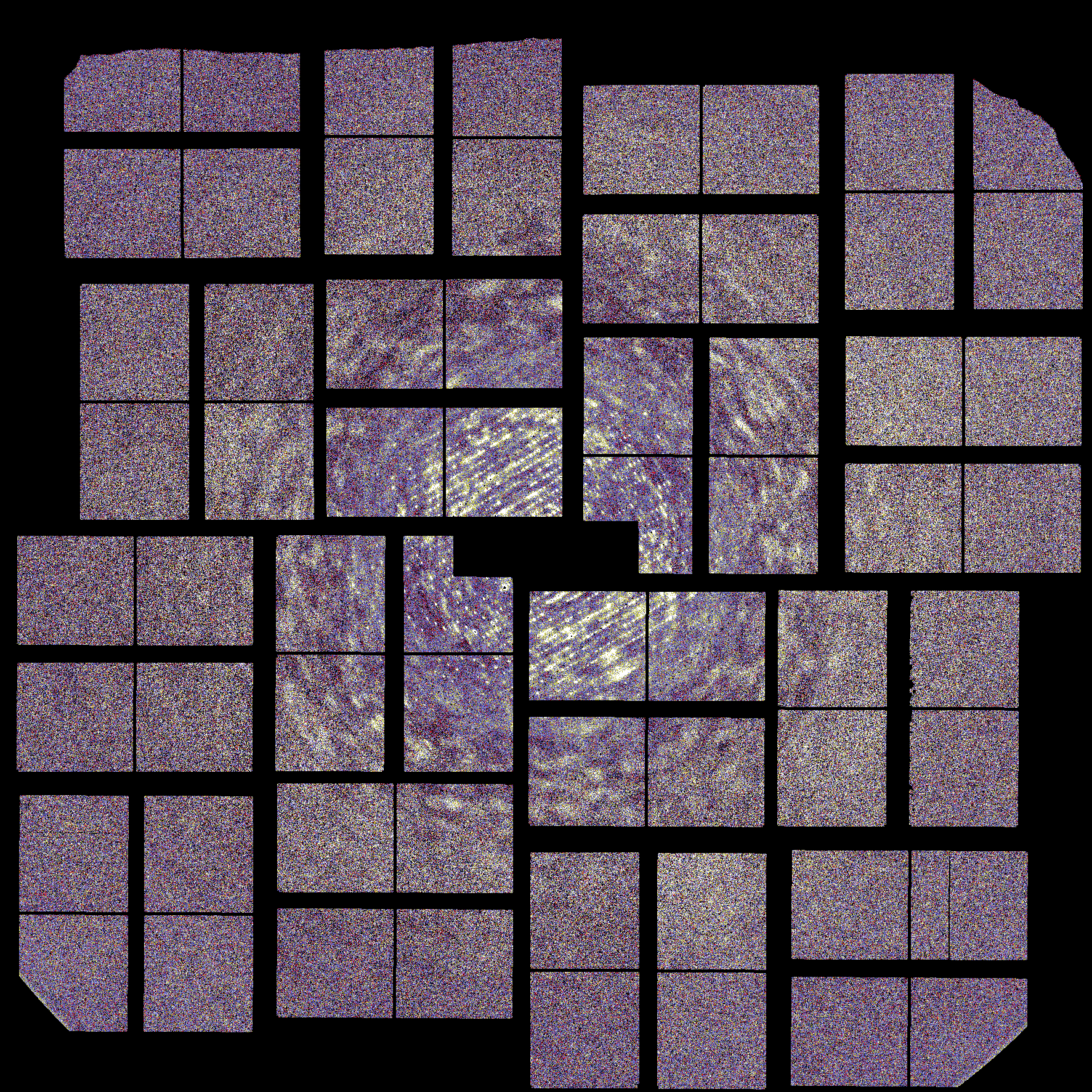}
    }
    \put(60,0){
      \includegraphics[width=5.7cm, keepaspectratio]{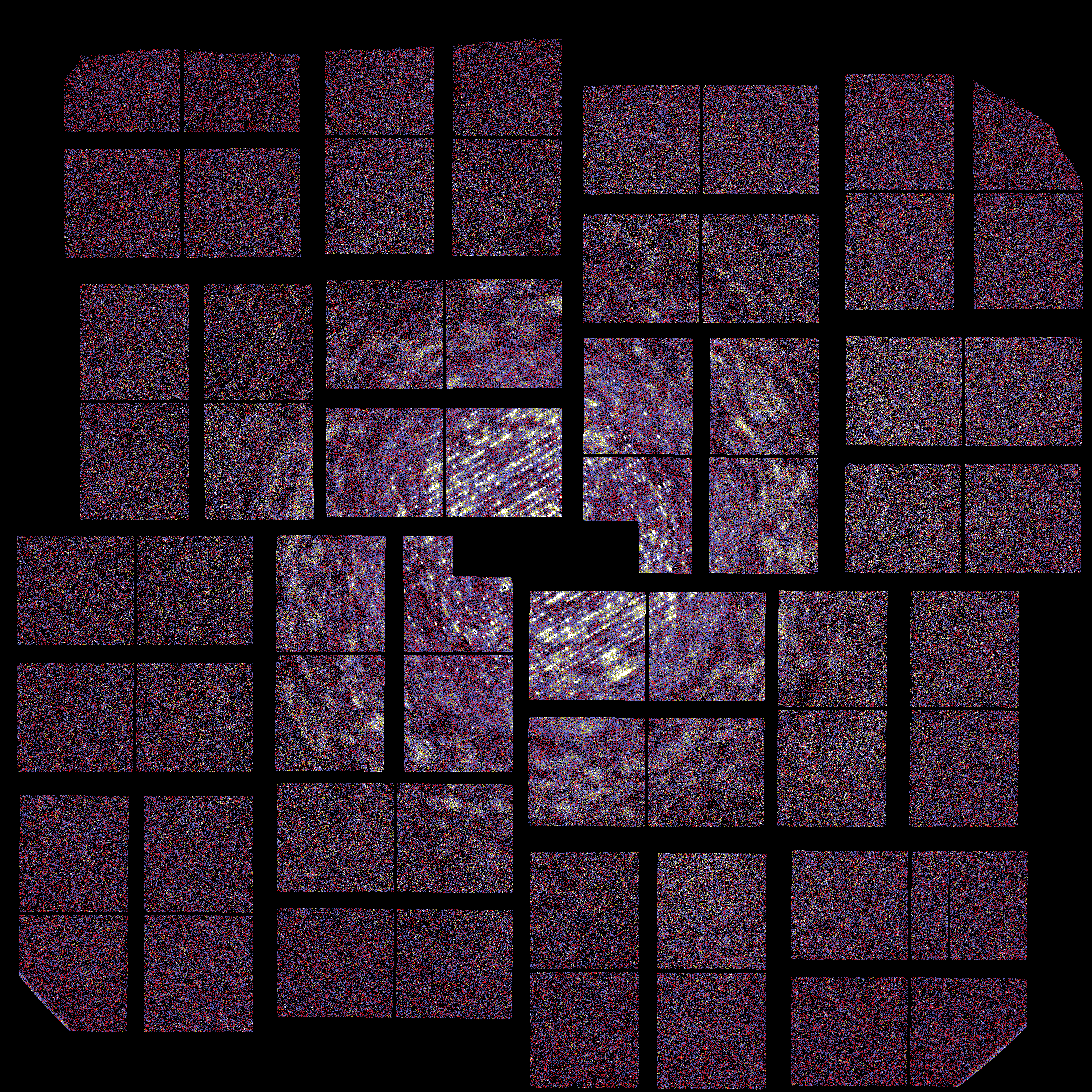}
    }
    \put(120,0){
      \includegraphics[width=5.7cm, keepaspectratio]{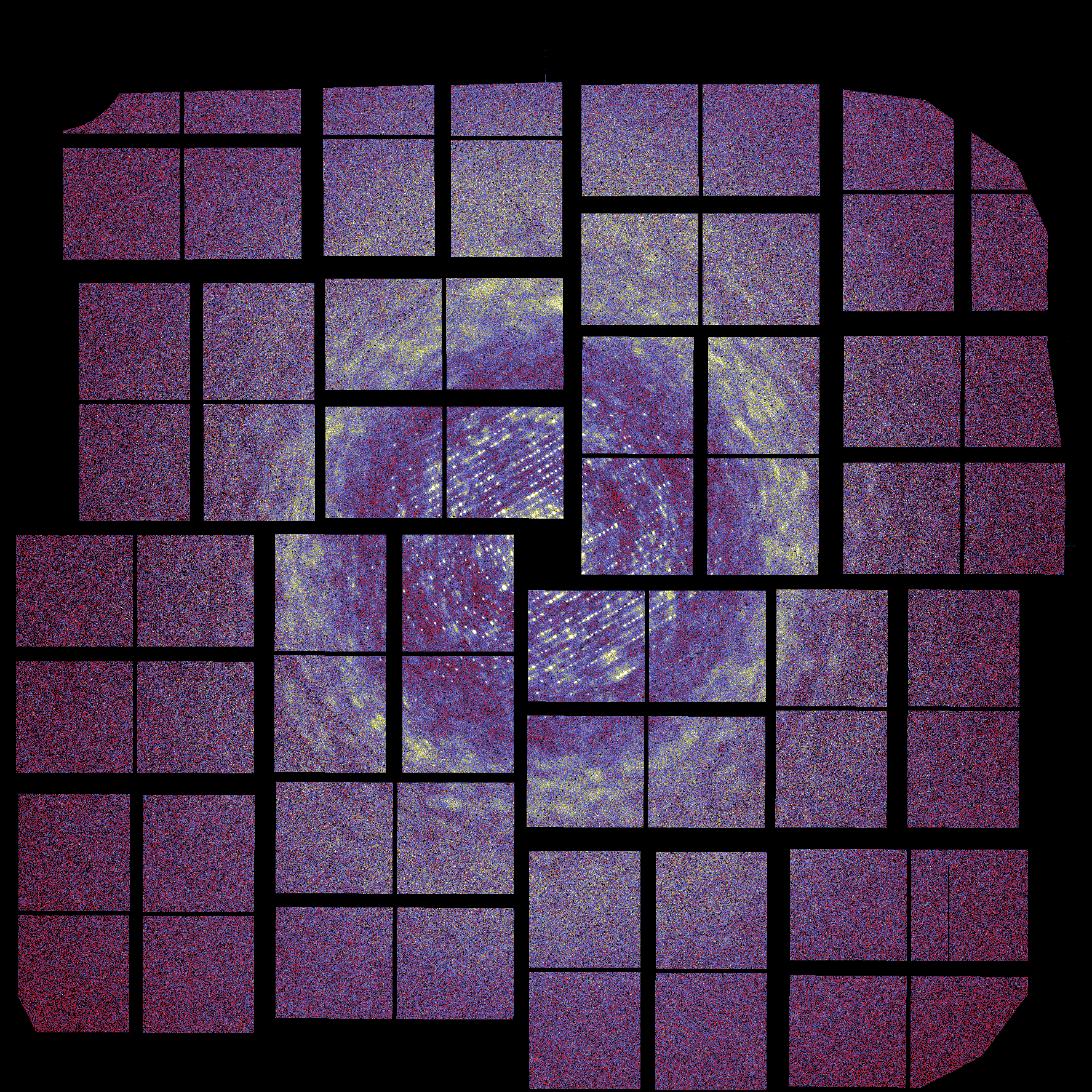}
    }
   \put(178,5){
      \frame{\includegraphics[width=2mm,height=4.5cm]{color_bar.png}}
    } 
    \put(178.5,1){\fontfamily{phv}\fontsize{8pt}{8pt}\selectfont 0}
    \put(178,56){\fontfamily{phv}\fontsize{8pt}{8pt}\selectfont 600}
    \put(3,53){\textcolor{white}{\figfont(\textit{a})}}    
    \put(63,53){\textcolor{white}{\figfont(\textit{b})}}
    \put(123,53){\textcolor{white}{\figfont(\textit{c})}}
  \end{picture}
  \vspace{3mm}
  \caption{The pattern from Fig.~\ref{fig:snapshot} (a) after subtracting the
    background $\mu(k_x,k_y)$ as calculated from the intensity moments for non-discrete
    statistics (a) and discrete statistics (b), and after subtracting the scaled no-sample
    signal (c). The colour scale ranges from 0 to \SI{600}{\adu}, corresponding to 0 to 23
    photons. \textcopyright The Authors licensed under CC BY 4.0}
  \label{fig:snapshot-corrected}
  \end{adjustwidth}
\end{figure}

Background-corrected patterns, obtained by subtracting backgrounds $\mu(k_x, k_y)$ from
the pattern of Fig.~\ref{fig:snapshot} (a) are shown in
Fig.~\ref{fig:snapshot-corrected} for the cases of non-discrete statistics and
discrete statistics.  For discrete statistics the background estimates were calculated in
regions obtained by contouring the average pattern as calculated by
Eqn.~(\ref{eq:polarisation-average}), subtracting
this map from the polarisation-uncorrected pattern, before finally applying the
polarisation correction.  This way the variance in each region was calculated from the
detected counts, and was not affected by the polarisation correction factor.  Again it is
clear that the application of discrete statistics gives a more reasonable result.

The total photon count of the background-corrected pattern shown in
Fig.~\ref{fig:snapshot-corrected} (b) is \num{2.7e6} photons, which is only 2.6\% of the
total counts before background subtraction which is \num{1.03e8} photons.  Furthermore,
the Bragg peaks account for \num{0.58e6} photons.  This was found by summing the values in
pixels defined by the dilated mask mentioned above, which generously encompasses all Bragg
peaks and thus could be considered an overestimate.  It maybe somewhat surprising that the
continuous diffraction contains about 4.6 times the number of photons than Bragg counts.
The total scattering power of the asymmetric units does not change depending on whether
those units are arranged in a strictly periodic fashion or not, as can be seen from
Eqn.~(\ref{eq:diffraction}).  The continuous diffraction extends over a much larger area
of reciprocal space, which may account for the factor of 4.6.  However, the atomic
scattering factors are stronger at low $q$ and so one may expect a greater proportion of
the total scattering in the Bragg peaks, depending on how much data is missing at lowest
$q$.  Nevertheless, it is clear that the continuous diffraction is not weaker in total
than the Bragg diffraction.  Since it is not concentrated into narrow Bragg peaks but
spread over many pixels of the detector the signal to noise of the continuous diffraction
is lower than the Bragg data.  From Figs.~\ref{fig:estimates} (a) and (b) the noise, given
by $\bar{\sigma} = \sqrt{\bar{\mu}}$ is comparable to the signal.  Note however, that
individual speckles cover more than 100 pixels, so these are measured with higher signal
to noise.

\subsection{Statistics of an ensemble of patterns}
\label{sec:ensemble}
\begin{figure}
  \begin{adjustwidth}{-2in}{0in}
    \setlength{\unitlength}{1mm}
   \begin{picture}(160,70)(0,0)  
    \put(0,0){
      \includegraphics[width=8cm, keepaspectratio]{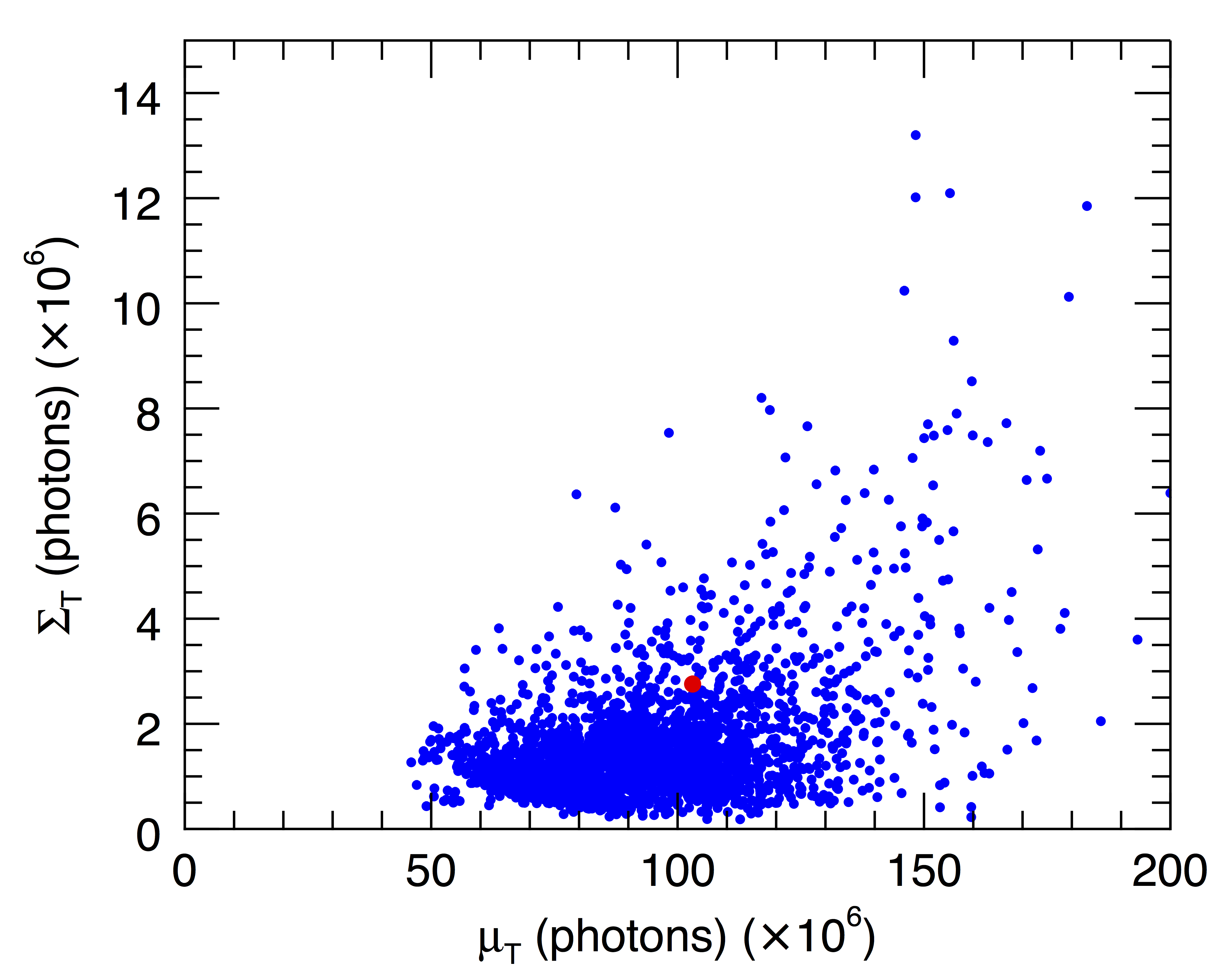}
    }
    \put(80,0){
      \includegraphics[width=8cm, keepaspectratio]{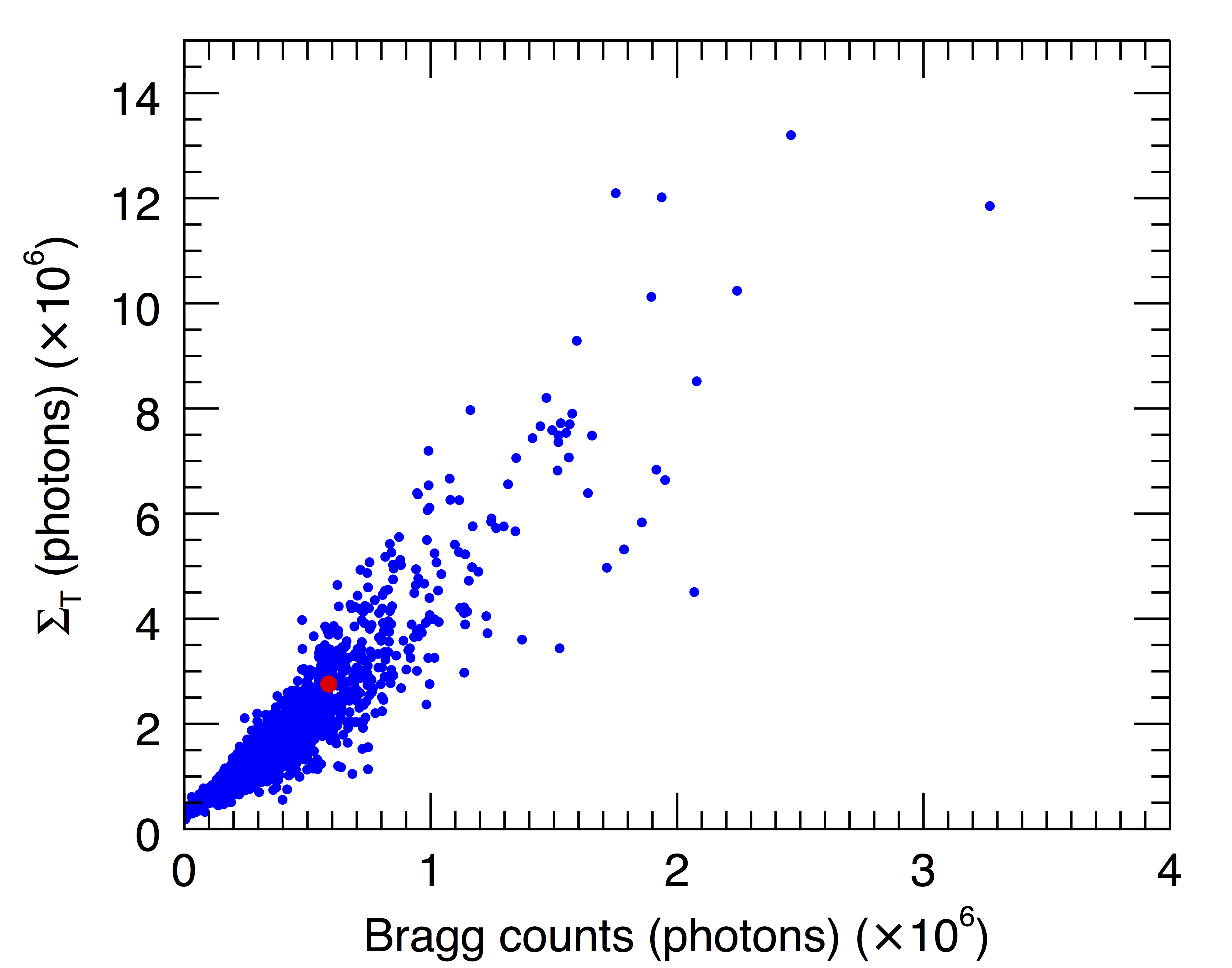}
    }
    \put(17,55){\figfont(\textit{a})}
    \put(97,55){\figfont(\textit{b})}
  \end{picture}
  \vspace{3mm}
  %  \setlength{\unitlength}{0.88mm}
  %  \begin{picture}(85,140)(0,0)  %Single column
  %   \put(0,70){
  %     \includegraphics[width=7.5cm, keepaspectratio]{chapman-fig8a.pdf}
  %   }
  %   \put(0,0){
  %     \includegraphics[width=7.5cm, keepaspectratio]{chapman-fig8b.pdf}
  %   }
  %   \put(17,128){\figfont(\textit{a})}
  %   \put(17,58){\figfont(\textit{b})}
  % \end{picture}
  \caption{Plots of the total continuous diffraction signal strength, $\bar{\Sigma}_T$, as
    a function of the total background strength $\bar{\mu}_T$ (a), and as a function of
    the total Bragg counts (b). The continuous diffraction signal is correlated with Bragg
    counts but not with the background. The continuous diffraction signal of the strongest
    patterns is about 5\% to 10\% of the background, and more than four times the strength
    of the Bragg signal.  The red point indicates the pattern shown in
    Fig.~\ref{fig:snapshot} (a). \textcopyright The Authors licensed under CC BY 4.0}
  \label{fig:ensemble-plots}
  \end{adjustwidth}
\end{figure}

The analyses described in Sec.~\ref{sec:single} can be repeated on the set of snapshot
diffraction patterns recorded in a serial crystallography experiment, in order to obtain
statistical measures of the experiment or to guide strategies of combining patterns into a
dataset (see Sec.~\ref{sec:analysis-3D}).  The PS II dataset reported by
Ayyer \emph{et al.} \cite{Ayyer:2016} consisted of \num{25585} snapshot patterns with Bragg peaks that could
be indexed to obtain the orientation of the diffraction in the frame of reference of the
crystal lattice.  In that work, the strongest \num{2848} patterns were oriented and aggregated
in a 3D reciprocal-space array for phasing.  That subset was reanalysed to obtain
parameters $\mu(k_x, k_y)$ and $\Sigma(k_x, k_y)$ in regions of near constant photon
counts in the polarisation-uncorrected patterns.  The overall strengths of the background
$\bar{\mu}_T$ and the continuous diffraction signal $\bar{\Sigma}_T$ in the PS II patterns
was estimated for each pattern by summing the parameters over the entire pattern, weighted
by the areas of each region. In Fig.~\ref{fig:ensemble-plots} (a) the dependence of the
total signal $\bar{\Sigma}_T$ is plotted as a function of the background $\bar{\mu}_T$.
The strength of the diffraction is about 1\% to 5\% of the background.  It is not strongly
correlated to the background except that the very strongest diffraction signals coincide
with the very strongest background.  This trend may suggest that a portion the background
is inherent to the liquid jet, with higher pulse energies giving rise to both strong
background and strong diffraction. Atomic diffuse scattering
caused by disorder of the atoms in the
molecules may also contribute to some portion of the background, perhaps induced by the pulse
itself and building up during the course of the pulse \cite{Barty:2012}.  The pattern
shown in Fig.~\ref{fig:snapshot} (a) is indicated by the red dots
in Fig.~\ref{fig:ensemble-plots}, with typical signal and
background strengths.

A plot of the continuous diffraction signal as a function of the total Bragg counts is
given in Fig.~\ref{fig:ensemble-plots} (b), indicating a high degree of correlation.
As with the pattern discussed in Sec.~\ref{sec:single}, the continuous diffraction
strength is about four times that of the Bragg counts, on average.  The plot suggests that
the strength of the continuous diffraction depends on the volume of the crystal in the
same way as the total Bragg counts depends on the total number of unit cells contributing.
This strong degree of correlation also indicates that all crystals possess a similar
degree of disorder, such that the fraction of scattered counts in Bragg peaks versus
continuous diffraction is roughly constant.

\section{Statistics of the 3D Continuous Diffraction Intensities}
\label{sec:analysis-3D}
A 3D dataset of the continuously varying diffraction intensities of PS II was constructed
using the approach described by Yefanov \emph{et al.} \cite{Yefanov:2014} and Ayyer
\emph{et al.} \cite{Ayyer:2016}.  Briefly, in this
approach the orientation of each snapshot diffraction pattern was determined by indexing
its Bragg spots using the software CrystFEL \cite{White:2012}.  The pattern was then
interpolated onto the appropriate spherical surface (the Ewald sphere) in a 3D array of
reciprocal space, where the coordinates of the array were chosen to be parallel to the
reciprocal lattice axes.  Compared with the previous work \cite{Ayyer:2016}, the smoothed
background $\mu(q)$, interpolated onto the detector plane, was first subtracted from each
pattern, which was also scaled by $1/\Sigma_T$, before merging into the 3D volume.

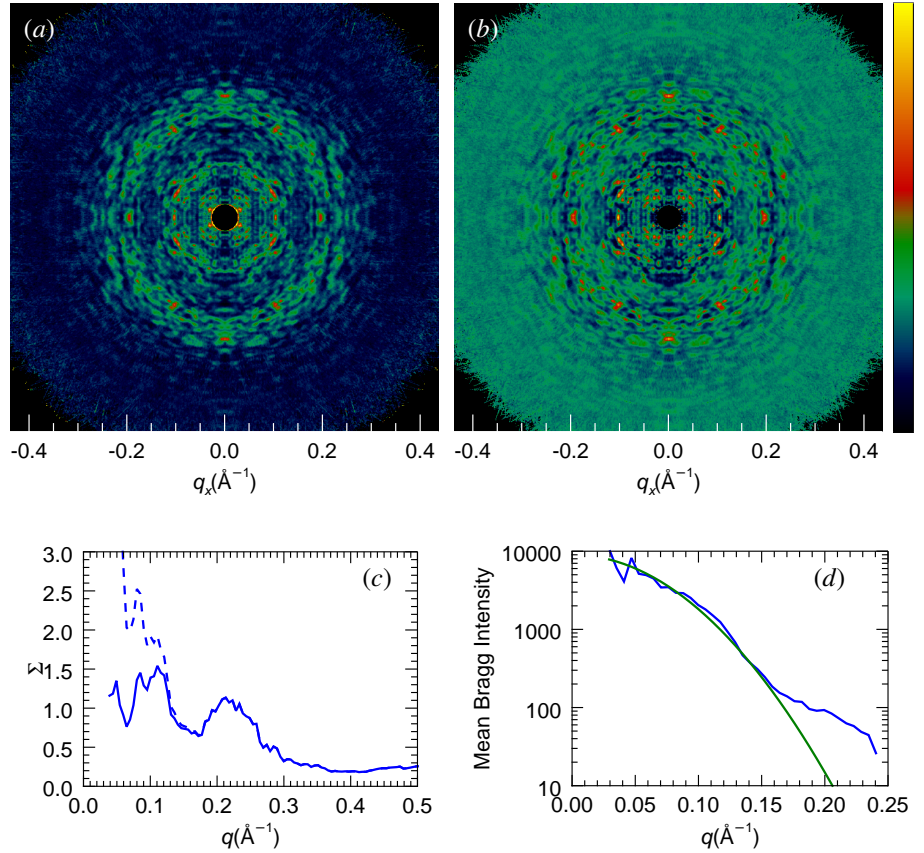
\begin{figure}
  \setlength{\unitlength}{1mm}
  \begin{picture}(120,113)(0,0) %1.5 Column width
    \put(0,46.5){
      \includegraphics[width=5.7cm, keepaspectratio]{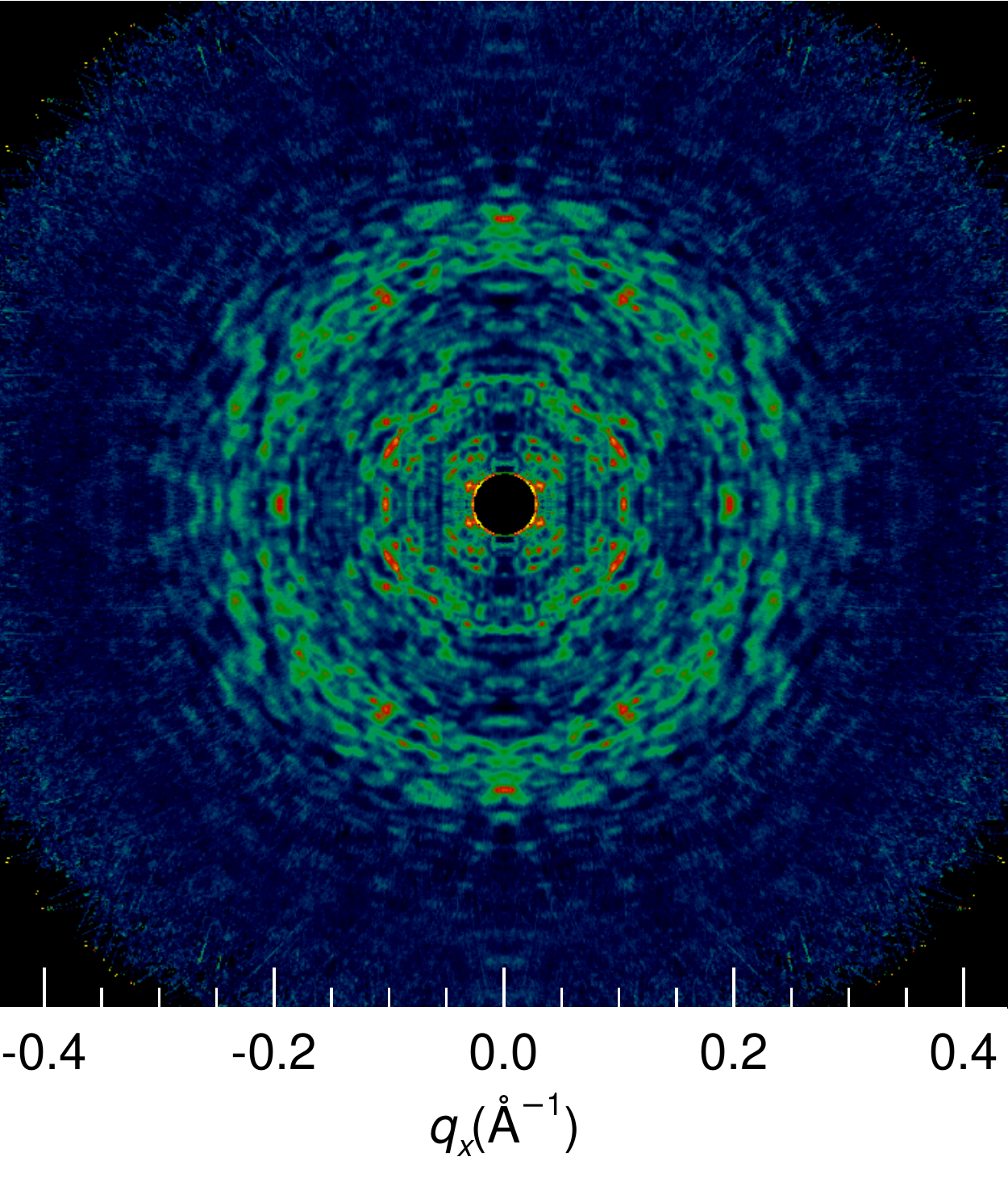}
    }
    \put(59,46.5){
      \includegraphics[width=5.7cm, keepaspectratio]{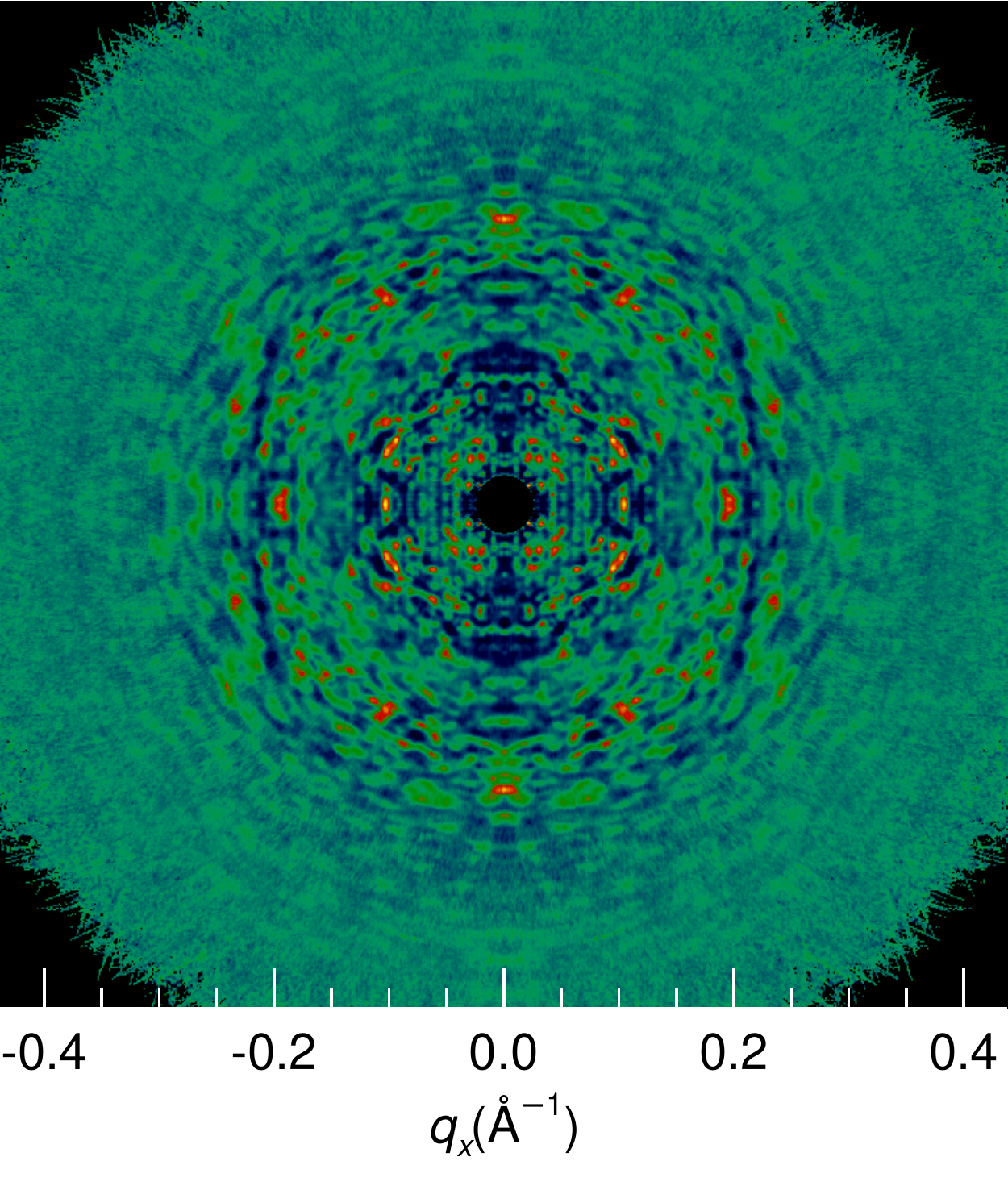}
    }
    \put(0,0){
      \includegraphics[width=6cm, keepaspectratio]{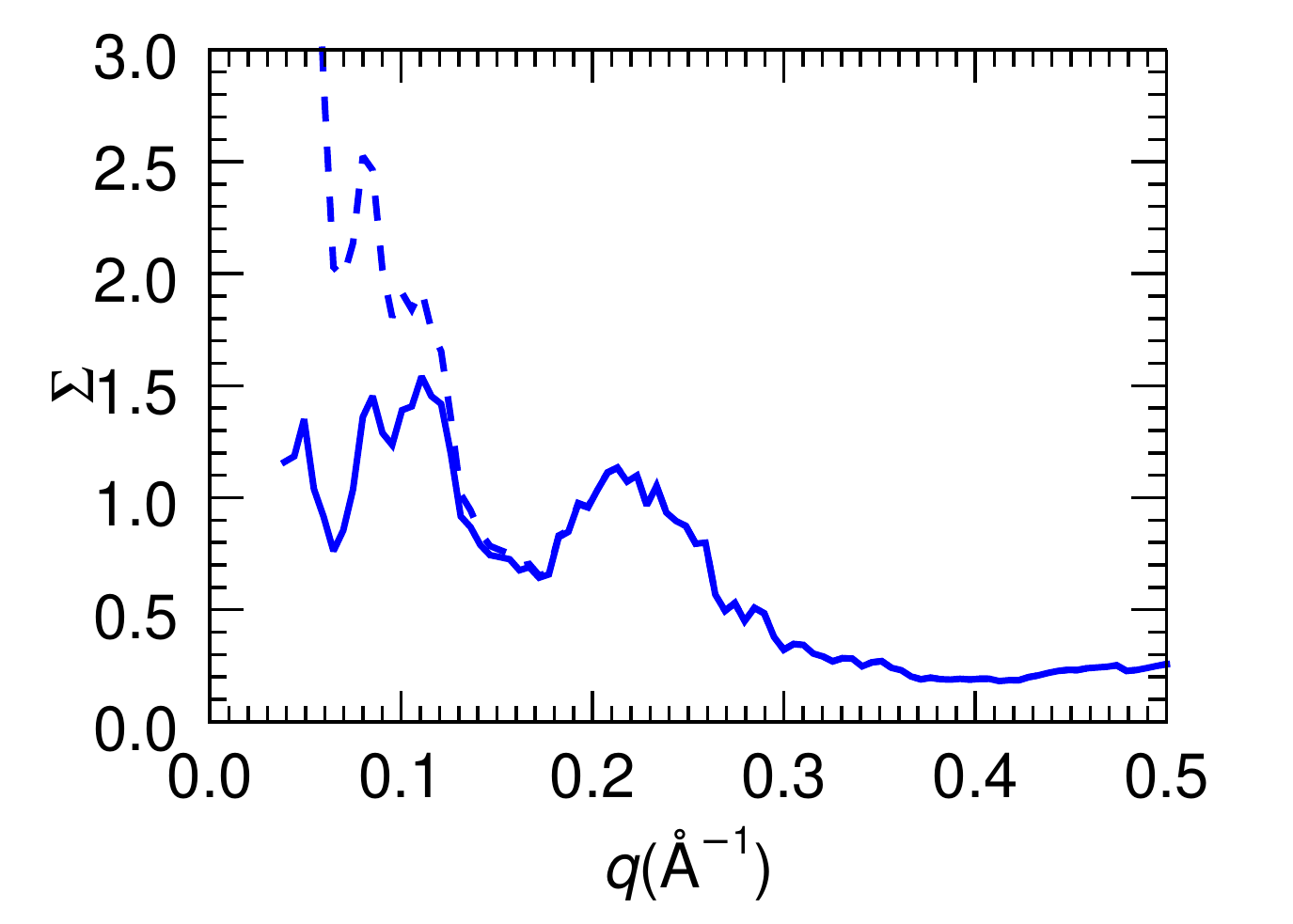}
    }
    \put(60,0){
      \includegraphics[width=6cm, keepaspectratio]{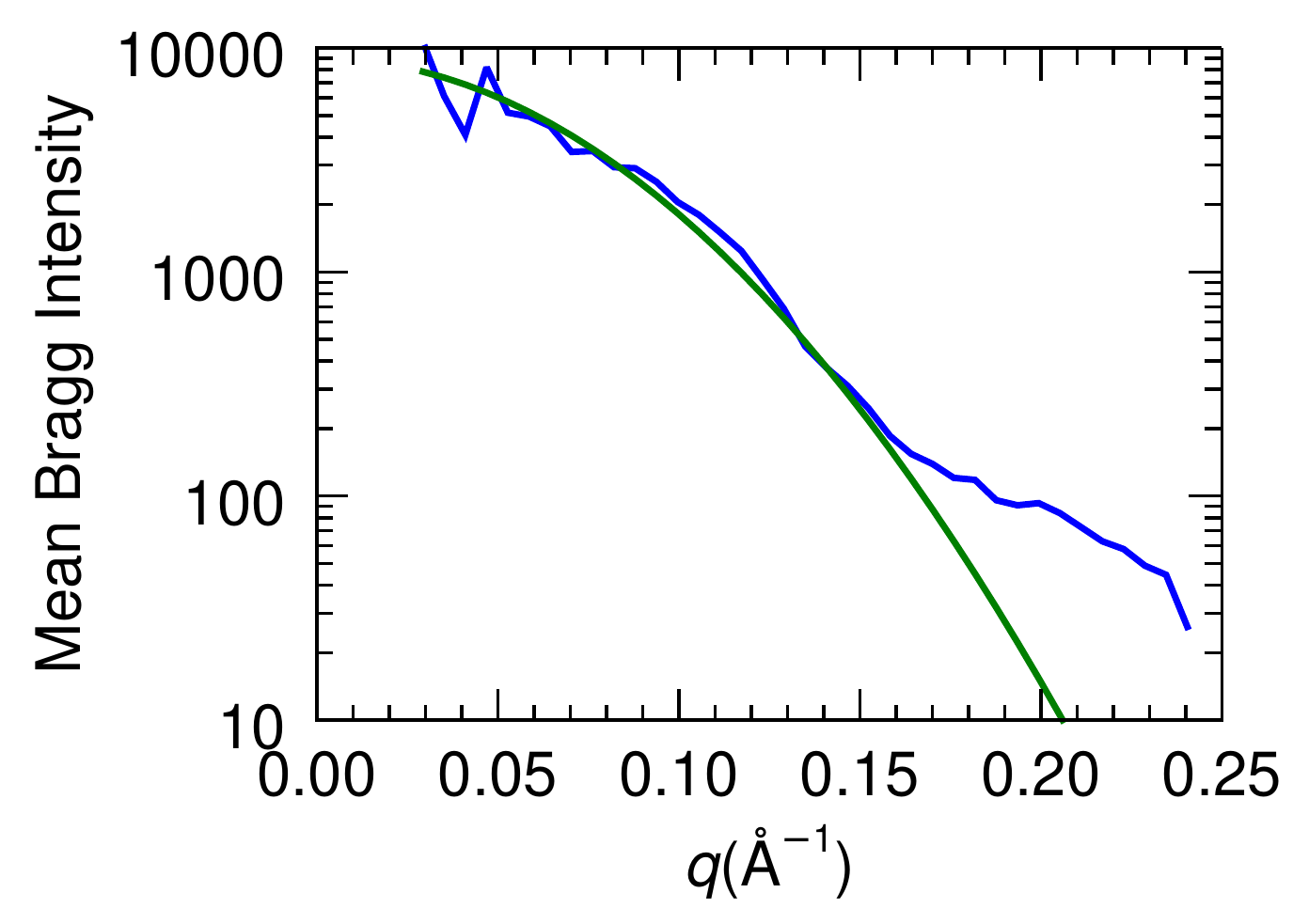}
    }
    \put(117.5,56.5){
      \frame{\includegraphics[height=5.7 cm,width=0.25 cm]{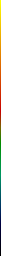}}}
    \put(3,109){\textcolor{white}{\figfont(\textit{a})}}   
    \put(62,109){\textcolor{white}{\figfont(\textit{b})}}  
    \put(48,36){\figfont(\textit{c})}
    \put(108,36) {\figfont(\textit{d})}
  \end{picture}
  \caption{(a) A central section of the merged volume of continuous diffraction intensities
    using the method of this paper, and (b) the previously published results. The
    central sections are normal to [101], chosen to avoid the centric planes. The colour
    scale is indicated to the right and varies from \num{-0.6} to 6 for (a) and \num{-100} to 250
    for (b).  (c) The scaling $\Sigma(q)$ obtained by fitting the distribution
    $p_{NW}(I, 4)$ to the merged continuous diffraction intensities in 3D shells of $q$.
    The dashed line gives the scaling corrected for the complementary Debye-Waller factor
    with $\sigma_\Delta = \SI{2.01}{\angstrom}$. (d) The scaling $\Sigma(q)$ obtained by
    fitting $p_{NW}(I, 1)$ to merged Bragg intensities in 3D shells of $q$ in blue and the
    fit to a Debye-Waller factor with $\sigma_\Delta = \SI{2.01}{\angstrom}$, in
    green. \textcopyright The Authors licensed under CC BY 4.0}
  \label{fig:merge}
\end{figure}

The \num{2514} strongest patterns were chosen based on the values of $\Sigma_T$.  After
merging the patterns into the 3D array, the symmetry operations of the point group 222
were then applied corresponding to summing the 3D intensity array with copies of itself
rotated about each of the three orthogonal axes of the crystal.  This symmetrisation
simply averages equivalent observations of intensities in order to increase the signal to
noise.  There is no loss of information in carrying out these operations since the crystal
exhibits this symmetry anyway and the averaging cannot be avoided.  We could also choose
to impose centrosymmetry, which loses any information pertaining to Bijvoet differences.
A map of the merged intensities in a central section normal to the [101] axis of the crystal
is given in Fig.~\ref{fig:merge} (a), which can be compared with the previously published
results in Fig.~\ref{fig:merge} (b) that were obtained by subtracting the radially-averaged
intensity from each pattern \cite{Ayyer:2016}.  Fig.~\ref{fig:merge} (b)
appears to show more detail and contrast at high resolution, at least with the chosen
colour scale. This may not be surprising, given that the no-crystal
patterns were typically fit before subtracting, resulting in high contrast and negative
intensities, neither of which accurately represents the incoherent sum of molecules in
several orientations. The new method avoids over-subtraction of
background and provides an improved scaling of the patterns.  

\begin{figure}
  \begin{adjustwidth}{-2in}{0in}
  \setlength{\unitlength}{1mm}
  \begin{picture}(160,70)(0,0)  
    \put(0,0){
      \includegraphics[width=8cm, keepaspectratio]{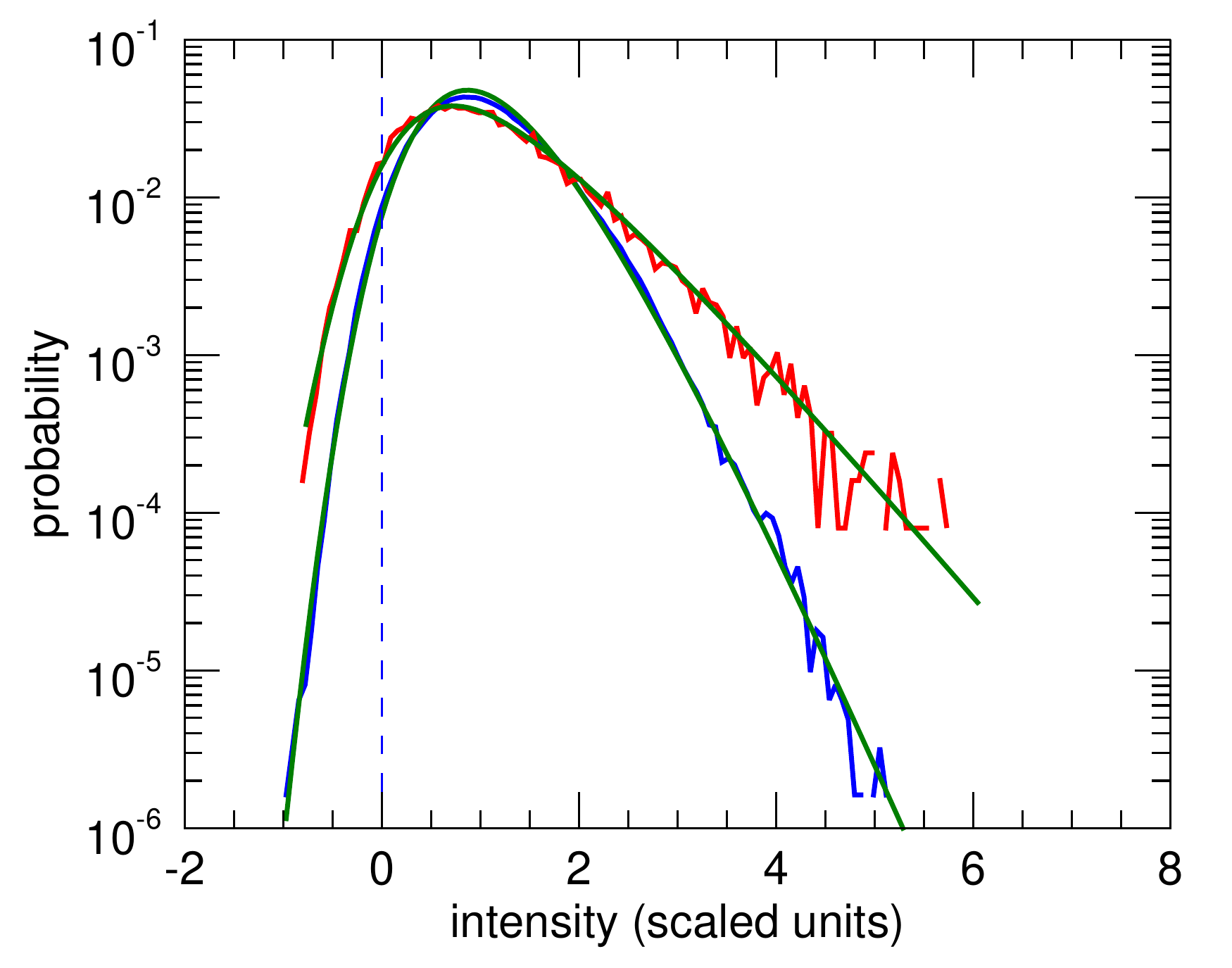}
    }
    \put(80,0){
      \includegraphics[width=8cm, keepaspectratio]{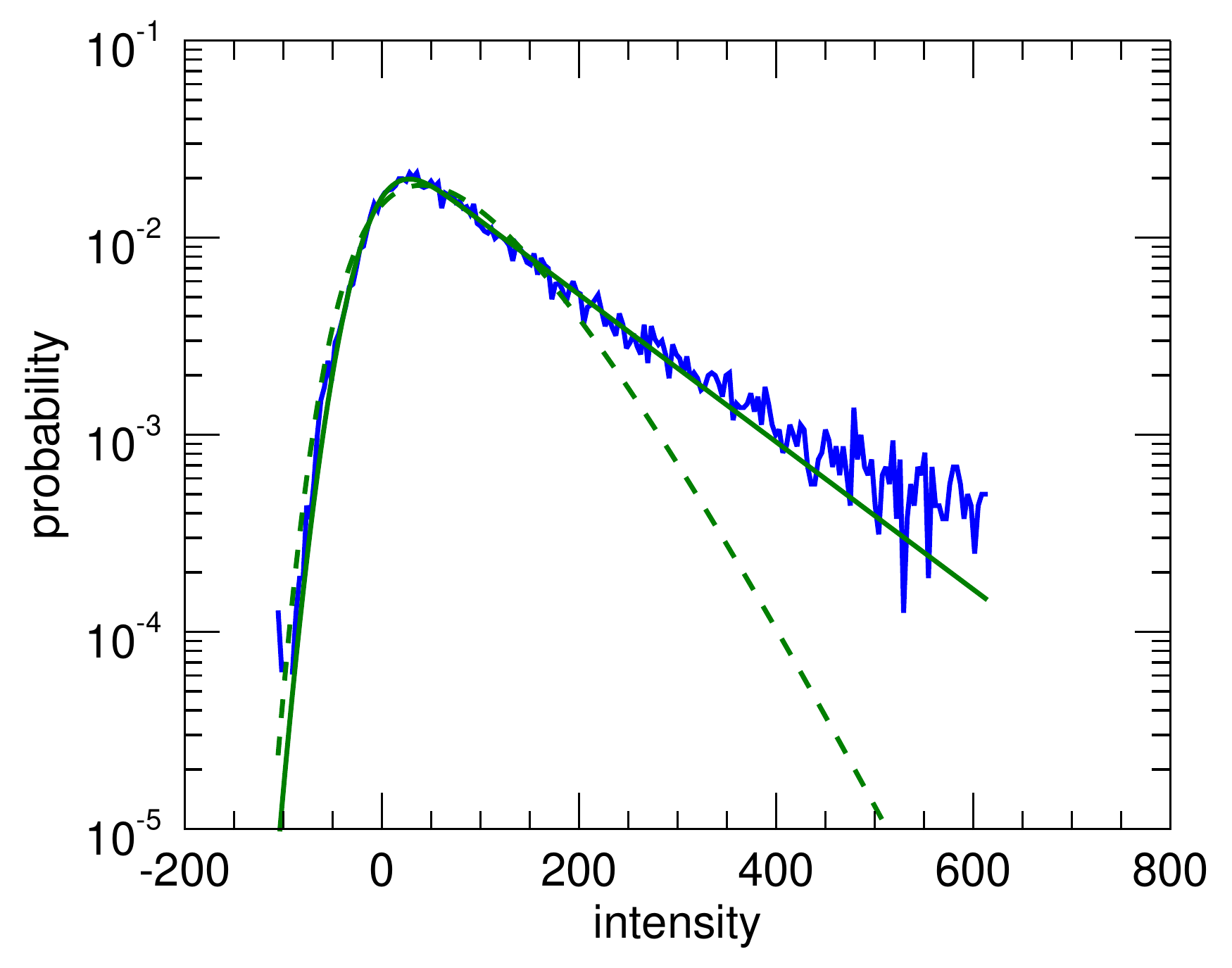}
    }
    \put(17,59){\figfont(\textit{a})}
    \put(97,59){\figfont(\textit{b})}
  \end{picture}
  % \setlength{\unitlength}{0.88mm}
   % \begin{picture}(85,140)(0,0)  %One column
  %   \put(0,67){
  %     \includegraphics[width=7.5cm, keepaspectratio]{chapman-fig10a.pdf}
  %   }
  %   \put(0,0){
  %     \includegraphics[width=7.5cm, keepaspectratio]{chapman-fig10b.pdf}
  %   }
  %   \put(17,126){\figfont(\textit{a})}
  %   \put(17,59){\figfont(\textit{b})}
  % \end{picture}
  \vspace{2mm}
 \caption{(a) Normalised histograms of intensities in the 3D continuous diffraction for
    voxels excluding central sections parallel to each of the two-fold rotation axes
    (blue) and for voxels lying on those planes (red), all contained within a shell
    $\SI{0.213}{\per\angstrom} < q < \SI{0.231}{\per\angstrom}$. The green lines are fits of
    $p_{NW}(I,4)$ and $p_{NW}(I,2)$. (b) Histogram of Bragg intensities in a shell between
    $\SI{1/6}{\per\angstrom} < q < \SI{1/5}{\per\angstrom}$ and excluding centric
    reflections (blue), with a fit of $p_{NW}(I,1)$ shown in green.  The green dashed line
    is a fit of $p_{NW}(I,4)$, showing that the Bragg intensities do not result from
    twinning. \textcopyright The Authors licensed under CC BY 4.0}
  \label{fig:histograms-3D}
  \end{adjustwidth}
\end{figure}

The statistics of the diffraction intensities in the 3D volume can be used to verify the
scaling and placement of the data, and to verify the number of independent orientations of
the rigid objects.  For PS II crystals, which have P$2_12_12_1$ symmetry, we expect that
non-centric continuous diffraction follows the the distribution $p_{NW}$ of Eqn.~(\ref{eq:pNW}) with a
twinning of $N = 4$ and that the zones perpendicular to each of the crystallographic
two-fold axes will be twinned with $N = 2$.  The intensities are expected to follow the
non-discrete Gamma distribution with a normally distributed background since they arise
from the sum of many scaled (and background subtracted) patterns.  Histograms of
intensities chosen from the shell lying between voxel radii of 170 and 185 ($\SI{0.213}{\per\angstrom} < q <
\SI{0.231}{\per\angstrom}$) are given in Fig.~\ref{fig:histograms-3D} (a), excluding the volume within 10 voxels of the three
orthogonal zones, and for only those voxels lying on the three orthogonal zones.  The
histograms are normalised to unity total, giving an experimental probability distribution.
It is immediately seen that the two distributions indeed are different, and fits of
Eqn.~(\ref{eq:pNW}) can be obtained for $N = 4$ and $N = 2$, respectively (green lines).  Furthermore
the fits were obtained for almost the same diffraction intensity mean, $\Sigma$, as expected.
The fitted parameters (in arbitrary units due to the scaling) were $\sigma = 0.30$ and $\Sigma = 1.05$
for the $N = 4$ ``non-centric'' intensities and $\sigma = 0.36$ and $\Sigma = 1.10$ for the $N = 2$ centric
intensities.  Although the residual background level given by $\mu$ was low, it was subtracted
to set this to zero. Thus there are some remaining negative intensities due to the
distribution of the noise. The average per-voxel signal to noise of the 3D intensities in
this shell is $\Sigma/\sigma = 3.5$, larger than that of the individual patterns due to signal
averaging.  With \num{2514} patterns included in this merged dataset, the multiplicity at
$q = \SI{0.22}{\per\angstrom}$
was 24.  The larger standard deviation for the centric intensities than for the centric
intensities could be attributed to the smaller sample size (\num{5.0e4} versus \num{4.9e6}
voxels). 

The voxels of the 3D array have a width of \SI{0.00125}{\per\angstrom}, which is larger than the
width of the detector pixels.  The largest diameter of the PS II dimer is \SI{178}{\angstrom},
and thus the largest extent of its autocorrelation is \SI{356}{\angstrom}.  The spacing for
Nyquist sampling of the diffraction intensities, which are equal to the Fourier transform
of the intensity pattern, is thus $\SI{1/356}{\per\angstrom} = \SI{0.0028}{\per\angstrom}$, giving a voxel
width relative to this of $w = 0.45$.  From Fig.~\ref{fig:coherence} (b) this should
not impact the coherence of the merged pattern or the intensity statistics.

The Bragg intensities obtained by processing all \num{25585} diffraction patterns using
CrystFEL, are found to follow a negative exponential distribution, as shown in
Fig.~\ref{fig:histograms-3D} (b) for a shell between $\SI{1/6}{\per\angstrom} < q <
\SI{1/5}{\per\angstrom}$, excluding 
centric reflections.  For this shell the intensities could be fit to the noisy Wilson
distribution $p_{NW}(I,1)$ with a mean signal $\Sigma = 116$ units and a background $\mu= -10.3$ units
with a standard deviation of $\sigma = 27.5$.  The expected distribution in this case is for $N =
1$ since there is no ambiguity of crystal orientation due to merohedry and hence no
effective twinning.  The intensities arise from coherent diffraction of the entire unit
cell, and there is only one instance of that unit cell contributing to the Bragg
intensities. The signal to noise in this shell is $\Sigma /\sigma = 4.2$ which is only moderately
greater than for the continuous diffraction even though ten times the number of patterns
contribute and the intensities are concentrated into Bragg peaks.  The distribution of
Bragg intensities is clearly different to the continuous diffraction as can be seen in the
slopes of the distributions in Fig.~\ref{fig:histograms-3D}.  Other values of $N$ do not fit as well to the
distributions of the continuous and Bragg intensities.  The fitting essentially amounts to
the early twin tests \cite{Rees:1980} which identified twinning based on comparing the form of
the cumulative distribution of intensities to the appropriate Gamma distributions.  The
fits here confirm that the continuous diffraction is indeed due to the incoherent sum of
four independent objects whereas the Bragg diffraction arises from a single untwinned
crystal.  This is supporting evidence that the continuous diffraction measured from the PS
II crystals does indeed arise from translational disorder of PS II dimers and not the
monomers, since there are four orientations of PS II dimers in the crystals but eight
orientations of monomers.  The statistics alone can not reveal if the four objects are
identical, but the symmetry of the continuous diffraction suggests that if they are
different then they are equally distributed over the four orientations.  The statistics
also suggest that the background-corrected continuous diffraction does not have a
significant contribution due to structural variability such as conformational disorder,
since the diffraction from many smaller sub-structures would give rise to intensities
approaching a Poisson distribution (the large $N$ limit of a Gamma distribution)
and such diffraction presumably would not be completely rotationally invariant as was the
Poisson-distributed background that was subtracted from each pattern.  Some degree of
orientational disorder of the rigid units is certainly possible, which would have the
effect of reducing the diffraction contrast with increasing $q$, due to the blurring of
speckles, as discussed in Sec.~\ref{sec:stats-mol}.

The scaling of the Bragg intensities as a function of $q$ is shown in Fig.~\ref{fig:merge}
(d), plotted on a log scale, for comparison with the continuous diffraction plotted in
Fig.~\ref{fig:merge} (c) on a linear graph.  This scaling predominantly follows the
familiar Wilson plot of Bragg intensities and the Debye Waller factor
$e^{-4\pi^2\sigma_\Delta^2 q^2}$ was fit with $\sigma_\Delta=\SI{2.01}{\angstrom}$, which can be
equated with an overall $B=8\pi^2\sigma_\Delta^2 = \SI{320}{\angstrom\squared}$.  That
is, this is the $B$ factor computed by attributing the reduction of Bragg intensity with $q$
to atomic displacement, whereas it is clear from the existence of the continuous
diffraction that the dependence of Bragg intensities with $q$ is mainly due to rigid body
displacements of the molecular complexes.  The effect of the complementary Debye Waller
factor $1-e^{-4\pi^2\sigma_\Delta^2 q^2}$ on the continuous diffraction is to suppress
intensity at values of $q < \SI{0.1}{\per\angstrom}$.  At higher photon momentum transfer than this, the
factor is greater than 0.8 and thus has little effect.  The mean intensity of the
continuous diffraction, corrected for this factor, is given in Fig.~\ref{fig:merge}
(c) as the dashed line.

\section{Comparison with Atomic Model}
\label{sec:comparison}
\begin{figure}
  \setlength{\unitlength}{1mm}
  \begin{picture}(130,100)(0,0)   %This should be a full-page figure - use the
                                                   % commented-out version below
    \put(0,60){
      \includegraphics[width=4cm, keepaspectratio]{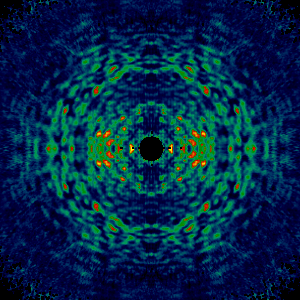}
    }
    \put(42,60){
      \includegraphics[width=4cm, keepaspectratio]{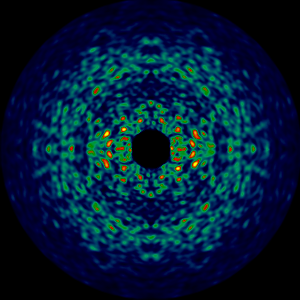}
    }
    \put(84,60){
      \includegraphics[width=4cm, keepaspectratio]{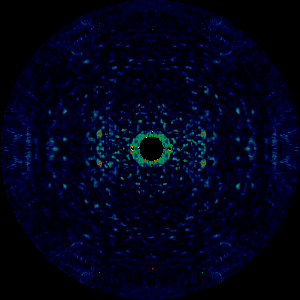}
    }
    \put(25,0){
       \includegraphics[width=7cm, keepaspectratio]{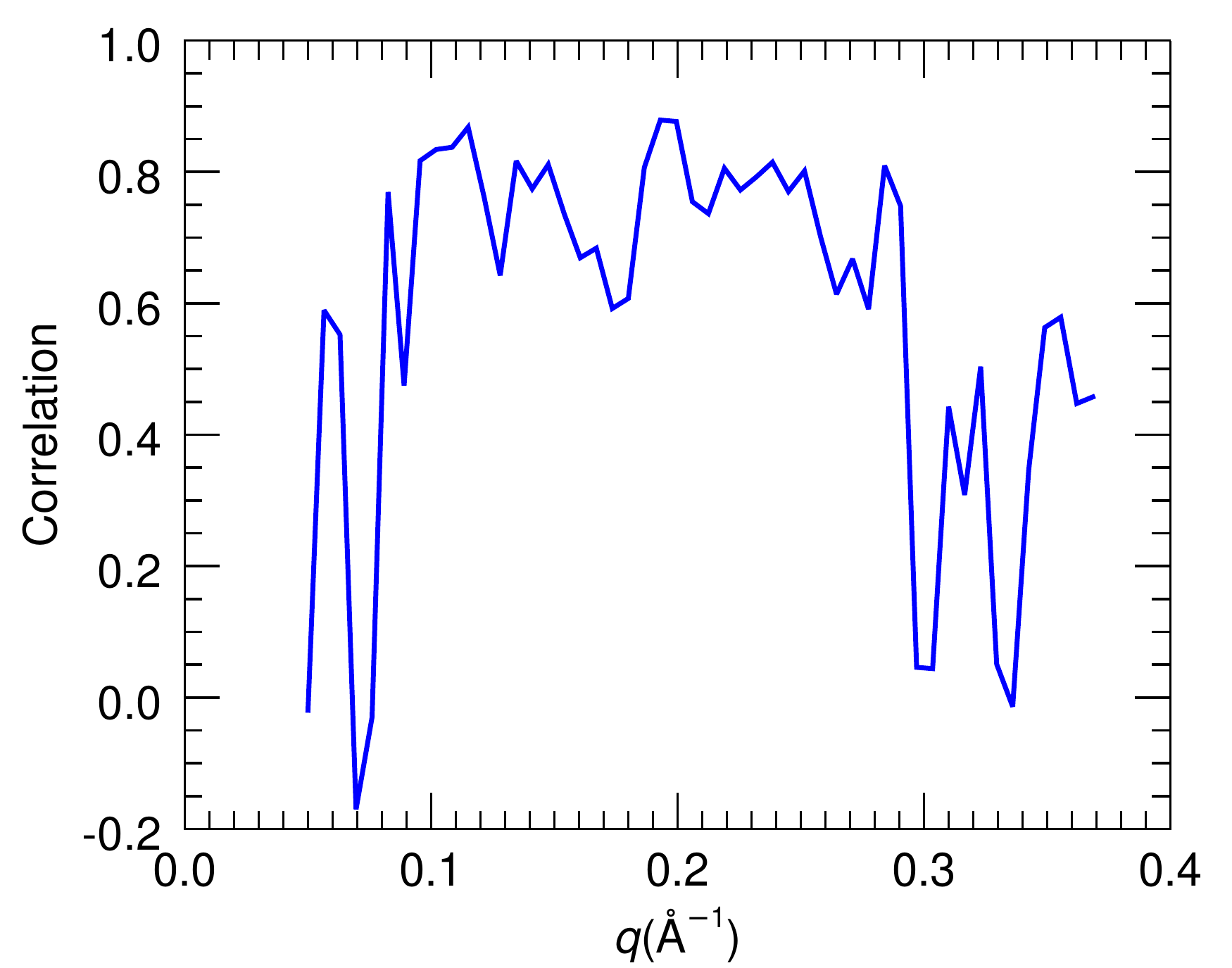}
    }
   \put(126,60){
      \frame{\includegraphics[height=4 cm,width=0.2 cm]{color_bar2.png}}}
    \put(3,96){\textcolor{white}{\figfont(\textit{a})}}
    \put(45,96){\textcolor{white}{\figfont(\textit{b})}}
    \put(87,96){\textcolor{white}{\figfont(\textit{c})}}
    \put(39,47){\figfont(\textit{d})}
  \end{picture}
  % \begin{picture}(180,120)(0,0)   %This should be a full-page figure
  %   \put(0,60){
  %     \includegraphics[width=5.7cm, keepaspectratio]{compare_merge4-fig2.png}
  %   }
  %   \put(59,60){
  %     \includegraphics[width=5.7cm, keepaspectratio]{compare_merge4-fig3.png}
  %   }
  %   \put(118,60){
  %     \includegraphics[width=5.7cm, keepaspectratio]{compare_merge4-fig4.png}
  %   }
  %   \put(55,0){
  %      \includegraphics[width=7cm, keepaspectratio]{compare_merge4-fig1.pdf}
  %   }
  %  \put(177,60){
  %     \frame{\includegraphics[height=5.7cm,width=0.2 cm]{wilson_3d_4-color-bar.png}}}
  %   \put(3,110){\textcolor{white}{\figfont(\textit{a})}}
  %   \put(62,110){\textcolor{white}{\figfont(\textit{b})}}
  %   \put(121,110){\textcolor{white}{\figfont(\textit{c})}}
  %   \put(69,47){\figfont(\textit{d})}
  % \end{picture}
  \caption{(a) A central section of the merged volume of continuous diffraction
    intensities, normal to the (010) lattice vector, compared with (b) the same section of
    the simulated continuous diffraction assuming a rotational disorder of \ang{1} RMS
    and a translational disorder of $\sigma_\Delta = \SI{2.01}{\angstrom}$.  (c) The
    difference of the experimental and simulated intensities, shown on the same colour
    scale as (a) and (b).  (d) Plot of the Pearson correlation in shells of $q$ between the
    experimental and simulated data. \textcopyright The Authors licensed under CC BY 4.0}
  \label{fig:central-section}
\end{figure}

As a final analysis of the continuous diffraction, we compare it with the continuous
diffraction of a disordered crystal of PS II as calculated from an atomic model.  For the
model we used atomic coordinates obtained by a refinement of a structure of the PS II
dimer to the electron density obtained by diffractive imaging \cite{Ayyer:2016}.  The
molecular transform $F(\boldsymbol{q})$ of the PS II dimer was calculated by summing diffracted
waves scattered from each atom on a 3D array of $q$ vectors spaced by \SI{0.0025}{\per\angstrom}
(twice that of the merged experimental data).  From this the square modulus
$|F(\boldsymbol{q})|^2$ was calculated before applying the rotation operations $\boldsymbol{R}_m$
of the point group of the crystal and incoherently summing the four equally-weighted sets
of intensities.  The 3D array was then multiplied by the factor
$1-e^{-4\pi^2\sigma_\Delta^2 q^2}$ with the previously determined value of
$\sigma_\Delta = \SI{2.01}{\angstrom}$.  It was found that the Pearson correlation between the
experimental and computed data for the volume within the shell
$\SI{0.088}{\per\angstrom} < q < \SI{0.29}{\per\angstrom}$ was \num{0.67}, compared with a
value of \num{0.55} 
obtained previously \cite{Ayyer:2016}.  An even higher degree of correlation of \num{0.77} was
obtained by blurring the computed intensities slightly by assuming rotational disorder of
the PS II dimers by \ang{1} RMS.  To simulate this disorder, the symmetrised intensities were rotated in
all three directions, by amounts chosen from a normal distribution with a width of
\ang{1}.  This was repeated 500 times and the results averaged.
Fig.~\ref{fig:central-section} displays the experimental and calculated intensities,
this time on one of the centric zones (normal to the 010 lattice vector), and the
difference, all on the same colour scale.  No manipulation of the background or scaling of
the data was made---that is, there are no fitted parameters other than the \ang{1}
rotational blurring.  A plot of the Pearson correlation coefficient computed in shells of
$q$ is also given in Fig.~\ref{fig:central-section}.  This reaches a maximum value of
\num{0.88}.  A very similar result was achieved by uniformly
convolving the computed data by a $4 \times 4 \times 4$ voxel kernel instead of applying
the rotational blurring.  The high degree of correlation confirms the origin of the
continuous diffraction and validates the approach of distinguishing the molecular
diffraction from structureless background.

\section{Discussion and Conclusions}
We have carried out an extensive analysis of the continuous diffraction arising from
translationally disordered crystals of PS II that was used previously for
macromolecular coherent diffractive imaging \cite{Ayyer:2016}.  That the diffraction could be
directly phased, and used to obtain a volume image of the electron density of the
PS II dimer, was certainly strong evidence that the continuous diffraction
originates from the incoherent sum of randomly displaced rigid objects (the PS II
dimers), but here that particular analysis was expanded with rigorous statistical tests to
gain a deeper understanding of the nature of the continuous diffraction and how to measure
it.  One of the most crucial aspects in treating the continuous diffraction is
distinguishing it from diffuse (i.e. structureless) background scattering.  Unlike Bragg
peaks which can easily be discriminated from a slowly-varying background in the
diffraction pattern, the continuous diffraction cannot be readily separated from such
background.  In the previous work \cite{Ayyer:2016} the background was simply estimated
from the radial average of the patterns.  Such background was fitted to each crystal
diffraction pattern with a result of maximising the contrast of the speckles of the
molecular diffraction that ultimately led to an over-subtraction and negative intensities.
In this work the statistics of the molecular diffraction intensities were exploited to
obtain estimates of their scaling and zero level.  In particular, the intensities in a
shell of reciprocal space are assumed to follow a ``noisy Wilson'' distribution, which is
that due to the sum of a random variables describing the structured signal and the
unstructured background, where the signal follows the familiar Gamma distribution of
Wilson statistics and the noise follows a normal distribution.  When photon counting is
considered, this corresponds to the sum of discrete random variables from a negative
binomial distribution and a Poisson distribution.

The statistics of the structured component of the continuous diffraction depend on the
number of independent objects contributing, or the number of modes in the speckle pattern.
There are four orientations of dimers in PS II crystals, and since the displacements of
each are random and uncorrelated the diffraction from each orientation adds incoherently,
giving rise to continous diffraction with the same point group symmetry as the Bragg intensities.
The statistics of the intensities thus do not follow the usual negative exponential of a
single object, where the most common intensity value is zero (in between speckles), but a
Gamma distribution that shows it is unlikely that zero intensity from one mode matches up
with zero intensity from other modes.  This reduction of speckle contrast must be taken
into account when estimating the zero level of the structured diffraction.  One way to
achieve this is to fit the expected distribution to histograms of the measured
intensities.  More conveniently it is possible to solve for the means of the signal and
background from the moments of the measured intensities using the formulae in Table~\ref{tab:1}.

Perhaps one of the surprising aspects of the analysis is that the continuous diffraction
accounts for the majority of the diffracted signal.  With the ability to partition photon counts into Bragg peaks,
molecular diffraction, and background we found that the continuous molecular diffraction
is about four times as strong as the Bragg diffraction.  This is due to the greater area
of diffraction space that the continuous diffraction covers, compared with the Bragg peaks
that only extend to a resolution of about \SI{5}{\angstrom}.  The molecules in the crystal
scatter the same number of photons whether those molecules are perfectly registered on a
lattice or if they are randomly displaced, and thus the diffraction counts beyond the
cut-off of the Bragg peaks should be similar to the case of the perfect crystal.  In the
large ensemble of patterns the total number of counts in the continuous diffraction is
found to be very strongly correlated with Bragg counts, showing that the strength of both
the continuous and Bragg diffraction depends on the size of the crystal.  Of course, since
those counts are not concentrated into sparse Bragg peaks the signal to background of the
continuous diffraction is lower than for Bragg peaks.  For example, if individual Bragg
peaks fit in a single pixel and were spaced on average by 10 pixels in each direction,
then the counts per pixel in the continuous diffraction would be about 1\% of the
equivalent Bragg signal.  Nevertheless, since individual speckles cover similar areas to
the spacings between Bragg peaks (depending on the size of the rigid unit compared to the
width of the unit cell), the total counts per speckle is similar to the Bragg counts and
the signal to background is found to be almost comparable to the Bragg signal.

The low numbers of photons per pixel in the continuous diffraction is of course one reason
why less attention is paid to it than the easily-measured Bragg peaks.  This also demands
a proper treatment of counting statistics when estimating the contributions due to signal
and background.  For example, in the case of continuous random variable the structureless
background is considered to follow the normal distribution.  This distribution has no
skew, and thus any skew in the distribution of measured intensities is a signature of
structured diffraction signal.  However, the Poisson distribution is skewed due to there
being no negative photon counts, and the skew is significant even at signal levels
approaching 100 counts.  It was found that only the use of discrete statistics gave rise
to a reasonable estimation of the background of individual snapshot patterns whereas the
non-discrete statistics resulted in over-estimation of structured diffraction signal, as
clearly illustrated in Figs.~\ref{fig:estimates} and ~\ref{fig:snapshot-corrected}.  The counting statistics are obviously modified by any scaling of the measured
intensities, such as the correction of the effect of linear X-ray polarisation.  In the
Poisson distribution the variance is equal to the mean, which is not the case if counts
are multiplied by some factor.  Regions of near-equal counts from which to determine
moments of the photon-counting distribution were therefore obtained by averaging the
polarisation corrected pattern in shells of $q$, and then reapplying the polarisation
factor.  Such an approach will not be valid for background due to fluorescence or which is
non-uniformly distributed across the detector face: the conditions of the experiment must
be carefully controlled to avoid such contamination.

We are currently exploring the effectiveness of the approach presented here on continuous diffraction
data recorded from other samples and with different types of detectors.  Until then, it is
premature to release software, but we list in Appendix~\ref{appendix:procedure} the steps
of the procedure used here to obtain the 3D array of continuous diffraction intensities such
as shown in Fig.~\ref{fig:merge} (a).

The fruit of the method presented here is clearly seen in the improved continuous
diffraction intensity maps that are obtained---compare for example Figs.~\ref{fig:merge}
(a) and (b).  The background subtracted from the individual patterns, as well as from the
merged 3D array, are all smooth functions that are rotationally symmetric (apart from the
polarisation factor).  Hence the manipulations do nothing to the speckles other than
locally alter their contrast.  The result shows a very high degree of correlation with the
continuous diffraction calculated from an atomic model, with an overall value of
$\mathrm{CC} = 0.77$, compared with a value of 0.55 previously reported \cite{Ayyer:2016}.
The signal is well distinguished from background even in the case of a strong background
that was more than 25 times the diffraction signal.  Stronger diffraction should arise
from bigger crystals, and it is worth exploring the quality of the continuous diffraction
with crystal size.  Our first demonstration of macromolecular diffractive imaging may have
been on the most challenging samples, but there may be an advantage of collecting diffraction from
small volumes.  Given the relative strength of signal and background the volume of each
crystal was about 4\% on average of the total probed volume of the jet, assuming the
background is due entirely to the jet (and not to pulse-induced disorder of the molecules
\cite{Barty:2012} or conformational variability \cite{Maia:2009}, for example).  The
maximum jet diameter was about \SI{5}{\micron} and the beam diameter 
was 1 to \SI{2}{\micron}, giving a total probed volume of up to about
\SI{20}{\micron\cubed}.  Thus the diffracting crystal volume was on average less than
\SI{1}{\micron\cubed} even though the crystals were visually more than 10 times this
volume.

The analysis presented here may prove useful to studies of protein dynamics, which have
examined continuous diffraction from crystals due to various kinds of differences of the
constituent molecules from the average 
\cite{Doucet:1987,Wall:1997,Perez:1996,VanBenschoten:2016}. Such measurements are 
usually not time resolved and hence cannot
distinguish static from dynamic disorder, but such measurements can be compared with
diffraction calculations based on molecular dynamics trajectories to hopefully gain
insights into protein motion and function. As shown here and in the previous work
\cite{Ayyer:2016} one can establish the origin of the continuous diffraction and account
for the dominant effects (such as translation of rigid units) prior to examining the
effects of correlated motions. For example, the autocorrelation function obtained by a
Fourier transformation of the continuous diffraction intensities reveals the shape and
size of rigid units.  Indeed, previous studies of lysozyme crystals determined from the
speckle size that the rigid unit was the size of a lysozyme molecule \cite{Perez:1996},
and considerations of the mechanical properties of protein crystals lead to the conclusion
that molecular translations and rotations (dependent on the elastic
and shear moduli repectively) in these bodies are inevitable \cite{Morozov:1986}.  Based on
the formalism of Morozov \cite{Morozov:1986}, a molecular displacement of $\sigma_\Delta
= \SI{2}{\angstrom}$ at room temperature for a molecule width of \SI{178}{\angstrom} implies a
Young's modulus of only \SI{0.01}{\giga\pascal}.  This is similar to soft
rubber, which is consistent with experience in handling these crystals.

The statistics of diffraction intensities, and in particular the speckle contrast directly
yields information of the number of independent modes contributing to the diffraction.
Together with the symmetry of the crystal, this can indicate whether such independent
objects correspond to asymmetric units in the crystal. Comparing centric and acentric
sections gives further evidence of the origin of the diffraction.  In all studies of
continuous diffraction it is imperative to accurately measure the diffraction at a
sufficient sampling in all three dimensions, and to remove the the background
that accompanies such measurements.  The approach given here is shown to be 
effective in extracting single-molecule diffraction from patterns arising from
translationally-disordered crystals and it is expected that it may similarly improve
measurements of continuous diffraction due to other kinds of disorder.

Although beyond the scope of this paper, we expect that the improved treatment of the
background should lead to a better structure determination from the continuous diffraction
through iterative phasing.  A better correlation with the simulated diffraction was found
by assuming a \ang{1} rotational disorder of the PS II dimers.  For the \SI{178}{\angstrom} diameter dimer,
this will cause a blurring of the structure at resolutions beyond \SI{3}{\angstrom}, but by using
methods of partially-coherent diffractive imaging \cite{Whitehead:2009} it should be possible
to take such blurring into account, to a resolution where speckles are no longer visible.
That the speckles do indeed appear visible to the edge of the detector at \SI{2}{\angstrom} resolution
suggests that with more measurements, structural information should be obtainable to at
least that resolution.

\clearpage
\section*{Appedix A: The ``noisy Wilson'' distribution}
\label{appendix:noisy-Wilson}
We consider non-discrete acentric diffraction intensities incoherently summed with
a background $I_B$ that is normally distributed with a mean $\mu$ and variance
$\sigma^2$, $I_B \sim \mathcal{N}(\mu,\sigma)$, with
\begin{equation}
  \label{eq:normal}
  p_B(I_B) = \frac{1}{\sqrt{2\pi}\sigma}\,\exp \left(-\frac{(I_B-\mu)^2}{2\sigma^2} \right).
\end{equation}
We refer to the distribution of this sum $I \sim \mathrm{Gamma}(N,\Sigma/N) +
\mathcal{N}(\mu,\sigma)$ as the
noisy Wilson distribution, with a probability distribution function found by
convolution of Eqns. (\ref{eq:3}) and (\ref{eq:normal}) as
% Use something like this for the non-preprint version
% \begin{multline}
%   \label{eq:pNW}
%   p_{NW}(I;N)=\frac{2^{(N-3)/2}
%     N^N}{\sqrt{\pi}\Gamma(N)}\,\frac{\sigma^{N-1}}{\Sigma^N}\,\exp\left(
%     -\frac{(I_B-\mu)^2}{2\sigma^2}\right)\\
%   h_N \left(\frac{N\sigma^2
%       -(I-\mu)\Sigma}{\sqrt{2} \sigma \Sigma} \right)
% \end{multline}
\begin{equation}
  \label{eq:pNW}
  p_{NW}(I;N)=\frac{2^{(N-3)/2}
    N^N}{\sqrt{\pi}\Gamma(N)}\,\frac{\sigma^{N-1}}{\Sigma^N}\,\exp\left(
    -\frac{(I_B-\mu)^2}{2\sigma^2}\right) \,
  h_N \left(\frac{N\sigma^2
      -(I-\mu)\Sigma}{\sqrt{2} \sigma \Sigma} \right)
\end{equation}
where
% Use something like this for the non-preprint version
% \begin{multline}
%   \label{eq:hN}
%   h_N(x)=\Gamma\left(\frac{N}{2}\right) {}_1F_1 \left(\frac{N}{2},\frac{1}{2},x^2
%   \right) \\
%   -2x\Gamma \left(\frac{N+1}{2}\right) {}_1F_1 \left(\frac{N+1}{2},\frac{3}{2},x^2\right)
% \end{multline}
\begin{equation}
  \label{eq:hN}
  h_N(x)=\Gamma\left(\frac{N}{2}\right) {}_1F_1 \left(\frac{N}{2},\frac{1}{2},x^2
  \right) 
  -2x\Gamma \left(\frac{N+1}{2}\right) {}_1F_1 \left(\frac{N+1}{2},\frac{3}{2},x^2\right)
\end{equation}
and ${}_1F_1$ is the confluent hypergeometric function.  For $N = 1$, 2, 4, 6, and 8, $h_N$ evaluates to
\begin{subequations}
\label{eq:h}
  \begin{align}
   &  h_1(x)=2 F_H(x) \\
   & h_2(x)=1 - 2x F_H(x) \\
   & h_4(x)=1+x^2-x(2x^2+3)F_H(x) \\
   & h_6(x)=\frac{1}{2}\left(4+9x^2+2x^4-x(15+20x^2+4x^4) F_H(x) \right ) \\
   & h_8(x)=\frac{1}{4}\Bigl(24+87x^2+40x^4+4x^6 \notag \\
   &  \qquad \qquad -x(105+210x^2+84x^4+8x^6)F_H(x) \Bigr)
  \end{align}
\end{subequations}
where $F_H(x)$ is proportional to the scaled complementary error function,
% Use something like this for the non-preprint version
% \begin{multline}
%   \label{eq:FH}
%   F_H(x)=\exp(x^2)\int_x^\infty \exp(-y^2)\,dy \\
%   =\frac{\sqrt{\pi}}{2} \exp(x^2) \left(1-\mathrm{Erf}(x)\right).
% \end{multline}
\begin{equation}
  \label{eq:FH}
  F_H(x)=\exp(x^2)\int_x^\infty \exp(-y^2)\,dy 
  =\frac{\sqrt{\pi}}{2} \exp(x^2) \left(1-\mathrm{Erf}(x)\right).
\end{equation}
Some plots of $p_{NW}(I)$ are given in Figs.~\ref{fig:distributions} (c) and (d).

Equations (\ref{eq:pNW}) and (\ref{eq:hN}) can be evaluated for non-integer values of $N$,
such as needed to account for partial coherence, by computing the series expansion of the
confluent hypergeometric function.

\section*{Appendix B: Procedure for processing diffraction data}
\label{appendix:procedure}
The following itemises the steps taken to process still diffraction patterns of crystals
recorded in random orientations or a series of orientations.  Here it is assumed that the
patterns are recorded with a common (unchanging) detector geometry and wavelength, the
incident beam is linearly polarised, and the unstructured background is radially symmetric
when corrected by the polarisation factor.  The procedure also assumes intensity data in units
of photon counts, either obtained using a
well-calibrated integrating detector or a photon counting detector.  
\begin{enumerate}
%  \item Calculate the polarization correction $P(k_x, k_y)$.
  \item Process the dataset using CrystFEL \cite{White:2012} to find the indexable
    patterns and the lattice orientation for each indexed pattern, as well as to create a
    set of merged Bragg intensities\label{itm:CrystFEL}.
  \item For each indexed pattern (calibrated and bad areas of the detector masked, but
    otherwise uncorrected for polarisation):
    \begin{enumerate}
      \item Mask Bragg peaks using a threshold filter or the \emph{peakfinder8} procedure
        from \emph{Cheetah} \cite{Barty:2014}. Dilate the mask by a kernel about 7 pixels
        wide. The intensities in the masked pixels
        are not included in any further analysis.  
        \item Create contours of $I_{av}$, Eqn.~(\ref{eq:polarisation}), at a spacing of about 1 photon count, and
          from these determine contiguous regions bounded by those contours.  
        \item Calculate the moments (mean and variance) of photon counts in each
          contiguous region to compute the parameters of the discrete noisy Wilson
          distribution $\bar{\mu}$, $\bar{\Sigma}$ from Eqns.~(\ref{eq:solutions-2}) with
          the appropriate value of $N$ (here $N=4$).
        \item Set the background $\mu(k_x, k_y)$ to values $\bar{\mu}$ for each region,
          then smooth with a square kernel of about 3 pixels width.\label{itm:background}
        \item Determine a scaling array $\Sigma(k_x, k_y)$ in a similar way to step
          \ref{itm:background}. \label{itm:scale} 
        \item Subtract the smoothed background of step \ref{itm:background} from the pattern.
        \item Normalise the pattern by dividing by the sum of $\Sigma(k_x, k_y)$ from step
          \ref{itm:scale} over a predetermined range of $|\boldsymbol{k}|$.
        \item Correct for polarisation by dividing by the polarisation factor $P(k_x,
          k_y)$, Eqn.~\ref{eq:polarisation}.
        \end{enumerate}
      \item Merge the corrected patterns into a 3D reciprocal space volume by
        interpolating each onto the appropriate Ewald sphere in the frame of reference of
        the crystal lattice \cite{Yefanov:2014}.  The spacing of voxels in the 3D array should be chosen to
        sufficiently sample the highest frequencies of the continuous diffraction,
        $\Delta q < 1/(2 w)$ where $w$ is the width of the rigid body.  
      \item Apply the symmetry operations of the point group of the diffraction to the 3D
        array (here, point group 222).
      \item Calculate moments of the scaled intensities in shells of $q$ from the 3D array
        and calculate parameters of the noisy Wilson distribution $\mu$, $\sigma$, and
        $\Sigma$ in those shells via Eqns.~(\ref{eq:solutions-1}), avoiding centric
        zones\label{itm:noisy-Wilson}.
      \item Construct $\mu(\boldsymbol{q})$ for all voxels of the 3D array, smooth it with
        a three-dimensional kernel of about 3 pixels wide, and then subtract this from the
        merged intensities, to account for residual background.
      \item Determine scaling and background of the merged Bragg intensities (from step
        \ref{itm:CrystFEL}) in shells of $q$ by applying the same procedure as per step
        \ref{itm:noisy-Wilson} but with the appropriate value of $N$ (here $N=1$). Fit the
        Debye-Waller factor to the values $\Sigma(q)$ for these Bragg intensities to obtain $\sigma_\Delta$.
      \item Correct the continuous diffraction intensities by dividing by the
        complementary Debye-Waller factor $1-\exp(-4 \pi^2 \sigma_\Delta^2 q^2)$.
\end{enumerate}

\section*{Acknowledgements}
We acknowledge support through program oriented funds of the Helmholtz Association to DESY
and through the Gottfried Wilhelm Leibniz Program of the DFG.  We thank Petra Fromme and Rick
Kirian for motivating discussions

\bibliographystyle{ieeetr}
\bibliography{chapman-stats.bib}

\end{document}